\documentclass[preprint,onecolumn,nofootinbib]{revtex4-1}
\pdfoutput=1

\usepackage[colorlinks=true,linkcolor=blue,urlcolor=blue,filecolor=black,citecolor=red,pdfstartview=FitV,pdftitle={},pdfsubject={},pdfkeywords={},pdfpagemode=None,bookmarksopen=true]{hyperref}
\usepackage{graphicx}
\usepackage{subfigure}
\usepackage{amsmath}
\usepackage{amsfonts}
\usepackage{amssymb,ulem}
\usepackage{color}%
\usepackage{dcolumn}
\usepackage[utf8]{inputenc}
\newcommand{\f}{\begin{equation}}
\newcommand{\ff}{\end{equation}}
\newcommand{\fa}{\begin{eqnarray}}
\newcommand{\ffa}{\end{eqnarray}}

\begin{document}
\title{Chaotic motion of the charged test particle in a Kerr-MOG black hole with explicit symplectic algorithms}
\author{Zhen-Meng Xu$^{1}$}
\author{Da-Zhu Ma$^{2}$}
\thanks{mdzhbmy@126.com, Corresponding author}
\author{Wen-Fu Cao$^{3}$}
\author{Kai-Li$^{1}$}
\affiliation{\small{$^1$~School of Mathematics and Statistics, Hubei Minzu University, Enshi 445000, China}}
\affiliation{\small{$^2$~College of Intelligent Systems Science and Engineering, Hubei Minzu University, Enshi 445000, China}}
\affiliation{\small{$^3$~School of Physics and Technology, University of Jinan, Jinan, 250022, China}}

\begin{abstract}
The Kerr-MOG black hole has recently attracted significant research attention and has been extensively applied in various fields. To accurately characterize the long-term dynamical evolution of charged particles around Kerr-MOG black hole, it is essential to utilize numerical algorithms that are high-precision, stable, and capable of preserving the inherent physical structural properties. In this study, we employ explicit symplectic algorithms combined with the Hamiltonian splitting technique to numerically solve the equations of motion for charged particles. Initially, by decomposing the Hamiltonian into five integrable components, three distinct explicit symplectic algorithms ($S2$, $S4$, and $PR{K_6}4$) are constructed.  Numerical experiments reveal that the $PR{K_6}4$ algorithm achieves superior accuracy. Subsequently, we utilize Poincaré sections and the Fast Lyapunov Indicator (FLI) to investigate the dynamic evolution of the particle. Our numerical results demonstrate that the energy $E$, angular momentum $L$, magnetic field parameter $\beta$, black hole spin parameter $a$, and MOG parameter $\alpha$ all significantly influence the particle's motion. Specifically, the chaotic region expands with increases in $E$, $\beta$, or $\alpha$, but contracts with increases in $a$ or $L$. Furthermore, when any two of these five parameters are varied simultaneously, it becomes evident that $a$ and $L$ predominantly dictate the system's behavior. This study not only offers novel insights into the chaotic dynamics associated with Kerr-MOG black holes but also extends the application of symplectic algorithms in strong gravitational field.
\end{abstract}
\maketitle

\section{Introduction}
It is widely recognized that Einstein’s general theory of relativity, a cornerstone of modern physics, has successfully elucidated numerous astronomical and gravitational phenomena. Nevertheless, it faces limitations in explaining key aspects of dark matter and dark energy, unifying quantum mechanics with gravity, resolving the singularities of black holes, and accounting for the origin of the universe’s accelerated expansion and the formation of large-scale structures. In light of these limitations, various modified theories have been proposed to extend the applications of general relativity by introducing new field variables or modifying its equations. These include the Modified Newtonian Dynamics (MOND) theory, the Scalar-Tensor-Vector Gravity (STVG) theory, and other alternative frameworks. For instance, Camilloni et al. \cite{Camilloni:2023wyn} investigated magnetospheric models and black hole jet emissions within the context of Modified Gravity (MOG). Additionally, Jai-akson et al. \cite{Jai-akson:2017ldo} examined the potential characteristics of black hole mergers in the framework of Einstein-Maxwell-dilaton theory. 

As is well known, over 70\% of the universe is composed of dark energy, which serves as the primary driver behind the accelerated expansion of the universe \cite{Caldwell2009,Peebles:2002gy}. In addition, dark matter is estimated to account for at least 20\% of the total matter content in the universe. Despite extensive efforts, no direct evidence of dark matter has been found. However, the existence of dark matter can be inferred from various phenomena, including galaxy rotation curves, gravitational lensing, and cosmic microwave background radiation. This has irritated the development of alternative theories that attempt to explain astronomical observations without the addition of dark matter. Milgrom \cite{Milgrom:1983ca} first proposed the MOND theory, which modifies the laws of gravity to account for astrophysical phenomena without the need for dark matter. MOND has successfully explained galaxy rotation curves and the dynamics of low-surface-brightness galaxies. However, its applicability on larger scales, such as galaxy clusters and cosmological contexts, remains limited, preventing it from fully replacing dark matter models. Subsequently, Moffat \cite{Moffat:2005si} proposed an alternative theory that does not assume the existence of dark matter, known as the STVG theory, or namely the MOG theory. This framework has been widely employed to address gravitational phenomena that cannot be fully explained by standard general relativity or conventional dark matter models. Under the assumption that dark matter does not exist in the universe, the STVG theory not only accounts for the dynamics of galaxies \cite{Moffat:2013sja,Moffat:2013uaa,Brownstein:2005dr,Moffat:2014pia}, but also provides explanations for phenomena within the solar system, the growth of large-scale structures, and the cosmic microwave background (CMB) sound power spectrum data \cite{Moffat:2014bfa,Moffat:2014asa}. In the STVG theory, the introduction of a vector field and a variable gravitational constant modifies the gravitational action, leading to non-standard gravitational effects experienced by objects such as particles orbiting near black holes. Moffat \cite{Moffat:2015kva} demonstrated that the parameter $\alpha$ increases the shadow radius of MOG black holes. Lee and Han \cite{Lee:2017fbq} found that $\alpha$ influences the radius of the innermost stable circular orbit (ISCO) in Kerr-MOG black holes. Guo et al. \cite{Guo:2018kis} investigated the shadows cast by near-extremal Kerr-MOG black holes for varying values of the MOG parameter. Additionally, Hou et al. \cite{Hou:2022eev} explored the observational signatures of Kerr-Melvin black hole illuminated by accretion disks, revealing that the inner shadow and critical curves can be used to infer the presence of a non-degenerate magnetic field around the black hole. 

Moreover, dynamics of the charged and neutral particles in the vicinity of black holes constitutes a significant area of research in relativistic astrophysics. For instance, Battista and Esposito \cite{Battista:2022krl} conducted a comprehensive investigation into geodesic motion within Euclidean Schwarzschild geometry. Dutta et al. \cite{Dutta:2024rta} explored the dynamics of circular geodesics (for chargeless massive particles) and pulsating classical strings in the p-brane backgrounds. Dalui et al. \cite{Dalui:2018qqv,Dalui:2019umw} demonstrated that the presence of an event horizon induces chaotic behavior in the motion of particles. The Kerr-MOG black hole, a predicted solution in the MOG theory, significantly modifies the geometric structure of the classical Kerr spacetime by introducing an additional gravitational coupling parameter 
$\alpha$ \cite{Moffat:2015}. Recent observations of the shadow of Sgr ${A^ * }$ by the Event Horizon Telescope (EHT) have revealed a remarkable alignment with the predictions of General Relativity (GR), thereby placing stringent constraints on modified parameters such as $\alpha$ \cite{Vagnozzi:2023}. Within this modified framework, the motion of charged particles becomes exceedingly complex due to the nonlinear interplay between gravitational and electromagnetic fields, requiring advanced numerical methods to resolve chaotic dynamics.

Numerical simulations of chaotic motion for charged particles in Kerr-MOG spacetime encounter two primary challenges. At first, the highly nonlinear nature of the equations of motion renders analytical solutions unattainable. Secondly, traditional numerical methods often suffer from energy error accumulation during long-term integration, resulting in pseudo-chaotic phenomena \cite{MA2008216}. In order to solve these problems, symplectic algorithms have emerged as the preferred method for conservative systems due to their inherent preservation of system structure \cite{Sanz-Serna_1992,Marsden_West_2001}. 
Symplectic algorithms rigorously preserve energy and momentum conservation, thereby avoiding pseudo-chaotic phenomenon caused by energy drift in long-term integrations that plague traditional methods like Runge-Kutta methods \cite{SanzSerna:1994}. 
The symplectic algorithm introduced by Feng \cite{feng1984difference} represents a high-order implicit symplectic method rooted in the implicit midpoint approach, particularly well-suited for non-separable Hamiltonian systems. Meanwhile, Ruth \cite{Ruth:1983TNS} proposed an explicit symplectic algorithm that divides the Hamiltonian system into kinetic energy $T$ and potential energy $V$. Extensive research has demonstrated that explicit symplectic algorithms are generally incompatible with inseparable Hamiltonian systems, rendering implicit symplectic algorithms more appropriate for such cases. Implicit symplectic algorithms contain semi-implicit symplectic algorithms and implicit midpoint methods \cite{1996CeMDA..66..243L,SUZUKI1990319,PhysRevD.81.084045,PhysRevD.81.104037,PhysRevD.82.124040,2011GReGr..43.2185W,10.1093/mnras/stt1441,2013EPJC...73.2413M}. Despite their versatility in applying to any Hamiltonian system, implicit symplectic algorithms tend to be less computationally efficient compared with the explicit symplectic algorithms \cite{2013EPJC...73.2413M,Preto:2009cp}.
The explicit symplectic algorithms achieve iterative schemes by decomposing the Hamiltonian into an integrable component ${H_A}$ and a perturbative component ${H_B}$, and employing symmetric operators ($\Phi_h = e^{hL_{H_A}/2} \circ e^{hL_{H_B}/2} \circ e^{hL_{H_A}/2}$, where ${L_H}$ is the Lie derivative of the Hamiltonian, and $h$ is the step size) \cite{McLachlan_Quispel_2002}. Since then, the explicit symplectic algorithms have experienced rapid development. The force-gradient symplectic algorithm \cite{Wisdom:1991kb} adeptly circumvents the emergence of negative time coefficients through the incorporation of a force gradient operator. Ruth \cite{Ruth:1983TNS} and Chin \cite{CHIN1997344,2005CeMDA..91..301C} proposed the third-order and fourth-order force gradient symplectic algorithms, respectively. Sun et al. \cite{2011RAA....11..353S} developed a symplectic integrator incorporating third-order potential derivatives. Recently, Wu et al. \cite{Wu:2024ehd} have constructed an explicit symplectic integrator featuring adaptive time steps in curved spacetime, offering substantial utility for the study of Hamiltonian with anticipated separations.

In the realm of constructing symplectic integrators within black hole spacetimes, significant advancements have been made by various researchers. For example, Wang et al. \cite{Wang:2021gja,Wang:2021xww,Wang:2021yqk} constructed explicit symplectic integrators for Schwarzschild, Reissner-Nordström, and Reissner–Nordström-(anti)-de sitter black holes by dividing Hamiltonian systems into multiple integrable sub-Hamiltonian systems. In addition, Wu et al. \cite{Wu:2021rrd} introduced an explicit symplectic integrator for Kerr black hole, employing time transformation techniques as outlined in \cite{1997CeMDA..67..145M}. These contributions have significantly heightened interest in the construction of explicit symplectic integrators in curved spacetimes. Subsequent studies have leveraged these integrators to explore complex dynamical behaviors. Hu and Huang \cite{universe8070369} applied explicit symplectic integrators to investigate chaotic phenomena in a magnetized Brane-World spacetime. Zhou et al. \cite{2023AcASn..64...39Z} examined the chaotic motion of charged particles in a magnetized Schwarzschild black hole, while Cao et al. \cite{Cao:2024ihv} delved into the electromagnetic fields and chaotic dynamics of charged particles around hairy black holes in Horndeski gravity. Additionally, Lu and Wu \cite{Lu:2024srb} explored the influence of two quantum correction parameters on the chaotic dynamics of particles near modified Schwarzschild black hole under the renormalization group framework. These research works indicate that applying the symplectic algorithms to black hole models is feasible, which not only provides us with powerful numerical integration tools, but also helps to deepen our understanding of complex gravitational phenomena. Based on high-precision numerical solutions, we can discuss the long-term dynamic evolution of particles around the event horizons of black hole, as well as calculate the orbital dynamics of photons and the shadow of the black hole \cite{Wu:2024ehd}.

To investigate the chaotic dynamics in Kerr-MOG black hole, it is crucial to prioritize the accuracy and stability of numerical methods. Fortunately, the time transformation method offers a valuable tool for converting the Hamiltonian of Kerr-MOG black hole into a new form. This transformed Hamiltonian can be decomposed into several integrable components, making the explicit symplectic algorithms could be constructed. These algorithms generate high-precision numerical solutions, allowing for an accurate description of chaotic dynamics, as demonstrated in \cite{Ma:2014aha}. This paper is organized as follows, the description of the Hamiltonian system is presented in Section \ref{section2}, the construction of three explicit symplectic algorithms are given in Section \ref{section3}, effects of varying parameters on the motion of the charged particle near Kerr-MOG black hole are discussed in Section \ref{section4}, the final conclusions are shown in Section \ref{section5}.

\section{The black hole model}\label{section2}

In this section, we derive the black hole solution in the context of Kerr-MOG, which is based on the MOG theory. Next, we utilize the electromagnetic four-potential to obtain the external asymptotically uniform magnetic field surrounding the black hole. Finally, we describe the motion of a charged particle around Kerr-MOG black hole using Hamiltonian formalism.

\subsection{The Kerr-MOG black hole solution}

The action of MOG in the STVG framework is given by \cite{Mureika:2015sda}:
\begin{equation}
	S = {S_G} + {S_\phi } + {S_S} + {S_M},
\end{equation}
where the individual components are defined as:
\begin{eqnarray}
	&&
	S_G=\frac{1}{{16\pi }}\int {\frac{R}{G}} \sqrt { - g} {d^4}x,
	\
	\\
	&&
	S_\phi = - \frac{1}{{4\pi }}\int {\left[ {{{\cal K}} + {V({\phi _\mu })}} \right]} \sqrt { - g} {d^4}x,
	\
	\\
	&&
	S_S = \int \frac{1}{G} \Bigg[ \frac{1}{2} g^{\alpha \beta} \left( \frac{\nabla_\alpha G \nabla_\beta G}{G^2} + \frac{\nabla_\alpha \mu \nabla_\beta \mu}{\mu^2} \right)\nonumber\\
	\nonumber
	\\
	&&
	- \frac{V_G(G)}{G^2} - \frac{V_\mu(\mu)}{\mu^2} \Bigg] \sqrt{-g} \, d^4x.
	\
	\\
	&&
	S_M =  - \int {(\rho \sqrt {{u^\mu }{u_\mu }}  + Q{u^\mu }{\phi _\mu } + {J^\mu }{\phi _\mu })} \sqrt { - g} {d^4}x.
\end{eqnarray}

The gravitational action $S_G$ is a modified Einstein-Hilbert action where the gravitational coupling $G(x)$ is treated as a dynamical scalar field. For simplicity, we assume $G$ to be constant across spacetime coordinates and adopt the relation between $G$ and the Newton’s constant ${G_N}$:
\begin{eqnarray}
	G = G_N(1 + \alpha ),
\end{eqnarray}
where $\alpha$ is a dimensionless parameter quantifying deviations from general relativity. When $\alpha=0$, $G$ reduces to ${G_N}$. Here $R = {g^{\mu \nu }}{R_{\mu \nu }}$ denotes the Ricci scalar.

The vector field  action $S_\phi$ governs the dynamics of the Proca vector field ${\phi _\mu }$, where $\mu(x)$ denotes its position-dependent mass. The kinetic term ${{\cal K}}$, analogous to the electromagnetic field strength tensor in Maxwell’s theory, is defined as:
\begin{eqnarray}
	{{\cal K}} = \frac{1}{4}{B^{\mu \nu }}{B_{\mu \nu }},
\end{eqnarray}
where ${B_{\mu \nu }} = {\partial _\mu }{\phi _\nu } - {\partial _\nu }{\phi _\mu }$ represents the antisymmetric field strength tensor of $J^\mu = \kappa \rho_M u^\mu$. In addition, ${V({\phi _\mu })}$ is a potential term.

The scalar field action $S_S$ describes the dynamics of the scalar fields $G$ and $\mu$, including their self-interaction potentials $V_G(G)$ and $V_\mu(\mu)$. The field $G$ modulates gravitational strength, while $\mu$ determines the effective range of the vector field (e.g., $\mu \to 0$ yields a long-range force). For simplicity, we consider $\mu$ to remain constant in this study.

The matter action $S_M$ encodes interactions between matter and the vector field ${\phi _\mu }$. Here, ${u^\mu } = d{x^\mu }/d\tau$ is the four-velocity with $\tau$ being the proper time, $J^\mu = \kappa \rho_M u^\mu$ with $\rho_M$ being the density of matter and $\kappa=\sqrt {\alpha {G_N}}$. Moreover, $Q$ is the effective gravitational charge that couples the vector field ${\phi _\mu }$ with matter. At the weak field limit, the gravitational acceleration of STVG is corrected to $a(r)=-\frac{\mathcal{G}(r)M}{r^2}$, where the effective gravitational coupling strength $\mathcal{G}(r)=G_N\begin{bmatrix}1+\alpha-\alpha e^{-\mu r}(1+\mu r)\end{bmatrix}$. When the distance $r$ is much greater than the range of action of the vector field ($\mu r\gg1$), $\mathcal{G}(r)$ degenerates into $G = G_N(1 + \alpha )$. By matching and correcting the form of gravity and Newtonian acceleration, the coupling strength of the vector field is quantified as parameter $\alpha$, and $Q = \sqrt {\alpha {G_N}} M$ is derived.

By setting the potentials to zero ($V(\phi _\mu )= V_G(G)= V_\mu(\mu )=0$) and neglecting matter (${T_{M\mu \nu }}=0$), the field equations reduce to:
\begin{eqnarray}
	&&
	G_{\mu \nu } =  - 8\pi G T_{\phi \mu \nu },
	\label{eq:2}
	\\
	&&
	{\nabla_\nu }{B^{\mu \nu }} = 0,
	\label{eq:4v1}
	\\
	&&
	{\nabla _\sigma }{B_{\mu \nu }} + {\nabla _\mu }{B_{\nu \sigma }} + {\nabla _\nu }{B_{ \sigma \mu }} = 0,
	\label{eq:4}
\end{eqnarray}
where the energy-momentum tensor for ${\phi _\mu }$ is:
\begin{eqnarray}
	\label{eq:3}
	{T_{\phi \mu \nu }} =  - \frac{1}{{4\pi }}({B_\mu }^\alpha {B_{\nu \alpha }} - \frac{1}{4}{g_{\mu \nu }}{B^{\alpha \beta }}{B_{\alpha \beta }}).
\end{eqnarray}
The Kerr-like metric in modified gravity \cite{Moffat:2015}, derived from the gravitational field equations, takes the following form in Boyer-Lindquist coordinates $(t,r,\theta ,\varphi )$:
\begin{eqnarray}\label{9}
	d{s^2} = {g_{tt}}d{t^2} + {g_{rr}}d{r^2} + {g_{\theta \theta }}d{\theta ^2} + {g_{\varphi \varphi}}d{\varphi ^2} + 2{g_{t\varphi }}dtd\varphi. 
	\nonumber
	\\
\end{eqnarray}
with metric components:
\begin{eqnarray}\label{10}
	{g_{tt}} =  - (\frac{{\Delta  - {a^2}si{n^2}\theta }}{\Sigma }),{g_{rr}} = \frac{\Sigma }{\Delta },{g_{\theta \theta }} = \Sigma ,
\end{eqnarray}
\begin{eqnarray}\label{11}
	{g_{\varphi \varphi }} &=&\frac{{si{n^2}\theta }}{\Sigma }[{({a^2} + {r^2})^2} - {a^2}si{n^2}\theta \Delta ],\nonumber\\
	{g_{t\varphi  }} &=& \frac{{asi{n^2}\theta }}{\Sigma }[\Delta  - ({a^2} + {r^2})],
\end{eqnarray}
where,
\begin{eqnarray} \label{12}
	\Delta  &=& {r^2} - 2GMr + {a^2} + \alpha G{G_N}{M^2},\nonumber\\
	\Sigma  &=& {r^2} + {a^2}{\cos ^2}\theta.
\end{eqnarray}

The spacetime is characterized by mass $M$, spin $a$, and MOG parameter $\alpha$. The ADM mass ${M_\alpha }$ relates to $M$ via ${M_\alpha } = (1 + \alpha )M$. The horizons in the spacetime of Kerr-MOG black hole are identified by the roots of the equation,
\begin{equation}
	\Delta  = {r^2} - 2{M_\alpha }r + {a^2} + \frac{\alpha }{{(1 + \alpha )}}{M_\alpha }^2 = 0.
\end{equation}
As a result,
\begin{equation}
	{r_ \pm } = {M_\alpha } \pm \sqrt {\frac{{{M_\alpha }^2}}{{(1 + \alpha )}} - {a^2}}.
\end{equation}

Moreover, the extremal limit for Kerr-MOG black hole can be written as ${M_\alpha }^2 = (1 + \alpha ){a^2}$. The same as in \cite{Khan:2023nul}, by using the dimensionless processing $r \to r / G_N M$, $a \to a / G_N M$, $\Delta $ can be simplified to 
\begin{equation}
	\Delta  = {r^2} - 2(1 + \alpha )r + {a^2} + \alpha (1 + \alpha ).
\end{equation}

\subsection{Electromagnetic Four-Potential}
In rotating spacetime, the vector potential can be further generalized as follows,
\begin{equation}\label{17}
	{\phi _\mu } = \frac{{\sqrt \alpha  Mr}}{\Sigma }( - 1,0,0,asi{n^2}\theta ).
\end{equation}

An external asymptotically uniform magnetic field $B$ surrounding Kerr-MOG black hole is assumed to be perpendicular to the equatorial plane $\theta=\pi  /2$. The same as in \cite{Khan:2023nul}, an electromagnetic four-potential with two nonzero covariant components was proposed by using Wald’s method \cite{PhysRevD.10.1680},

\begin{eqnarray}\label{18}
	{A_t} = \frac{B}{2}{g_{t\varphi }} - \frac{{{Q_W}}}{2},{A_\varphi } = \frac{B}{2}{g_{\varphi \varphi }}.
\end{eqnarray}
Here, we set ${Q_W} = 2aBM$. Based on Eq. \eqref{11}, Eq. \eqref{18} can be simplified to

\begin{eqnarray}\label{19}
	{A_{\rm{t}}} &=& \frac{{aBsi{n^2}\theta [\Delta  - ({a^2} + {r^2})]}}{{2\Sigma }} - aB,\nonumber\\
	{A_\varphi } &=& \frac{{Bsi{n^2}\theta [{{({a^2} + {r^2})}^2} - {a^2}\Delta si{n^2}\theta ]}}{{2\Sigma }}.
\end{eqnarray}

\subsection{Hamiltonian system}

Consider a charged particle with charge $q$ orbiting a Kerr-MOG black hole in the presence of an external magnetic field in Eq. \eqref{19}, where we set ${p^\mu }{p_\mu } =  - {m^2}$. By employing the method of variable separation, the motion of the charged particle, as described by the Hamilton–Jacobi equation corresponding to the metric in Eq. \eqref{9}, can be analyzed \cite{Khan:2023nul}.

\begin{eqnarray}\label{20}
	{{H}} &=&  - \frac{{\partial S}}{{\partial \tau }}\nonumber \\
	&=& \frac{1}{2}{g^{\mu \nu }}({p_\mu } - q{A_\mu } + \widetilde q{\phi _\mu })({p_\nu } - q{A_\nu }+ \widetilde q{\phi _\nu })\nonumber\\
	&+& \frac{1}{2}{m^2}.
\end{eqnarray}
Here, $\widetilde q = \sqrt \alpha  m$ and ${m^2}$ represent gravitational test particle charge and the rest mass, respectively. ${p_\mu } = {{\partial S} \mathord{\left/{\vphantom {{\partial S} {\partial {x^\mu }}}} \right.\kern-\nulldelimiterspace} {\partial {x^\mu }}}$ is the four-momentum of particles. There are two motion constants, they are the energy and angular momentum of the test particle.
\begin{eqnarray}
	-{{E}} &=& {p_t} = {g_{tt}}\dot t + {g_{t\varphi }}\dot \varphi  + q{A_t} + \widetilde q{\phi _t},\\
	{{L}} &=& {p_\varphi } = {g_{t\varphi }}\dot t + {g_{\varphi \varphi }}\dot \varphi  + q{A_\varphi } + \widetilde q{\phi _\varphi }.
\end{eqnarray}
Where, the dot is the derivative with respect to the proper time $\tau$. The Hamiltonian in Eq. \eqref{20} is modified as follows,
\begin{eqnarray}\label{23}
	{{H}} = {{{H}}_p}(r,\theta ) +\frac{\Delta }{2\Sigma }p_r^2 + \frac{1}{2\Sigma }p_\theta ^2,
\end{eqnarray}
\begin{eqnarray}\label{24}
	{{{H}}_p}(r,\theta ) &=& \frac{1}{2}[{g^{tt}}{({{E}} + q{A_t} -\widetilde q{\phi _t})^2} \nonumber \\
	&+& {g^{\varphi \varphi }}{({{L}} - q{A_\varphi } + \widetilde q{\phi _\varphi })^2} \nonumber \\
	&-& 2{g^{t\varphi }}({{E}} + q{A_t} -\widetilde q{\phi _t})({{L}} -q{A_\varphi } + \widetilde q{\phi _\varphi }) \nonumber \\
	&+&1].
\end{eqnarray}
${{{H}}_p}(r,\theta )$ denotes the potential component of the Hamiltonian. Meanwhile, the motion of the charged test particle can be limited by the energy boundaries defined by $H=0$. Substituting Eqs. \eqref{10}-\eqref{12} and Eq. \eqref{17} into Eqs. \eqref{23} and \eqref{24}, the form of the Hamiltonian in the magnetized Kerr-MOG black hole can be explicitly expressed as
\begin{equation}\label{25}
	\begin{split}
		H_p(r,\theta) &= \frac{1}{2} \biggl\{ 1 + \frac{1}{\Delta [a^2 + 2r^2 + a^2 \cos(2\theta)]^3} \biggl[ \\
		&\quad \begin{aligned}
			& \Bigl\{ a^2 + 2\bigl[ r^2 - 2r(1+\alpha) + \alpha(1+\alpha) \bigr] + a^2 \cos(2\theta) \Bigr\} \\
			& \quad \times \Bigl[ a^2 \beta \Delta - \bigl( a^4 \beta + 2a^2 \beta r^2 + \beta r^4 - 2ar\alpha \\
			& \qquad - 2a^2 L \cot^2\theta \bigr) \csc^2\theta + 2L r^2 \csc^4\theta \Bigr]^2 \sin^6\theta \\
			& + 2 \Bigl[ -2r(Er - a\beta r + \alpha) + 2a^2(-E + a\beta)\cos^2\theta \\
			& \qquad + a\beta(2r - \alpha)(1+\alpha)\sin^2\theta \Bigr]^2 \\
			& \quad \times \Bigl[ - (a^2 + r^2)^2 + a^2 \Delta \sin^2\theta \Bigr] \\
			& + 4a(1+\alpha)(-2r + \alpha) \Bigl[ -2r(Er - a\beta r + \alpha) \\
			& \qquad + 2a^2(-E + a\beta)\cos^2\theta + a\beta(2r - \alpha)(1+\alpha)\sin^2\theta \Bigr] \\
			& \quad \times \Bigl[ 2L r^2 + 2a^2 L \cos^2\theta - (a^4 \beta + 2a^2 \beta r^2 \\
			& \qquad + \beta r^4 - 2ar\alpha) \sin^2\theta + a^2 \beta \Delta \sin^4\theta \Bigr]
		\end{aligned} \biggr] \biggr\}.
	\end{split}
\end{equation}

Here, $\beta  = qB$.

\section{Construction of the explicit symplectic integrators} \label{section3}

In this section, we first provide a concise overview of symplectic algorithms, followed by the construction of the explicit symplectic algorithms for Kerr-MOG black hole.

\subsection{The symplectic algorithms}

Hamiltonian systems are described by generalized coordinates 
$q$ and generalized momenta $p$, and their equations of motion are determined by the Hamiltonian function $H\left( {p,q} \right)$. The construction of symplectic algorithms begins with the discretization of Hamiltonian systems, and thus, the core of symplectic algorithms lies in designing symplectic difference schemes. The common symplectic difference schemes include explicit symplectic methods (such as the explicit Euler methods and explicit Runge-Kutta methods) and implicit symplectic methods (such as the implicit midpoint methods). These schemes employ specific numerical discretization techniques to ensure that the symplectic structure is preserved at each iteration step.

For a harmonic oscillator model with one degree of freedom, the Hamiltonian is
\begin{eqnarray}\label{50}
	H\left( {p,q} \right) = \frac{{{p^2} + {q^2}}}{2}.
\end{eqnarray}
It is easy to obtain the differential equation of motion $\dot q = p$, $\dot p =  - q$.
If we use the first-order Euler method, the iterative formula from step $n-1$ to step $n$ is 
\begin{eqnarray}\label{51}	
	{q_n} &=& {q_{n - 1}} + h  \cdot {p_{n - 1}},\nonumber \\
	{p_n} &=& {p_{n - 1}} - h  \cdot {q_{n - 1}}.
\end{eqnarray}
The Poisson bracket is $\left\{ {{p_n},{q_n}} \right\} = 1 + {h ^2}$.
However, if we use the first-order symplectic algorithm, 
\begin{eqnarray}\label{52}	
	{q_n} &=& {q_{n - 1}} + h  \cdot {p_{n - 1}},\nonumber \\
	{p_n} &=& {p_{n - 1}} - h  \cdot {q_{n}}.
\end{eqnarray}
We have $\left\{ {{p_n},{q_n}} \right\} = 1$. 
It indicates that the former destroys the symplectic structure of the system, but the latter does not. The other methods such as the Runge-Kutta methods also do not maintain the symplectic structure of the system. As mentioned before, the symplectic algorithm is the most effective numerical integration tool for solving Hamiltonian systems. Not only does it maintain conservation of energy, momentum, and angular momentum during long-term numerical integration, but it also satisfies Liouville's theorem, thereby strictly preserving the symplectic geometry structure of the phase space in Hamiltonian systems \cite{Wang:2021yqk}. 

The Hamiltonian system in Eq. \eqref{50} can  be decomposed into two integrable subsystems,  \( H_1 \) and \( H_2 \), such that
\begin{align}
	H &= H_1 + H_2, & 
	H_1 &= \frac{p^2}{2}, & 
	H_2 &= \frac{q^2}{2}.
\end{align}
The canonical equations for each sub-Hamiltonian system are
\begin{eqnarray}
	H_1: \dot{q} &=& \dfrac{\partial H_1}{\partial p} = p, 
	\dot{p} = -\dfrac{\partial H_1}{\partial q} = 0.\\
	H_2: \dot{q} &=& \dfrac{\partial H_2}{\partial p} = 0, 
	\dot{p} = -\dfrac{\partial H_2}{\partial q} = -q.
\end{eqnarray}
Given initial conditions $(p_0,q_0)$, the analytical solutions for $H_1$ and $H_2$ at any time $t$ are
\begin{eqnarray}\label{88}
	H_1: p(t) = p_0, q(t) = q_0 + t \cdot p_0.
\end{eqnarray}
\begin{eqnarray}\label{89}
	H_2: p(t) = p_0 - t \cdot q_0, q(t) = q_0.
\end{eqnarray}
The Hamiltonian vector field is defined as
\begin{equation}
	X_H = \frac{\partial H}{\partial p}\frac{\partial}{\partial q} - \frac{\partial H}{\partial q}\frac{\partial}{\partial p},
\end{equation}
and its phase flow is represented via the exponential mapping,
\begin{equation}
	\exp(hX_H):\left[p(t),q(t)\right]\longrightarrow\left[p(t+h),q(t+h)\right].
\end{equation}
Since \(\exp(hX_{H_1})\) and \(\exp(hX_{H_2})\) have explicit analytical solutions, the original system in Eq. \eqref{50} can be solved by composing these sub-Hamiltonian solutions. Applying the Baker-Campbell-Hausdorff (BCH) formula \cite{Wu2003}, we derive
\begin{equation}
	\exp(\frac{h}{2}X_{H_{1}})\circ \exp(hX_{H_2})\circ \exp(\frac{h}{2}X_{H_{1}})=\exp[hX_{H}+\mathcal{O}(h^3)].
\end{equation}
The symmetric composition $\exp(\frac{h} {2}X_ {H_{1}})\circ \exp(hX_{H_2})\circ \exp(\frac{h} {2}X_ {H_ {1}})$ can be used as an approximation of the exact phase flow $\exp(hX_H)$. Therefore, numerical algorithms can be designed to 
\begin{equation}
	\begin{split}
		(p_n, q_n) &= \exp\left(\tfrac{h}{2}X_{H_1}\right) \circ \exp(hX_{H_2}) \\
		&\quad \circ \exp\left(\tfrac{h}{2}X_{H_1}\right)(p_{n-1}, q_{n-1}).
	\end{split}
\end{equation}
This is a second-order explicit symmetric symplectic algorithm, and according to Eqs. \eqref{88} and \eqref{89}, its specific iterative form is
\begin{eqnarray}
	q_{n+1/2} &=& q_{n-1} + \tfrac{h}{2}p_{n-1}, \nonumber\\
	p_n &=& p_{n-1} - hq_{n+1/2}, \nonumber\\
	q_n &=& q_{n-1} + \tfrac{h}{2}p_n.
\end{eqnarray}

The Hamiltonian system in Eq.\eqref{50} can be directly decomposed into two integrable components, facilitating the construction of explicit symplectic algorithms of second-order, fourth-order, or even higher orders. However, for the more complex Hamiltonian systems, merely decomposing them into two subsystems may not lead to analytical solutions, necessitating a breakdown into three, four, or even more subsystems. Consequently, the Kerr-MOG model discussed in this paper requires decomposition into five sub-Hamiltonians. 

\subsection{The explicit symplectic algorithms in Kerr-MOG black hole}

As mentioned in the introduction, in cases where Hamilton can be explicitly decomposed, the explicit symplectic algorithm is generally used. The same as in \cite{Wang:2021gja}, a splitting technique is used here. For the Hamiltonian in Eq. \eqref{23}, it can be split to five parts
\begin{equation}\label{111}
	H = {H_1} + {H_2} + {H_3} + {H_4} + {H_5}.
\end{equation}
Where,
\begin{eqnarray}
	{H_1} &=& {H_p}(r,\theta ),\nonumber \\
	{H_2} &=& \frac{{{a^2} + \alpha (1 + \alpha )}}{{2\Sigma }}p_r^2,\nonumber \\
	{H_3} &=& \frac{{{r^2}p_r^2}}{{2\Sigma }},\nonumber \\
	{H_4} &=& \frac{{ - r(1 + \alpha )}}{\Sigma }p_r^2,\nonumber \\
	{H_5} &=& \frac{{p_\theta ^2}}{{2\Sigma }}.
\end{eqnarray}
However, this method requires that every sub-Hamiltonian system be analytically solvable. For the Hamiltonian in Eq. \eqref{111}, ${H_1}$ has an analytical solution, but it is extremely difficult to obtain the analytical solutions for the other four parts. In order to solve this problem, the time transformation function $d\tau  = g(r,\theta )dw$ is adopted \cite{Wu:2021rrd,1997CeMDA..67..145M}. Here, $\tau $ is the proper time, and $w$ is a newly introduced virtual coordinate time. The method preserves the original dynamics by extending the system into a larger phase space. This approach is foundational in handling time-dependent systems, constrained dynamics, and regularization problems, maintaining the physical behavior while expanding analytical flexibility \cite{Hairer2006}. By means of the time-transformation function ${\rm{g}}\left( {r,\theta } \right) = \frac{{\Sigma }}{{{r^2}}}$, a new Hamiltonian is obtained,
\begin{eqnarray}\label{28}
	{{\cal H}} = \frac{\Sigma }{{{r^2}}}{(H_1+p_0)} + \frac{\Delta }{{2{r^2}}}p_r^2 + \frac{1}{{2{r^2}}}p_\theta ^2.
\end{eqnarray}
Where, $p_0=-H$, ${{\cal H}}$ is limited by the constraint ${{\cal H}}=0$. As the $\Sigma$ function is eliminated in the denominators of the second and third terms in the new Hamiltonian ${{\cal H}}$, ${{\cal H}}$ is integrable. Thus, ${{\cal H}}$ can also be decomposed into the following five parts,
\begin{eqnarray}\label{29}
	{{\cal H}} = {{{\cal H}}_1} + {{{\cal H}}_2} + {{{\cal H}}_3} + {{{\cal H}}_4} + {{{\cal H}}_5}.
\end{eqnarray}
Each of the subsidiary Hamiltonian systems is as follows,
\begin{eqnarray}
	{{{\cal H}}_1} &=& \frac{\Sigma }{{{r^2}}}{(H_1+p_0)},\nonumber\\
	{{{\cal H}}_2} &=& \frac{{{a^2} + \alpha (1 + \alpha )}}{{2{r^2}}}p_r^2,\nonumber\\
	{{{\cal H}}_3} &=& \frac{p_r^2}{2},\nonumber\\
	{{{\cal H}}_4} &=& \frac{{-(1 + \alpha )}}{r}p_r^2,\nonumber\\
	{{{\cal H}}_5} &=& \frac{{p_\theta ^2}}{{2{r^2}}}.
\end{eqnarray}
The equations of motion for the sub-Hamiltonian ${{{\cal H}}_1}$ in terms of the new coordinate time are expressed as
\begin{eqnarray}
	\frac{{d\tau }}{{dw}} &=& \frac{{\Sigma }}{{{r^2}}} = {\rm{g}}\left( {r,\theta } \right),\nonumber\\
	\frac{{d{p_r}}}{{dw}} &=&  - \frac{{\partial {{{\cal H}}_1}}}{{\partial r}} = {P_r}(r,\theta ),\nonumber\\
	\frac{{d{p_\theta }}}{{dw}} &=&  - \frac{{\partial {{{\cal H}}_1}}}{{\partial \theta }} = {P_\theta }(r,\theta ).
\end{eqnarray}
The equations of motion for the rest sub-Hamiltonians are given as following,
\begin{eqnarray}
	{{{\cal H}}_2}:\frac{{dr}}{{dw}} &=& \frac{{{a^2} + \alpha (1 + \alpha )}}{{{r^2}}}{p_r},\nonumber\\
	\frac{{d{p_r}}}{{dw}} &=& \frac{{{a^2} + \alpha (1 + \alpha )}}{{{r^3}}}{p_r}^2.\nonumber\\
	{{{\cal H}}_3}:\frac{{dr}}{{dw}} &=& {p_r},\frac{{d{p_r}}}{{dw}} = 0.\nonumber\\
	{{{\cal H}}_4}:\frac{{dr}}{{dw}} &=& \frac{{ - 2(1 + \alpha )}}{r}{p_r},\nonumber\\
	\frac{{d{p_r}}}{{dw}} &=& \frac{{ -(1 + \alpha )}}{{{r^2}}}{p_r}^2.\nonumber\\
	{{{\cal H}}_5}:\frac{{d\theta }}{{dw}} &=& \frac{p_\theta }{{{r^2}}},\frac{{d{p_r}}}{{dw}} = \frac{{p_\theta }^2}{{{r^3}}}.
\end{eqnarray}
The analytic solutions for each of the five splitting parts are explicitly expressed in terms of the new time variable $w$. Given the initial values of $({r_0},{\theta _0},{p_{{r_0}}},{p_{{\theta _0}}})$, the analytic solution for each part can be written as
\begin{eqnarray}
	\tau (w) &=& {\tau _0} + w{\rm{g}}\left( {{r_0},{\theta _0}} \right).\nonumber\\
	{{{\cal H}}_1}:{p_r}(w) &=& {p_{{r_0}}} + w\left( { - \frac{{\partial {{{\cal H}}_1}}}{{\partial r}}} \right),\nonumber\\
	{p_\theta }(w) &=& {p_{{\theta _0}}} + w\left( { - \frac{{\partial {{{\cal H}}_1}}}{{\partial \theta }}} \right).\nonumber\\
	{{{\cal H}}_2}:r(w) &=& {\left\{ {{r_0}^2 + \frac{{2[{a^2} + \alpha (1 + \alpha )]w{p_{{r_0}}}}}{{{r_0}}}} \right\}^{\frac{1}{2}}},\nonumber\\
	{p_r}(w) &=& \frac{{{p_{{r_0}}}}}{{{r_0}}}{\left\{ {{r_0}^2 + \frac{{2[{a^2} + \alpha (1 + \alpha )]w{p_{{r_0}}}}}{{{r_0}}}} \right\}^{\frac{1}{2}}}.\nonumber\\
	{{{\cal H}}_3}:r(w) &=& {r_0} + w{p_{{r_0}}}.\nonumber\\
	{{{\cal H}}_4}:r(w) &=& {\left\{ {\frac{{{{[{r_0}^2 - 3(1 + \alpha )w{p_{{r_0}}}]}^2}}}{{{r_0}}}} \right\}^{\frac{1}{3}}},\nonumber\\
	{p_r}(w) &=& {p_{{r_0}}}{\left[ {\frac{{{r_0}^2 - 3(1 + \alpha )w{p_{{r_0}}}}}{{{r_0}^2}}} \right]^{\frac{1}{3}}}.\nonumber\\
	{{{\cal H}}_5}:\theta (w) &=&{\theta _0} + \frac{{w{p_{{\theta _0}}}}}{{{r_0}^2}}\nonumber\\
	{p_r}(w) &=& {p_{{r_0}}} + \frac{{w{p_{{\theta _0}}}^2}}{{{r_0}^3}}.
\end{eqnarray} 

To solve the Hamiltonian in Eq. \eqref{28}, we construct a second-order explicit symplectic integrator $S_2^{{\cal H}}(h)$ with a fixed time step $h$, defined as a symmetric combination of two first-order solvers $\chi _h$ and $\chi _h^ *$,
\begin{eqnarray}
	S_2^{{\cal H}}(h) = {\chi _{{h \mathord{\left/{\vphantom {h 2}} \right.\kern-\nulldelimiterspace} 2}}} \circ \chi _{{h \mathord{\left/{\vphantom {h 2}} \right.\kern-\nulldelimiterspace} 2}}^ *,
\end{eqnarray}
where,
\begin{eqnarray}
	{\chi _h} = {{{\cal H}}_5}(h) \circ {{{\cal H}}_4}(h) \circ {{{\cal H}}_3}(h) \circ {{{\cal H}}_2}(h) \circ {{{\cal H}}_1}(h), \\
	\chi _h^ *  = {{{\cal H}}_1}(h) \circ {{{\cal H}}_2}(h) \circ {{{\cal H}}_3}(h) \circ {{{\cal H}}_4}(h) \circ {{{\cal H}}_5}(h).
\end{eqnarray}
Here, $\chi _h$ and $\chi _h^ *$ are the approximate solution operators constructed by sequentially combining the exact evolution operators of the five decomposed Hamiltonian subsystems. This algorithm updates position and momentum alternately through leapfrog schemes, maintaining the symplectic structure in phase space and achieving second-order accuracy \cite{Wu:2021rrd}.

To achieve higher precision, we employ Yoshida's coefficient construction method \cite{YOSHIDA1990262}, which symmetrically composes three second-order symplectic operators $S_2^{\mathcal{H}} $ with time-step scaling coefficients $ \gamma = 1/(2 - \sqrt[3]{2}) $ and $ \delta = 1 - 2\gamma $. 
Then, a fourth-order symplectic integrator is constructed as
\begin{eqnarray}
	S_4^{{\cal H}}(h) = S_2^{{\cal H}}(\gamma h) \circ S_2^{{\cal H}}(\delta h) \circ S_2^{{\cal H}}(\gamma h).
\end{eqnarray}
By canceling lower-order error terms, this method reduces the global error to $\mathcal{O}(h^4) $, making it suitable for medium-to-long-term high-precision simulations.

For further error optimization, we adopt $PR{K_6}4$ algorithm proposed by Zhou \cite{Zhou_2022}. $PR{K_6}4$ achieves fourth-order accuracy and minimizes error constants through a symmetric composition of six pairs of first-order operators $\chi_h$ and $\chi_h^* $, along with optimized time coefficients. It is defined as
\begin{eqnarray}
	PRK_64&=&\chi_{c_1h}\circ\chi_{c_2h}^*\circ\chi_{c_3h}\circ\chi_{c_4h}^*\circ\chi_{c_5h}\circ\chi_{c_6h}^*\nonumber\\&&\circ\chi_{c_7h}^*\circ\chi_{c_8h}\circ\chi_{c_9h}^*\circ\chi_{c_{10}h}\circ\chi_{c_{11}h}^*\circ\chi_{c_{12}h}
\end{eqnarray}
Here, the time coefficients \cite{Zhou_2022} are
\begin{eqnarray}
	\nonumber
	&&c_1=c_{12}= 0.079203696431196,   \\ \nonumber
	&&c _2=c_{11}= 0.130311410182166,   \\ \nonumber
	&&c_3=c_{10}= 0.222861495867608,    \\ \nonumber
	&&c_4=c_9=-0.366713269047426,    \\ \nonumber
	&&c_5=c_8= 0.324648188689706,   \\ \nonumber
	&&c_6=c_7=0.109688477876750.   \nonumber
\end{eqnarray}

It has been reported that $PR{K_6}4$ method is an optimized method, it can significantly reduce the discretization errors compared with other non optimized methods \cite{Zhou_2022}. Unlike conventional fourth-order symplectic algorithms, $PR{K_6}4$ contains multiple time coefficients and requires the combination of as many sub Hamiltonian systems as possible. For simplicity, $S_2^{{\cal H}}(h)$ and $S_4^{{\cal H}}(h)$ are denoted as $S2$ and $S4$, respectively.

To evaluate $S2$, $S4$ and $PR{K_6}4$, we perform numerical experiments with parameters  $h=1$, $E=0.995$, $L=4.6$, $\beta= 4\times {10^{- 4}}$, $a=0.5$, $\alpha=0.2$, and initial conditions $\theta=\pi/2$, $p_r=0$, and $p_{\theta}$ from Eq. \eqref{28} over ${10^{7}}$ integration time. The results are shown in Figs. \ref{fig:subfig1a}-\ref{fig:subfig1b}, the left panel (a) is for $r=11$, and the right panel (b) is for $r=110$. It is found that $S2$, $S4$ and $PR{K_6}4$ have good performance in preserving the Hamiltonian error $\Delta {{\cal H}}$. Specifically, $S4$ achieves two orders of magnitude higher accuracy than $S2$, but two orders lower than $PR{K_6}4$. Therefore, $PR{K_6}4$ is selected for subsequent calculations due to its superior performance.

\section{Dynamics of the charged particle} \label{section4}

\begin{figure*}[htbp]
	\center{
		\subfigure{\includegraphics[scale=0.25]{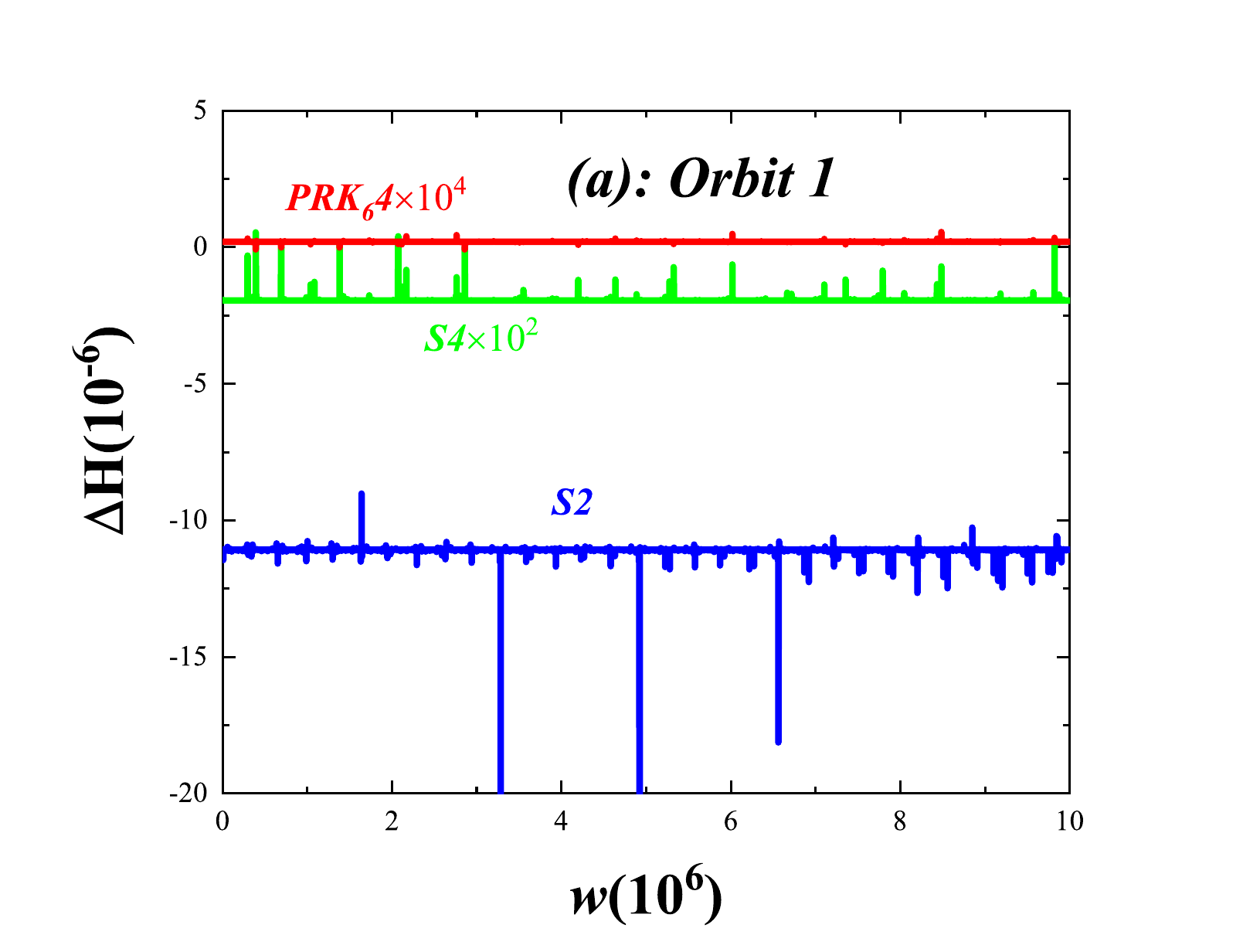} \label{fig:subfig1a}}
		\subfigure{\includegraphics[scale=0.25]{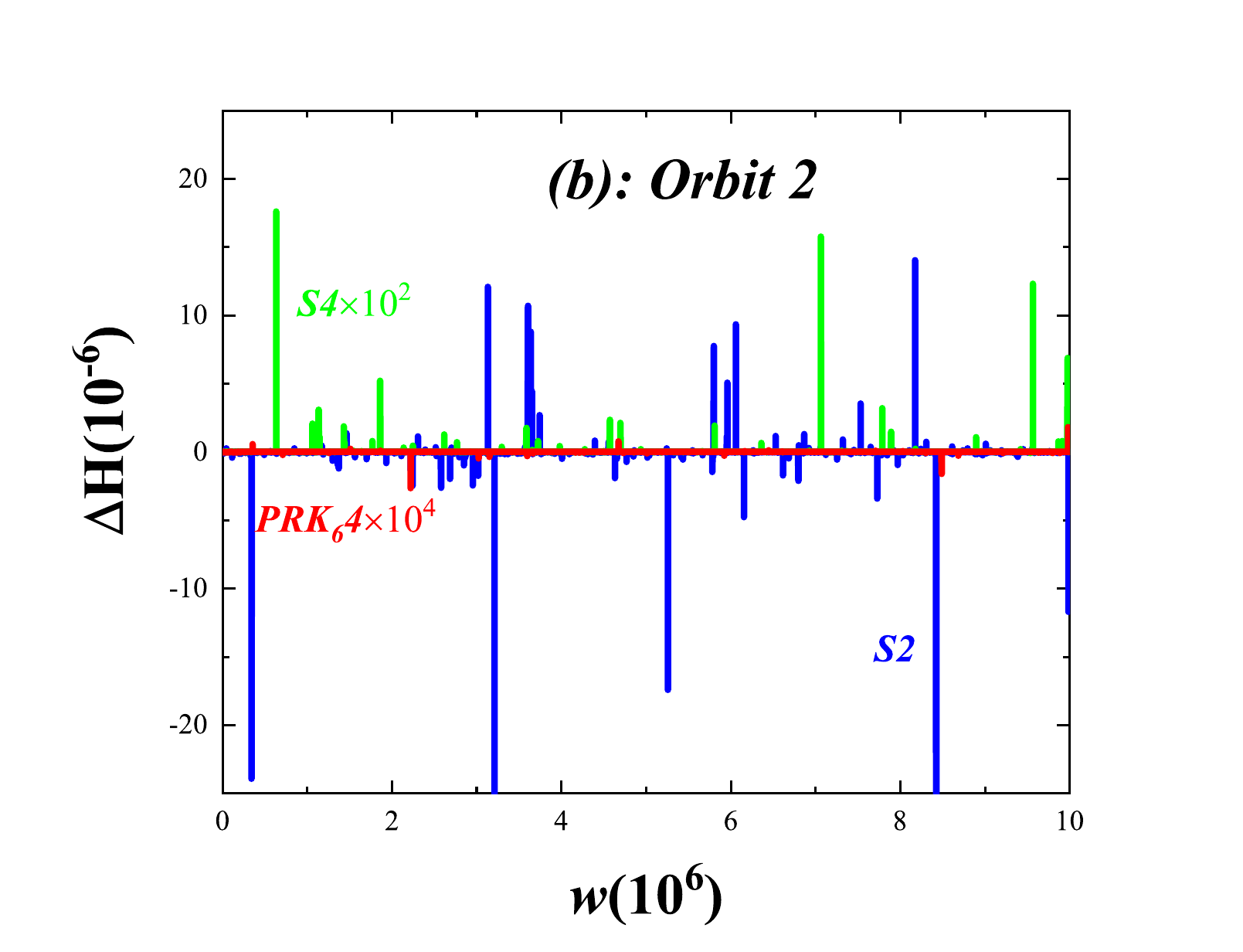} \label{fig:subfig1b}}
		\subfigure{\includegraphics[scale=0.25]{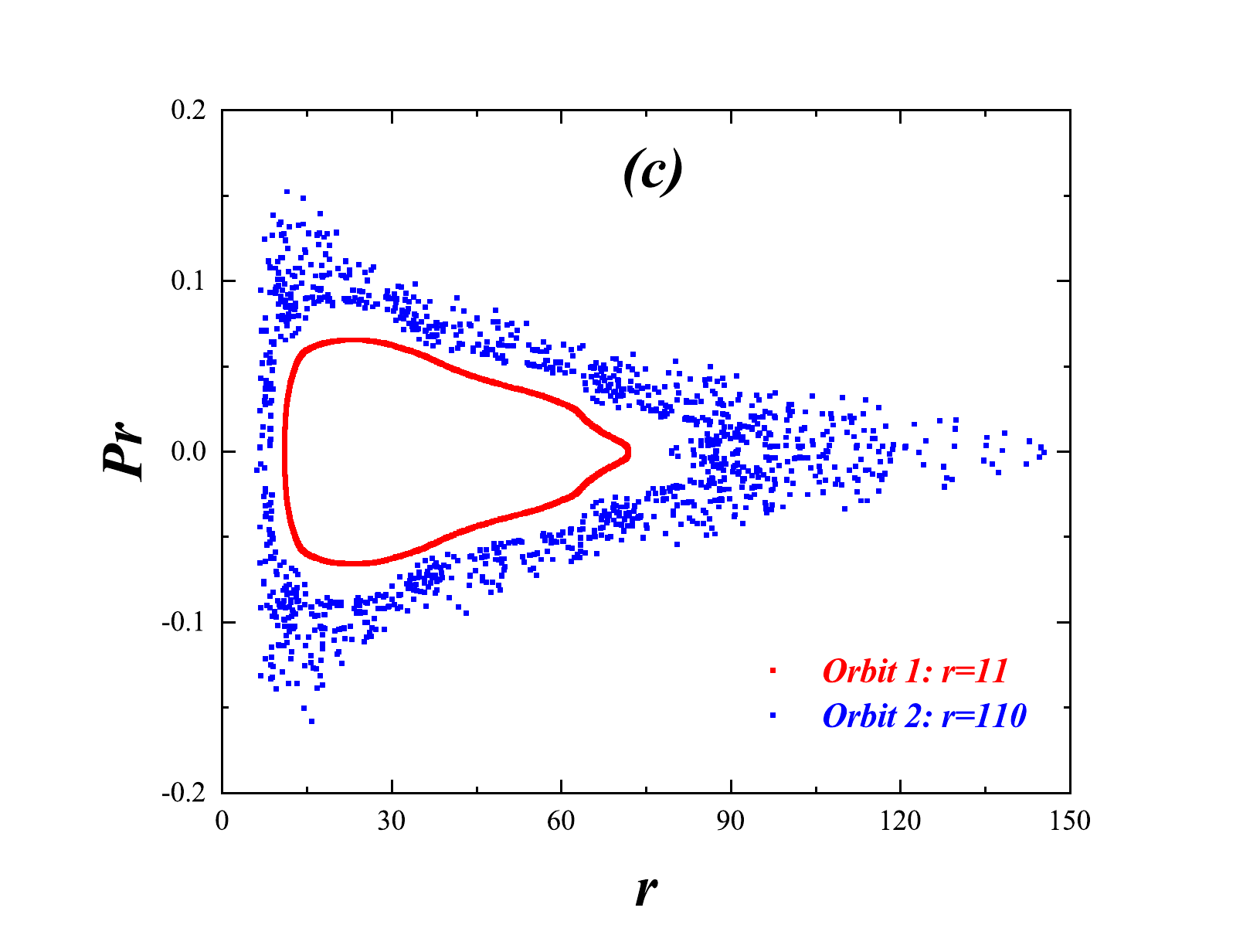} \label{fig:subfig1c}}
		\subfigure{\includegraphics[scale=0.25]{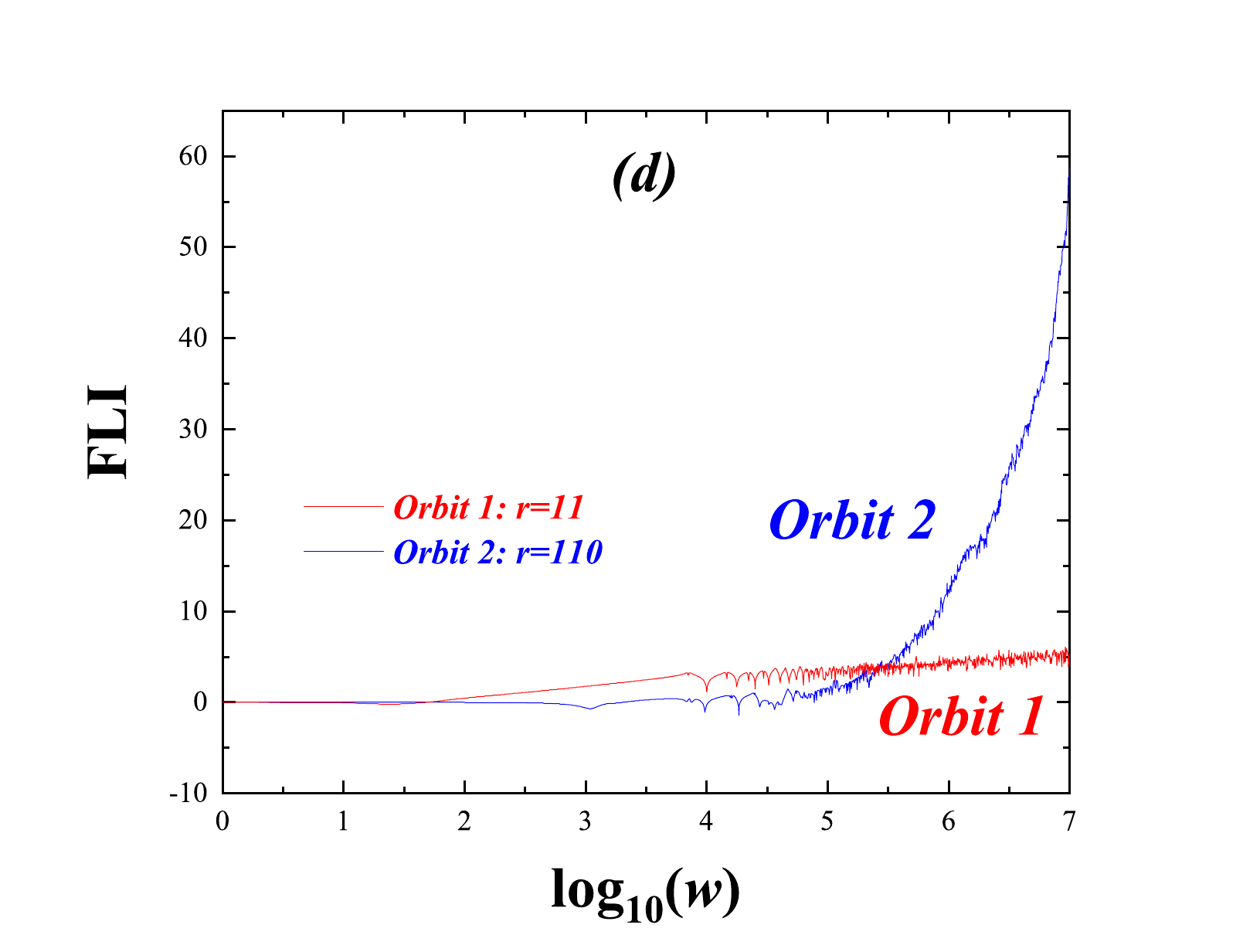} \label{fig:subfig1d}}
		\caption{\label{fig:1} \textbf{(a) and (b)} are the Hamiltonian error $\Delta{{\cal H}}$ for $S2$, $S4$, and $PR{K_6}4$. Here, the step size is $h=1$, the other parameters are $E=0.995$, $L=4.6$, $\beta = 4 \times {10^{ - 4}}$, $a=0.5$, and $\alpha  = 0.2$. \textbf{(a)} is related to the orbit 1, and \textbf{(b)} deals with the orbit 2. \textbf{(c)} Poincaré sections of the two orbits at $\theta = \pi/2$ and $p_\theta>0$. \textbf{(d)} FLIs of the two orbits.}}
\end{figure*}

Many chaos indicators have been proposed in literature, such as the Poincaré section method, the Lyapunov exponent, FLI, the 0-1 indicator, and spectral analysis. It has been reported that the Poincaré section method and FLI are simple and easy to implement methods \cite{MA2008216,Ma:2019ewq}, so they are used in this paper. Especially, the Poincaré section method is particularly suitable for studying four-dimensional conservative systems with two degrees of freedom. For example, the phase space structures of the orbits 1 and 2 in Figs. \ref{fig:subfig1a}-\ref{fig:subfig1b} can be depicted by using the Poincaré sections on the two-dimensional $r-p_r$ plane. These points in panel $(r,p_r)$ are obtained through linear interpolation on the section $\theta = \pi/2$. As shown in Fig. \ref{fig:subfig1c}, the red points form a closed curve, indicating that the orbit 1 is regular. In contrast, the blue points are randomly distributed, which implies that the orbit 2 is chaotic.

FLI is a widely used tool for distinguishing between regular and chaotic orbits, and it overcomes the limitations of the Poincaré section method in three-dimensional dynamical systems. Moreover, it distinguishes regular and chaotic orbits more quickly and sensitively than other indicators \cite{Ma:2019ewq}. By using the two nearby-orbit method \cite{2000CeMDA..78..167F,Wu:2006rx}, we have $FLI = \log_{10} \frac{d(w)}{d(0)}$, where $d(w)$ and $d(0)$ represent the distances between two nearby orbits at time $w$ and the initial time, respectively. As shown in Fig. \ref{fig:subfig1d}, for the regular orbit 1, FLI increases linearly with time; while for the chaotic orbit 2, FLI increases exponentially with time.

It is easy to conclude from Fig. \ref{fig:1} that the initial position of the particle is very important for the dynamical evolution. For example, under the same conditions, $r=11$ corresponds to an ordered orbit, while $r=110$ is associated with the chaotic orbit. Besides, the energy $E$, angular momentum $L$, magnetic field parameter $\beta$, black hole spin parameter $a$, and MOG parameter $\alpha$ may influence dynamics of the system, too. These situations will be discussed in the next subsection. 

\subsection{The impact of a single parameter}
In order to detect the impact of changing the parameters on the shift from regular to chaotic regimes, effects of $E$, $L$, $\beta$, $a$, and $\alpha$ on dynamics of the particle will be analyzed in this section.

\begin{figure*}[htbp]
	\center{
		\subfigure{\includegraphics[scale=0.25]{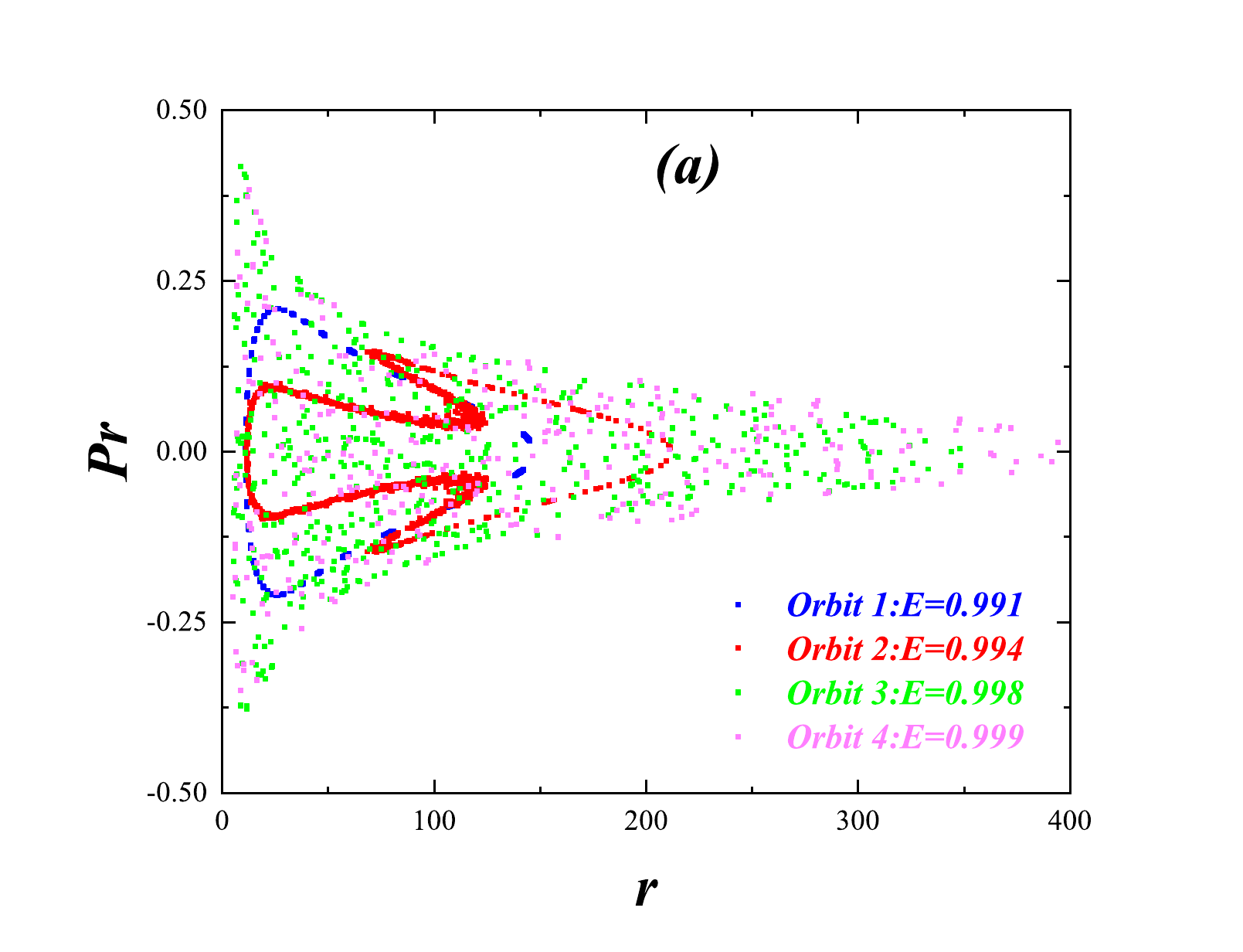} \label{fig:subfig2a}}
		\subfigure{\includegraphics[scale=0.25]{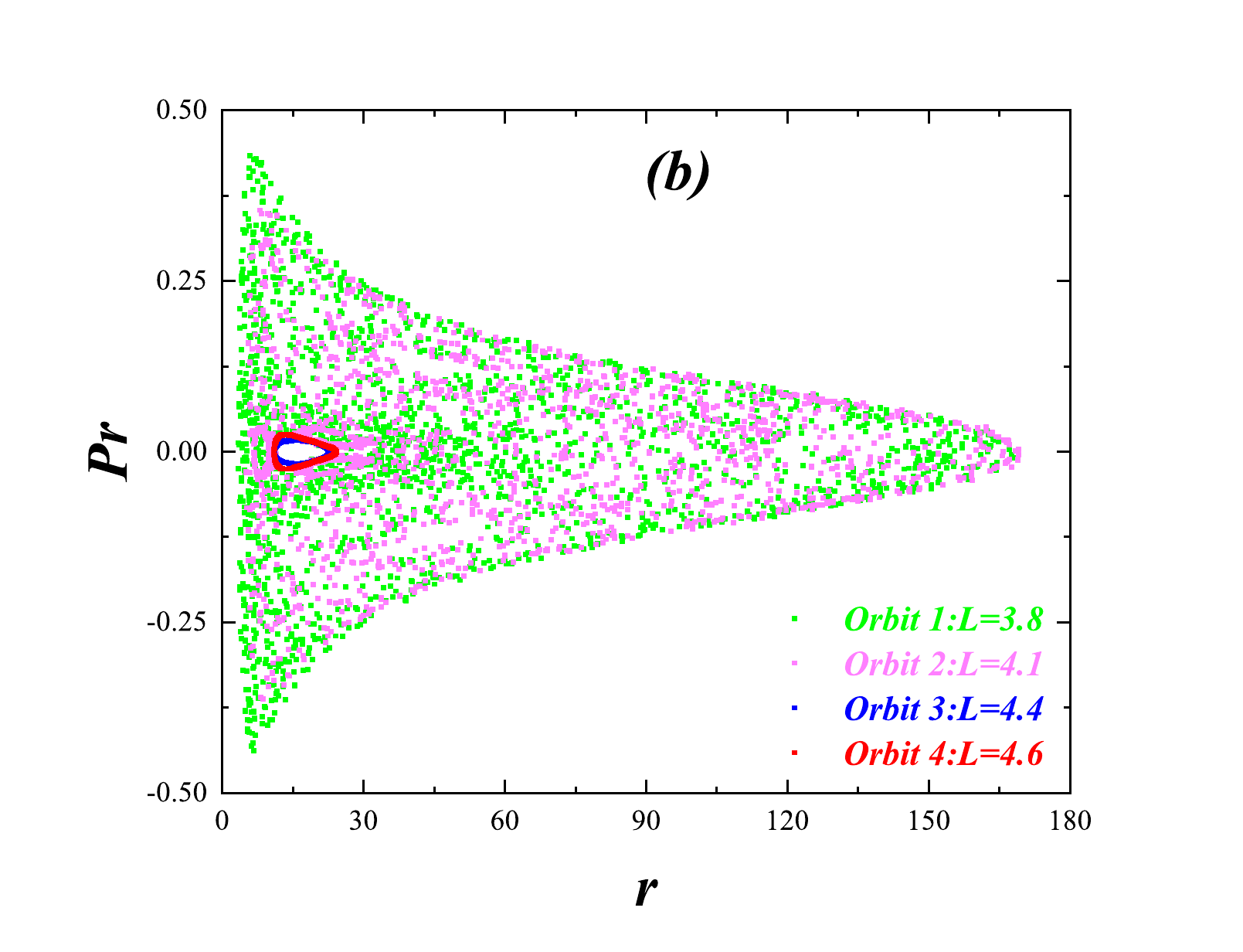} \label{fig:subfig2b}}
		\subfigure{\includegraphics[scale=0.25]{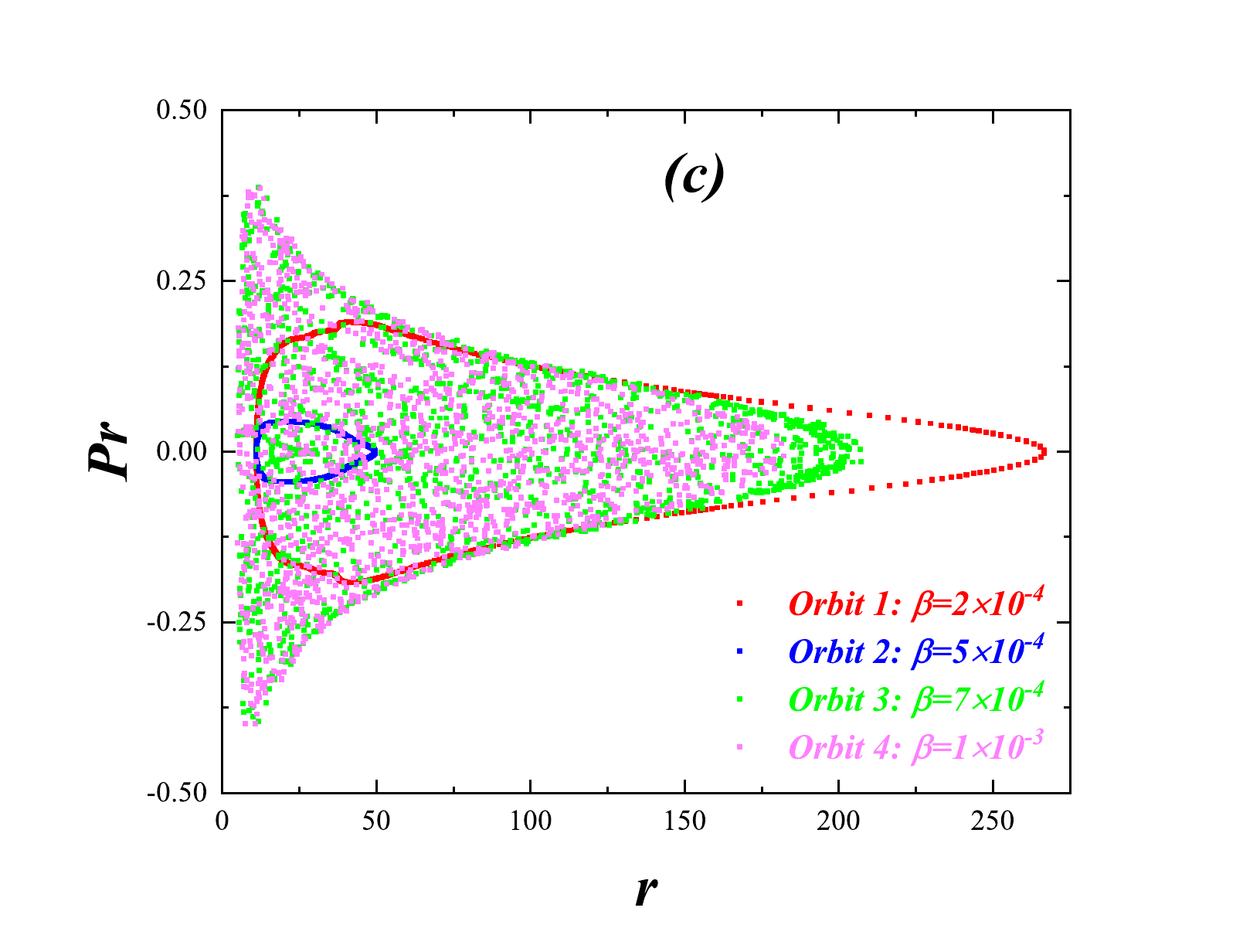} \label{fig:subfig2c}}
		\subfigure{\includegraphics[scale=0.25]{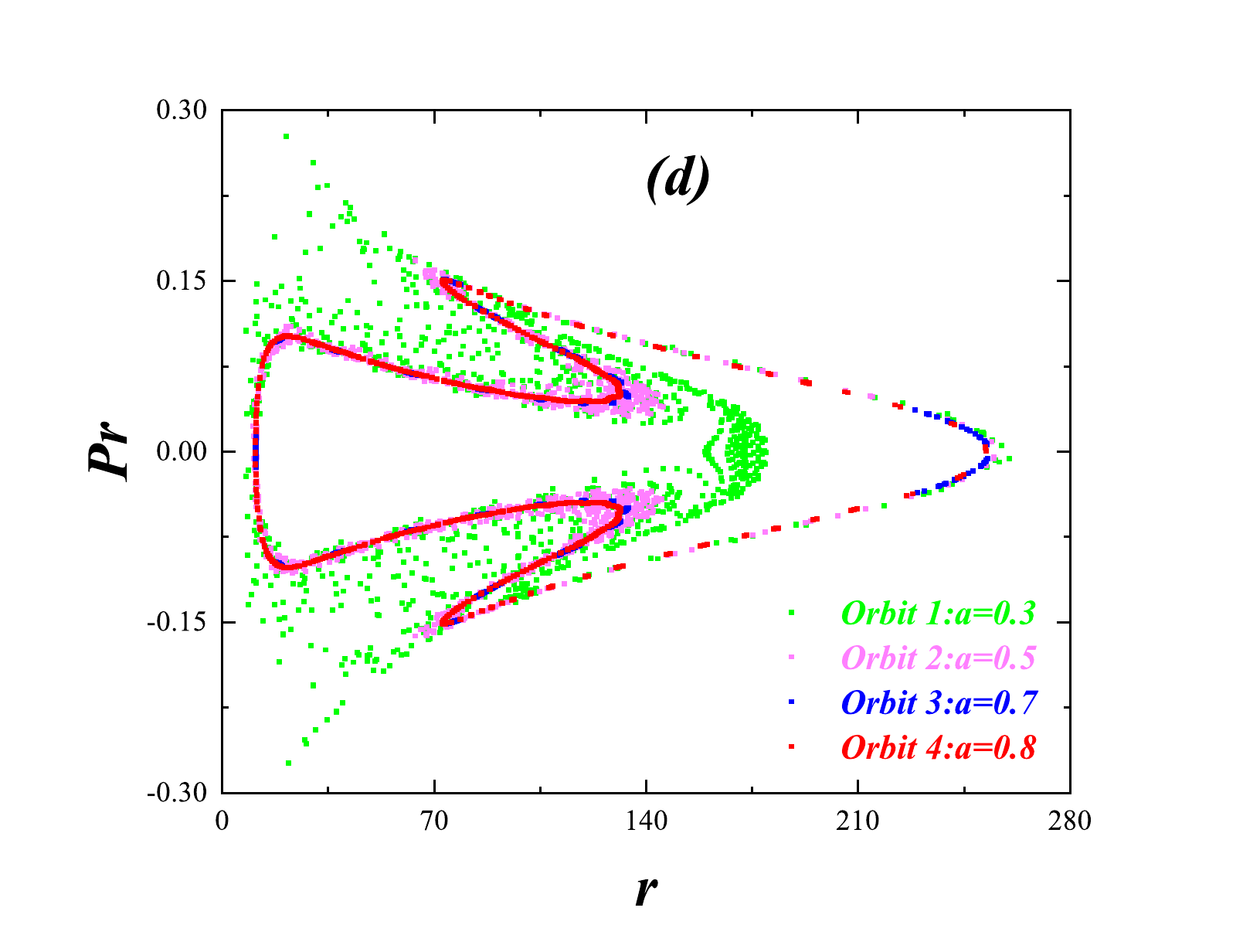} \label{fig:subfig2d}}
		\subfigure{\includegraphics[scale=0.25]{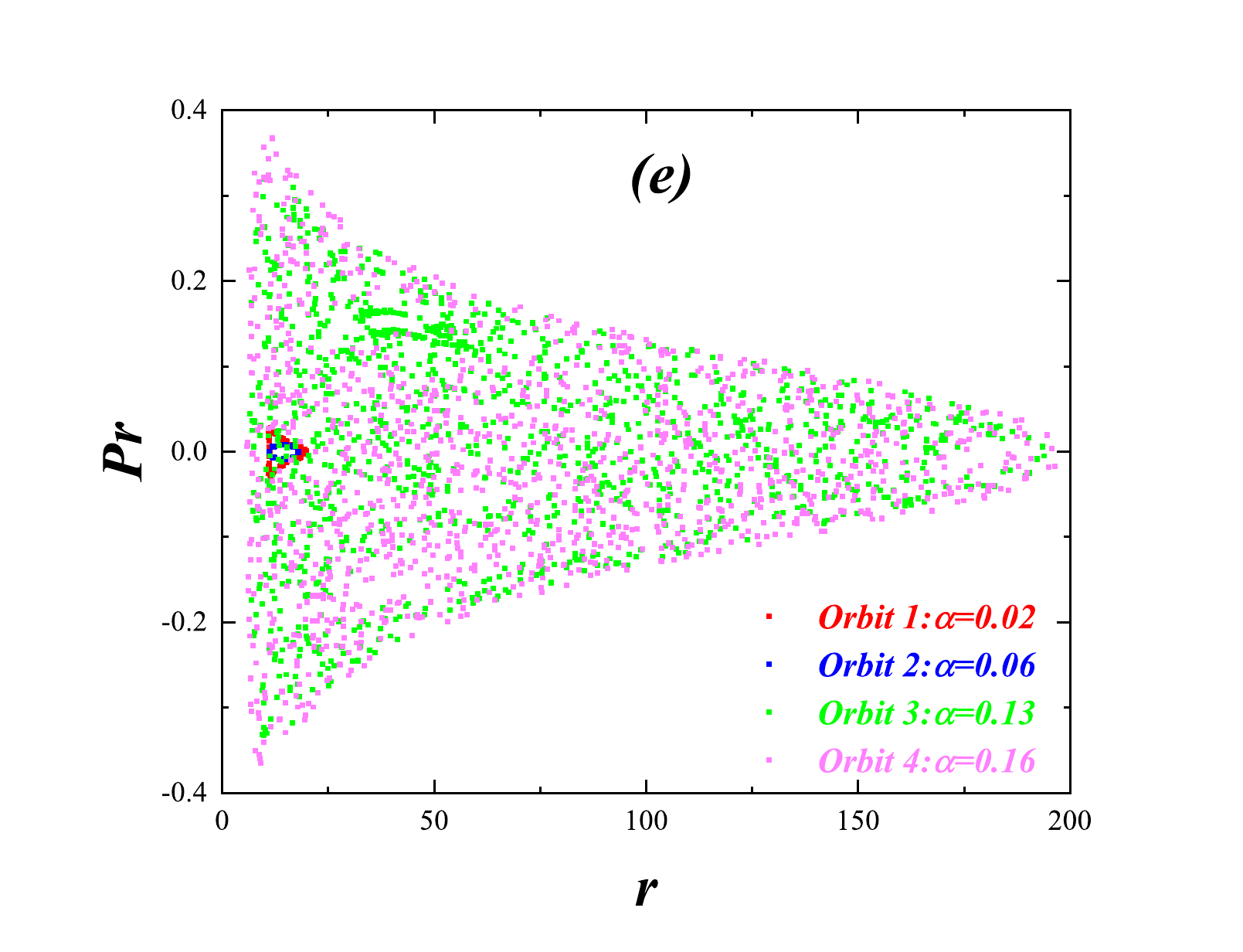} \label{fig:subfig2e}}
		\caption{\label{fig:2} Poincaré sections for different orbits at $r=11$ with varying parameters. Here, each figure contains four different motion trajectories. \textbf{(a)} $L=4.6$, $\beta= 4 \times{10^{- 4}}$, $a=0.5$ and $\alpha = 0.2$. \textbf{(b)} $E=0.995$, $a=0.5$, $\alpha= 0.08$ $\beta= 8.9\times {10^{- 4}}$. \textbf{(c)} $E=0.995$, $L=4.6$, $a=0.5$ and $\alpha = 0.2$. \textbf{(d)} $E=0.995$, $L=4.6$, $\beta= 3\times{10^{- 4}}$ and $\alpha = 0.2$. \textbf{(e)} $E=0.996$, $L=4.6$, $\beta= 8.9\times {10^{- 4}}$ and $a=0.5$.}}
\end{figure*}

(1) In case of r=11. At first, we set $L=4.6$, $\beta= 4\times {10^{- 4}}$, $a=0.5$ and $\alpha=0.2$, and let $E\in [0.990,0.999]$. $E=0.991$, $E=0.994$, $E=0.998$ and $E=0.999$ represent four different orbits 1, 2, 3 and 4, respectively. The results from Poincaré section are shown in Fig. \ref{fig:subfig2a}, it is found that the orbit 1 is in order, the orbit 2 is quasi-periodic, the orbits 3 and 4 are chaotic. This tells us that the system has a transition from order to chaos as $E$ increases. Next, if $E$ is invariant but $L$ is variant, the situation is different. For example, $E=0.995$, $\beta= 8.9\times {10^{- 4}}$, $a=0.5$ and $\alpha=0.08$, let $L\in [3.8,4.6]$. The results are given in Fig. \ref{fig:subfig2b}, $L=3.8$, $L=4.1$, $L=4.4$ and $L=4.6$ denote four trajectories. We find that the orbits 1 and 2 are chaotic, the orbits 3 and 4 are in order. Therefore, the system gradually shifts from chaotic to ordered regions with an increase of $L$. In Fig. \ref{fig:subfig2c}, $E=0.995$, $L=4.6$, $a=0.5$ and $\alpha=0.2$. $\beta= 2\times {10^{- 4}}$, $\beta= 5\times {10^{- 4}}$, $\beta= 7\times{10^{- 4}}$ and $\beta= 1\times{10^{-3}}$ correspond to four different orbits. When $\beta= 2\times {10^{- 4}}$ or $\beta= 5\times {10^{- 4}}$, the system is regular, while for $\beta=7\times{10^{- 4}}$ or $\beta=1\times{10^{-3}}$, the system is chaotic. Thus, as $\beta$ increases, the motion of the particle changes from order to chaos. In addition, the role of $a$ is considered in Fig. \ref{fig:subfig2d}. Here, $E=0.995$, $L=4.6$, $\beta= 3\times {10^{- 4}}$, $\alpha=0.2$ and $a\in [0.3,0.9]$. When $a=0.3$ and $a=0.5$, the two orbits are chaotic, but when $a=0.7$ and $a=0.8$, they are orderly. Consequently, as $a$ increases, the particle trajectories shift from chaos to order. Finally, the impact of parameter $\alpha$ is discussed in Fig. \ref{fig:subfig2e}. Here, $E=0.996$, $L=4.6$, $\beta=8.9\times {10^{- 4}}$, $a=0.5$ and $\alpha  \in [0,0.2]$. For $\alpha=0.02$ or $\alpha=0.06$, the orbit is regular, but for $\alpha=0.13$ or $\alpha=0.16$, the orbit is chaotic. It reveals that the motion of the particle transitions from order to chaos as $\alpha$ increases. 

\begin{figure*}[htbp]
	\center{
		\subfigure{\includegraphics[scale=0.25]{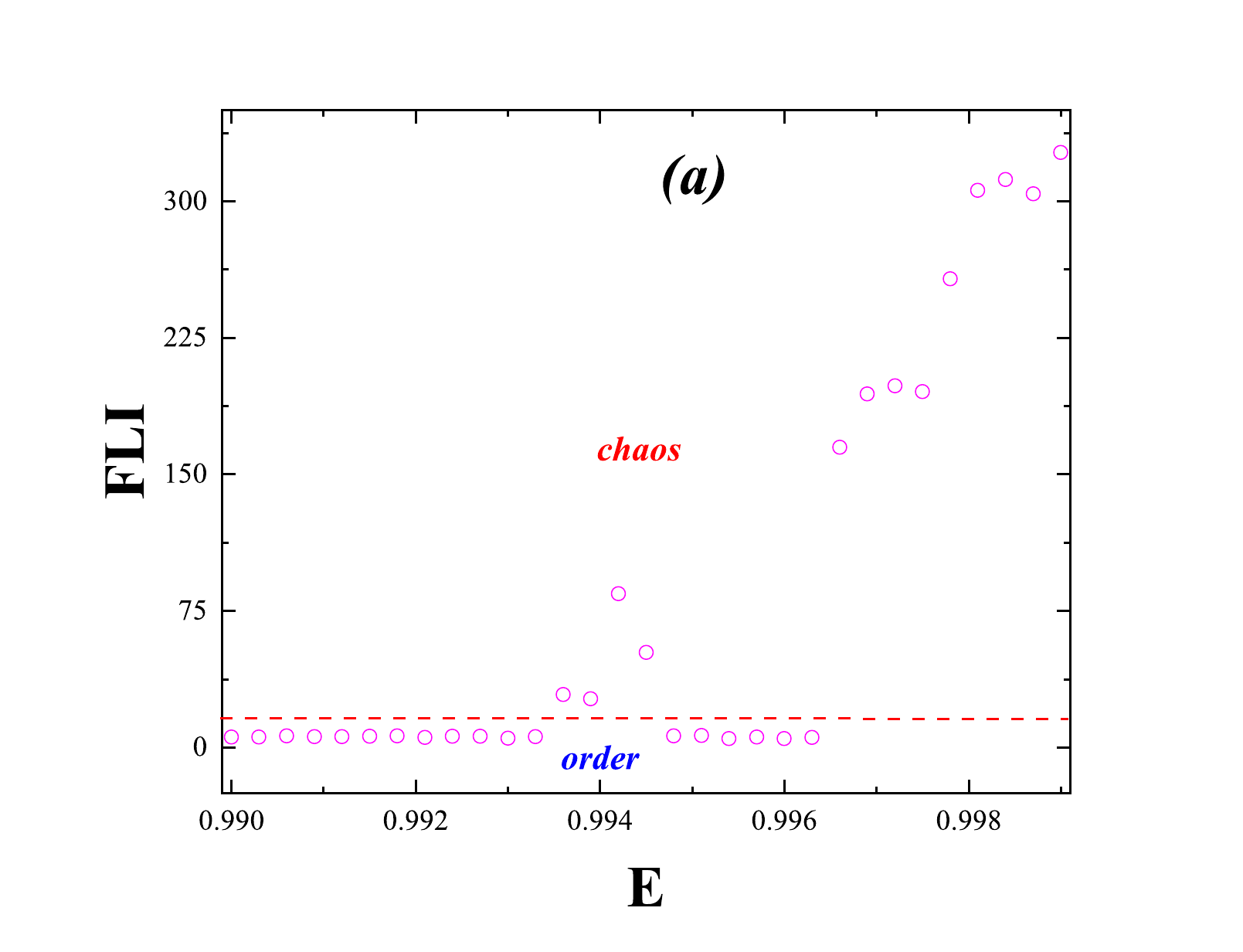} \label{fig:subfig3a}}
		\subfigure{\includegraphics[scale=0.25]{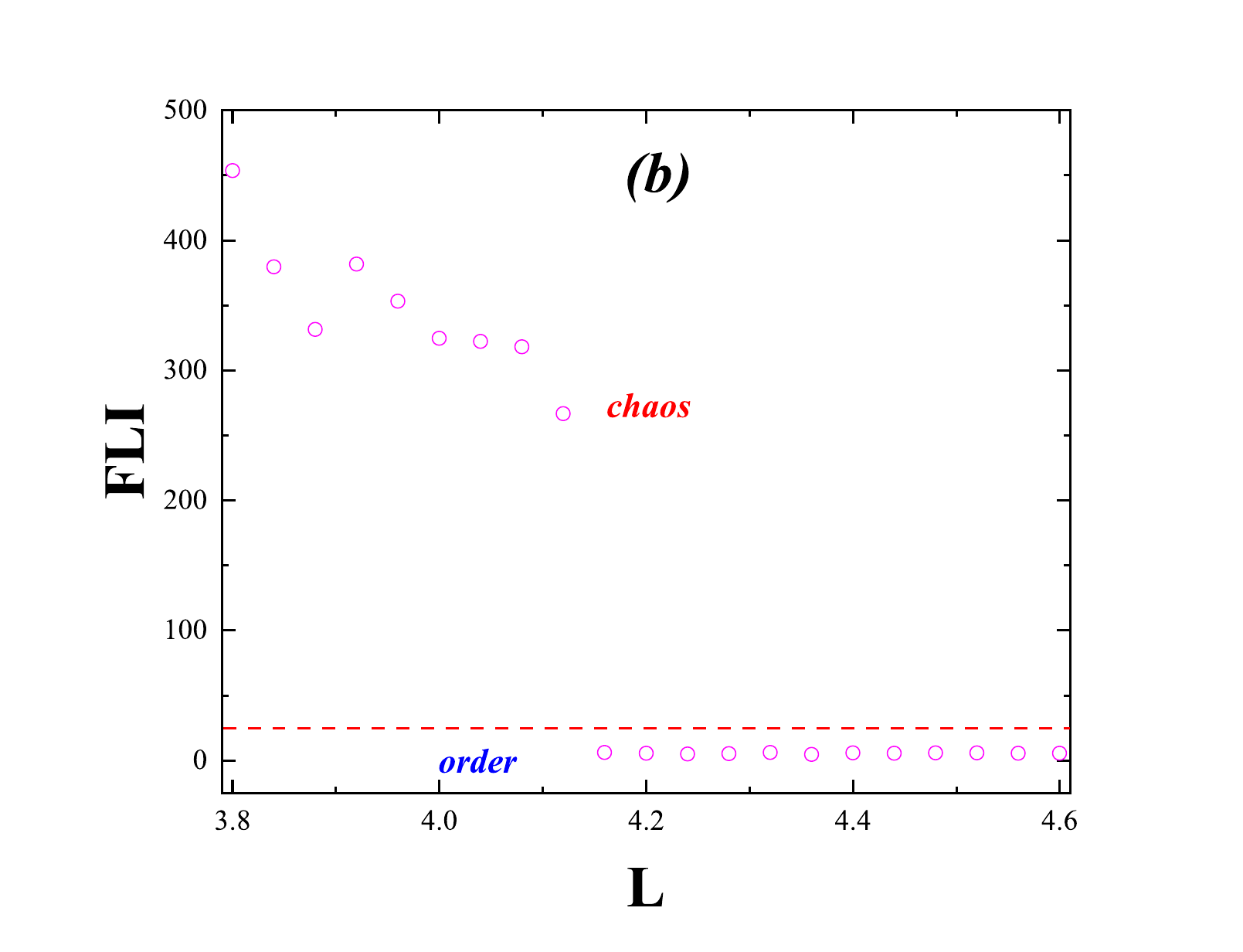} \label{fig:subfig3b}}
		\subfigure{\includegraphics[scale=0.25]{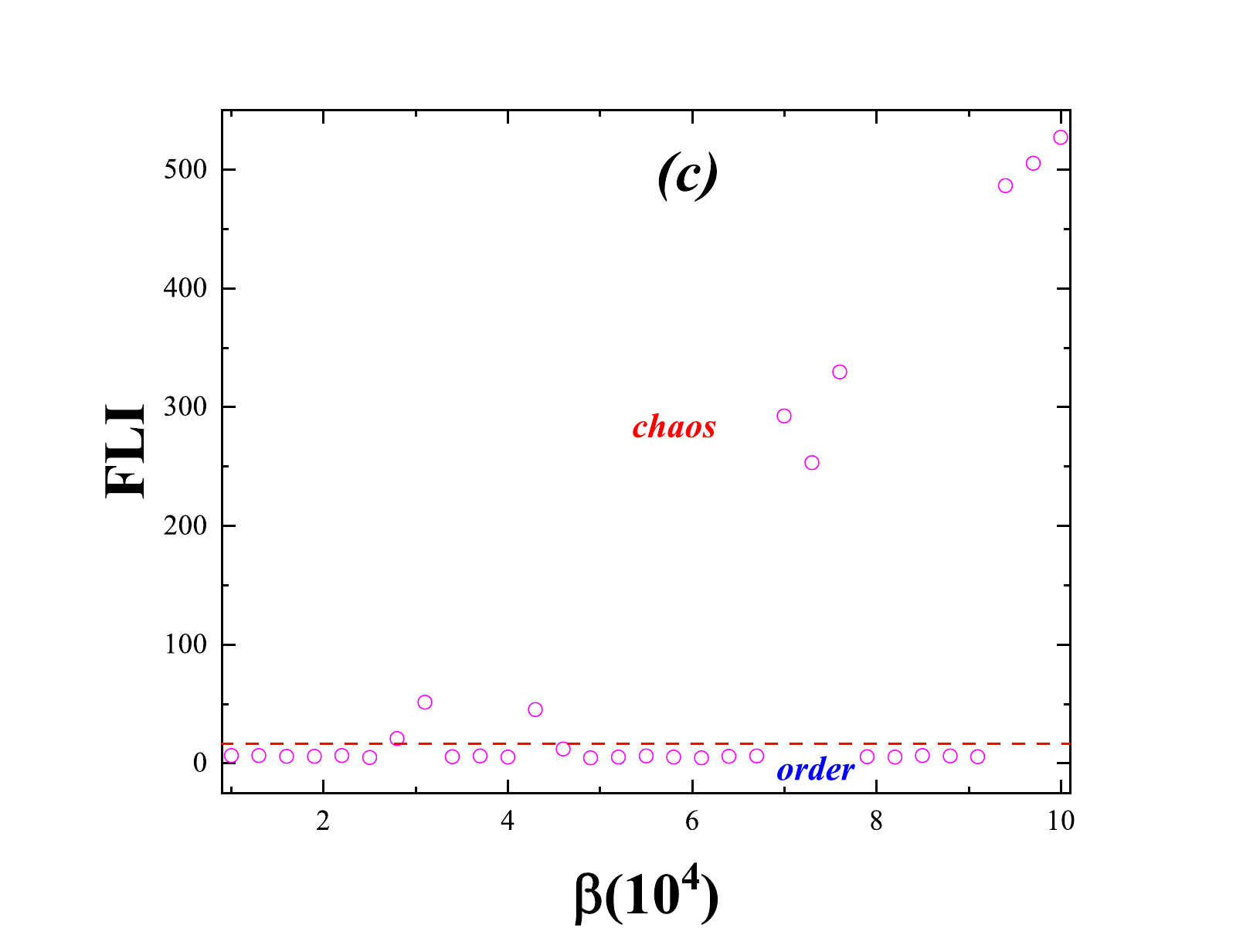} \label{fig:subfig3c}}
		\subfigure{\includegraphics[scale=0.25]{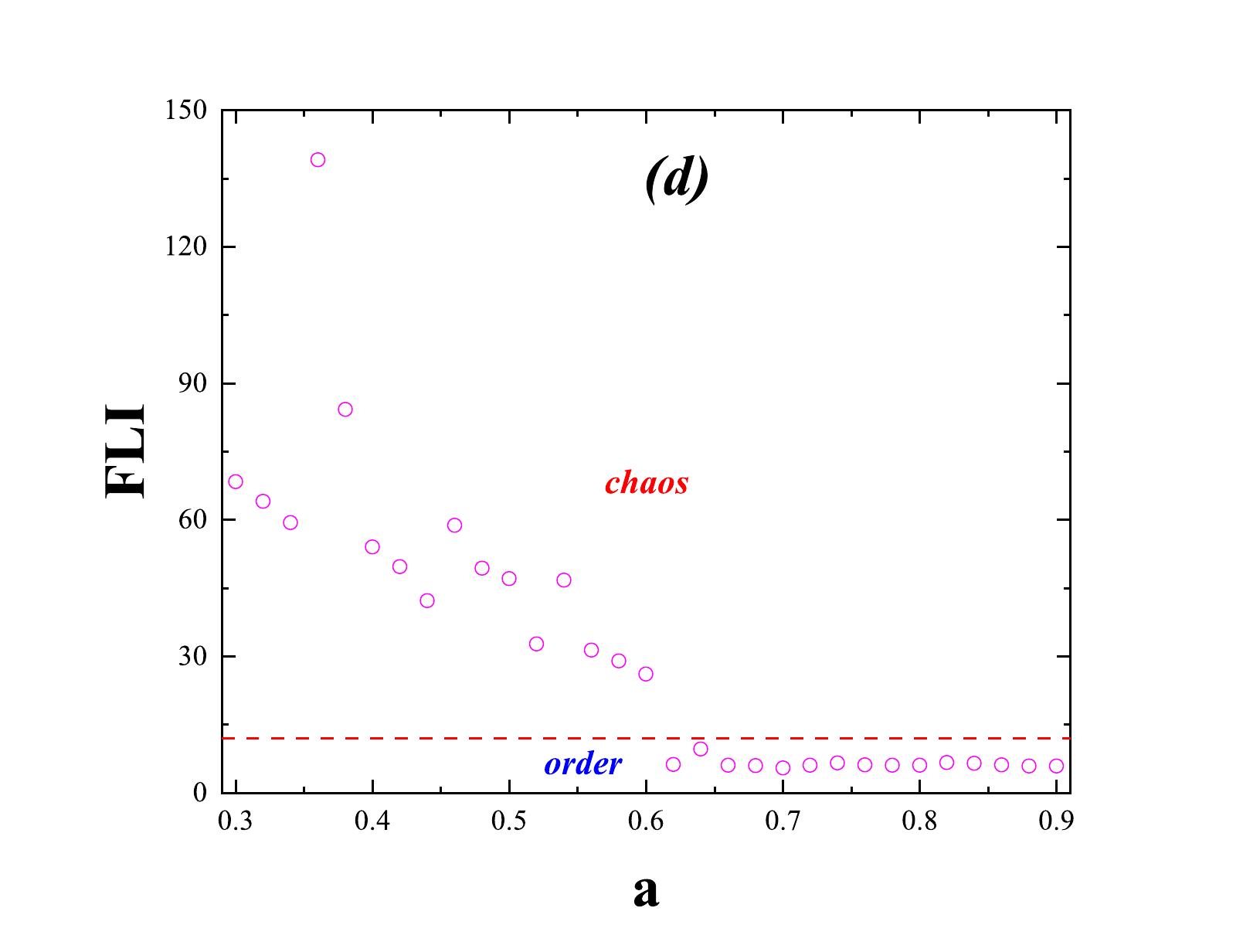} \label{fig:subfig3d}}
		\subfigure{\includegraphics[scale=0.3]{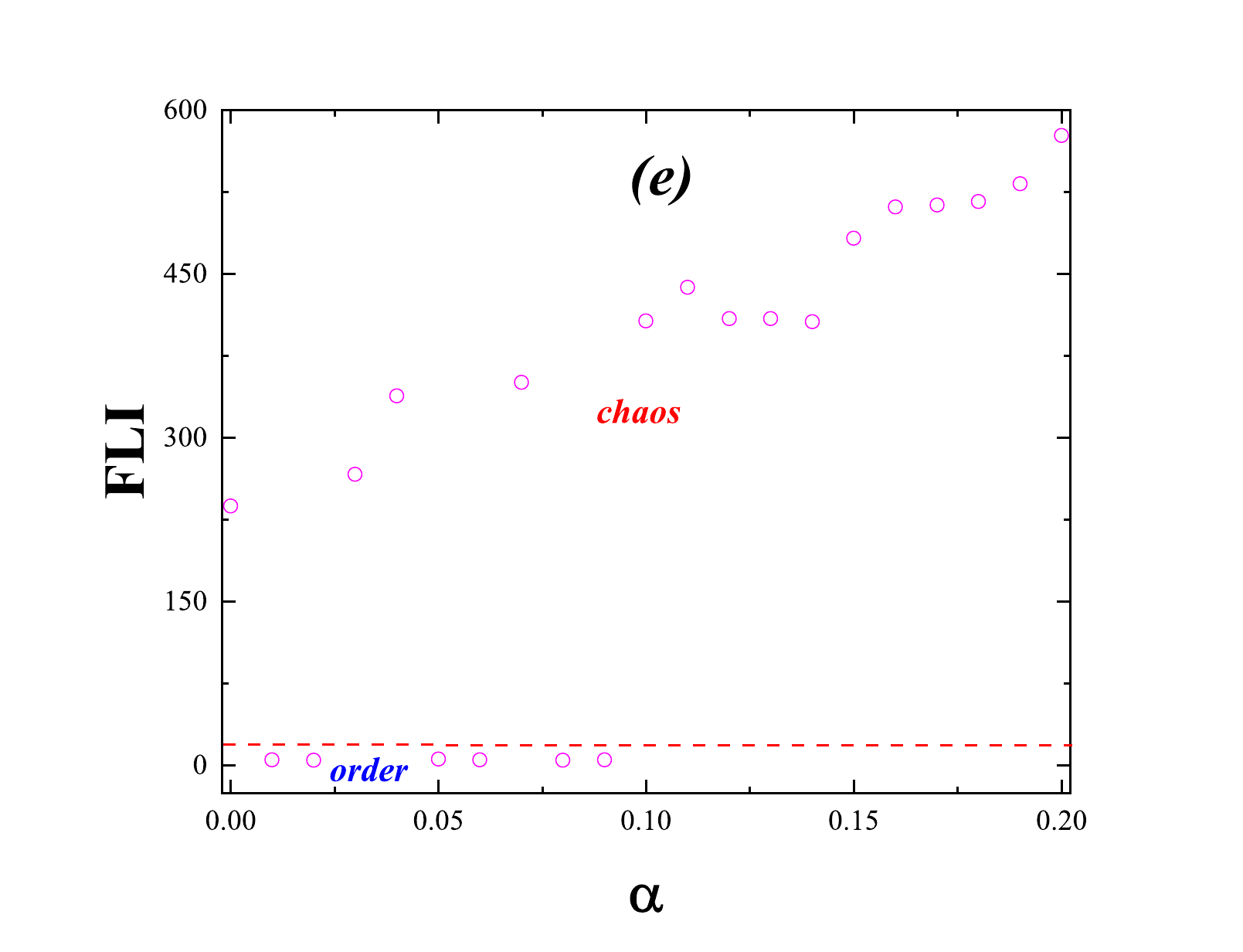} \label{fig:subfig3e}}
		\caption{\label{fig:3} The same as Figure 2, but for FLI. Here, $r=11$, the integration time is $w = {10^7}$.}}
\end{figure*}

The conclusions from Poincaré sections can be validated by FLI. The dependence of FLI on various parameters after the integration time reaches $w = {10^7}$ is shown in Fig. \ref{fig:3}. Here, each circle represents a distinct orbit, and the values of parameters in Figs. \ref{fig:subfig3a}-\ref{fig:subfig3e} are the same as in Figs. \ref{fig:subfig2a}-\ref{fig:subfig2e}. We estimate the FLIs of 30 trajectories in Fig. \ref{fig:subfig3a}, the $FLIs\ge 8$ and $FLIs < 8$ correspond to the chaotic and regular orbits. The transition from order to chaos is observed clearly, $E=0.9933$ and $E=0.9966$ are the two thresholds. For $E \le 0.993$, the system is in an ordered state, but for $E \ge 0.9969$, the system is in a chaotic state. However, when $0.9933 \le E \le 0.9966$, the dynamic evolution is very complex. For $0.9933 \le E \le 0.9945$, it is in a chaotic region; for $0.9948 \le E \le 0.9966$, it is in an ordered region. Overall, as the energy increases, the system transitions from order to chaos, but there is a interval between the completely chaotic region and ordered region. The orbits in interval are chaotic, but they are surrounded by the outside order region, they are still stable. This has been theoretically analyzed in \cite{2017RCD....22...54G}. Next, the FLIs of 20 trajectories are plotted in Fig. \ref{fig:subfig3b}, the $FLIs \ge 20$ indicate the orbits are chaotic, while the $FLIs < 20$ indicate the orbits are regular. It can be seen that $L=4.16$ is the threshold value, when $L$ is below the threshold, the system is orderly; when $L$ is above the threshold, the system is chaotic. In Fig. \ref{fig:subfig3c}, where $\beta$ is allowed to vary within the range $[1 \times {10^{ - 4}},1 \times {10^{ - 3}}]$. The FLIs of 30 different trajectories show that the transition from regular to chaotic regimes occurs at $\beta=2.8 \times {10^{ - 4}}$, $\beta= 4.3 \times {10^{ - 4}}$, $\beta= 7 \times {10^{ - 4}}$, and $\beta= 9.4 \times {10^{ - 4}}$. In Fig. \ref{fig:subfig3d}, chaos occurs in $a \le 0.6$, whereas regular dynamics occur in $a > 0.6$. Thus, it is clear that the extent of chaos decreases as $a$ increases. As illustrated in Fig. \ref{fig:subfig3e}, when $\alpha < 0.1$, chaotic and ordered regions alternate; when $\alpha \ge 0.1$, the trajectories are chaotic. The reason for this has already been explained in previous \cite{2017RCD....22...54G}.  

\begin{figure*}[htbp]
	\center{
		\subfigure{\includegraphics[scale=0.25]{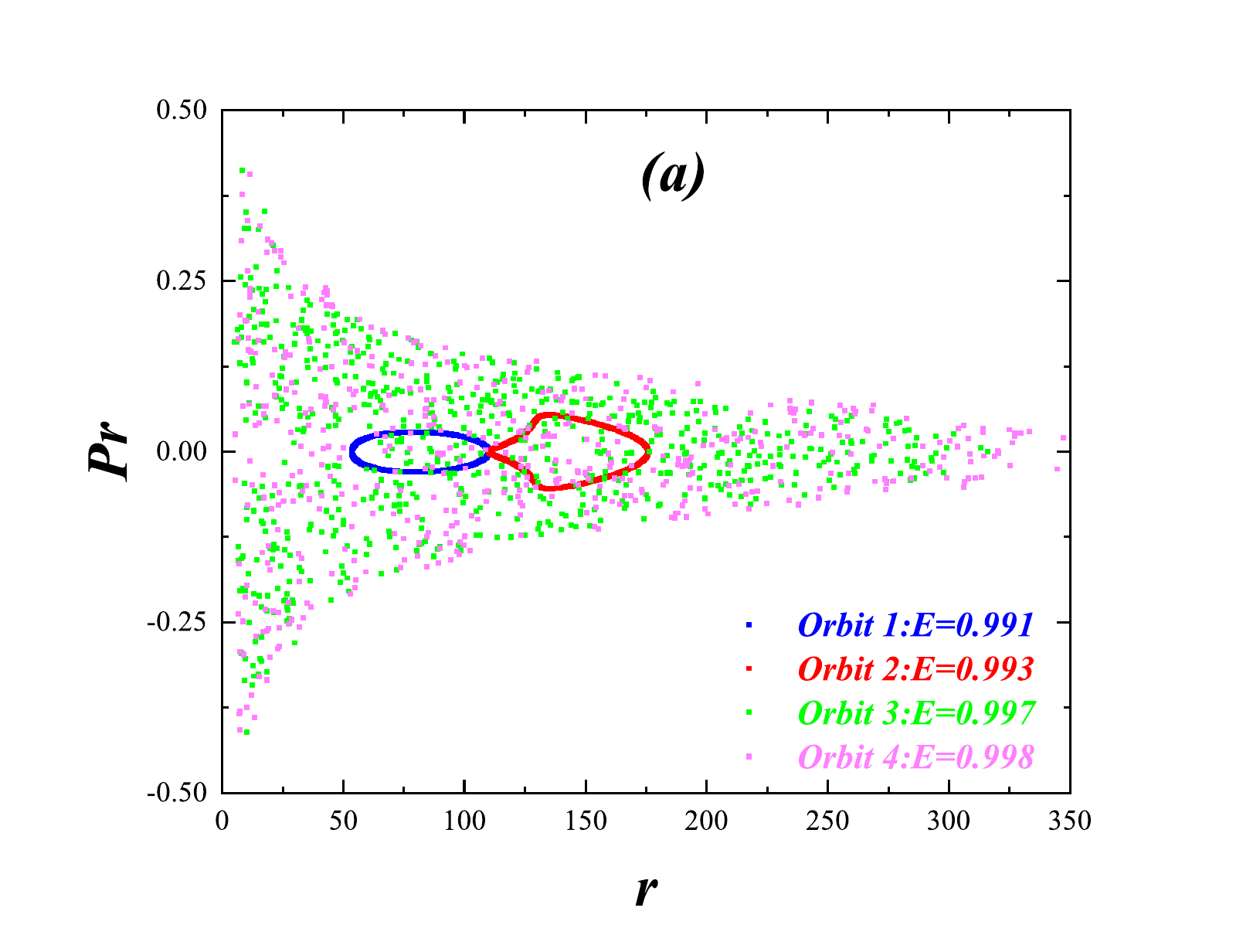} \label{fig:subfig4a}}
		\subfigure{\includegraphics[scale=0.25]{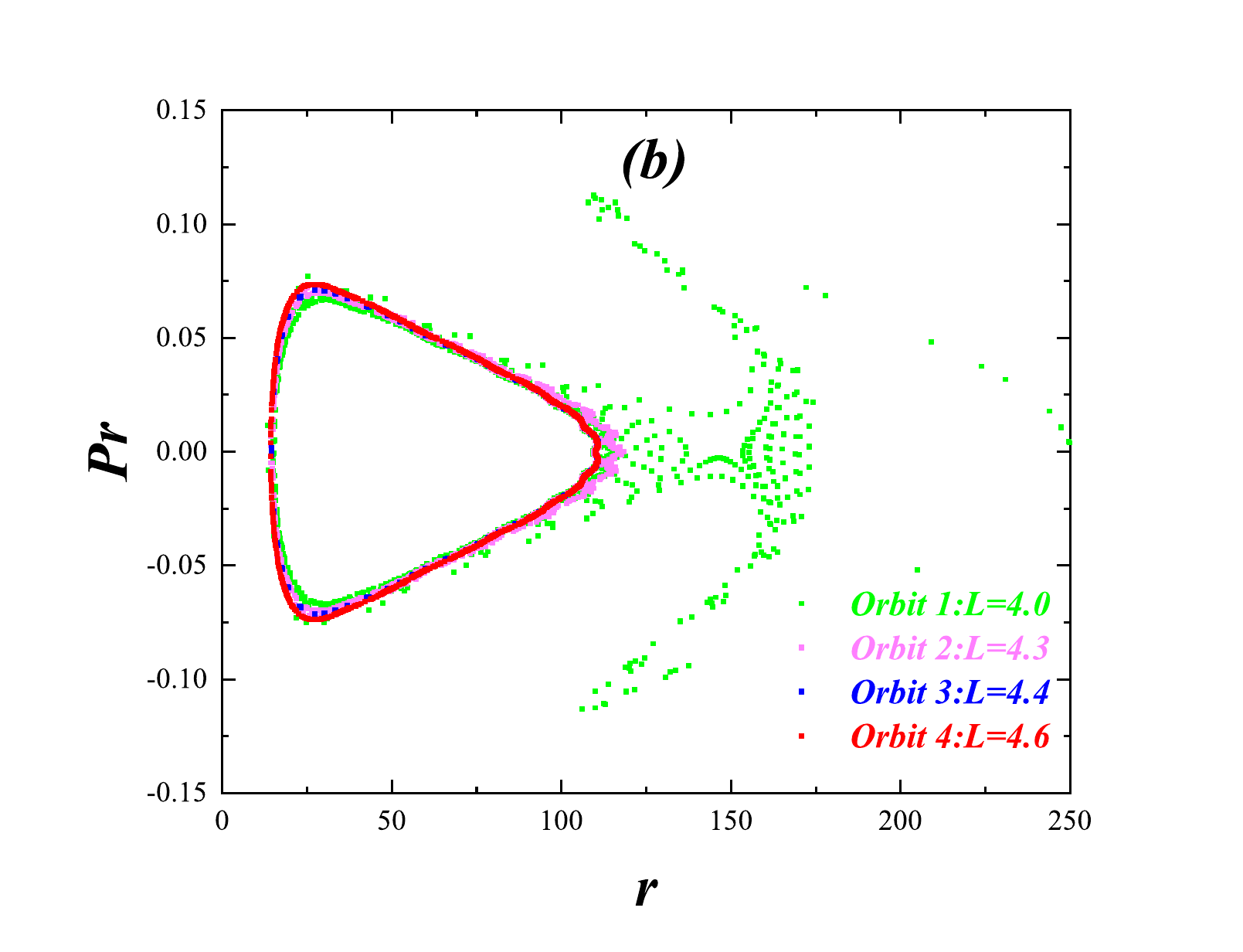} \label{fig:subfig4b}}
		\subfigure{\includegraphics[scale=0.25]{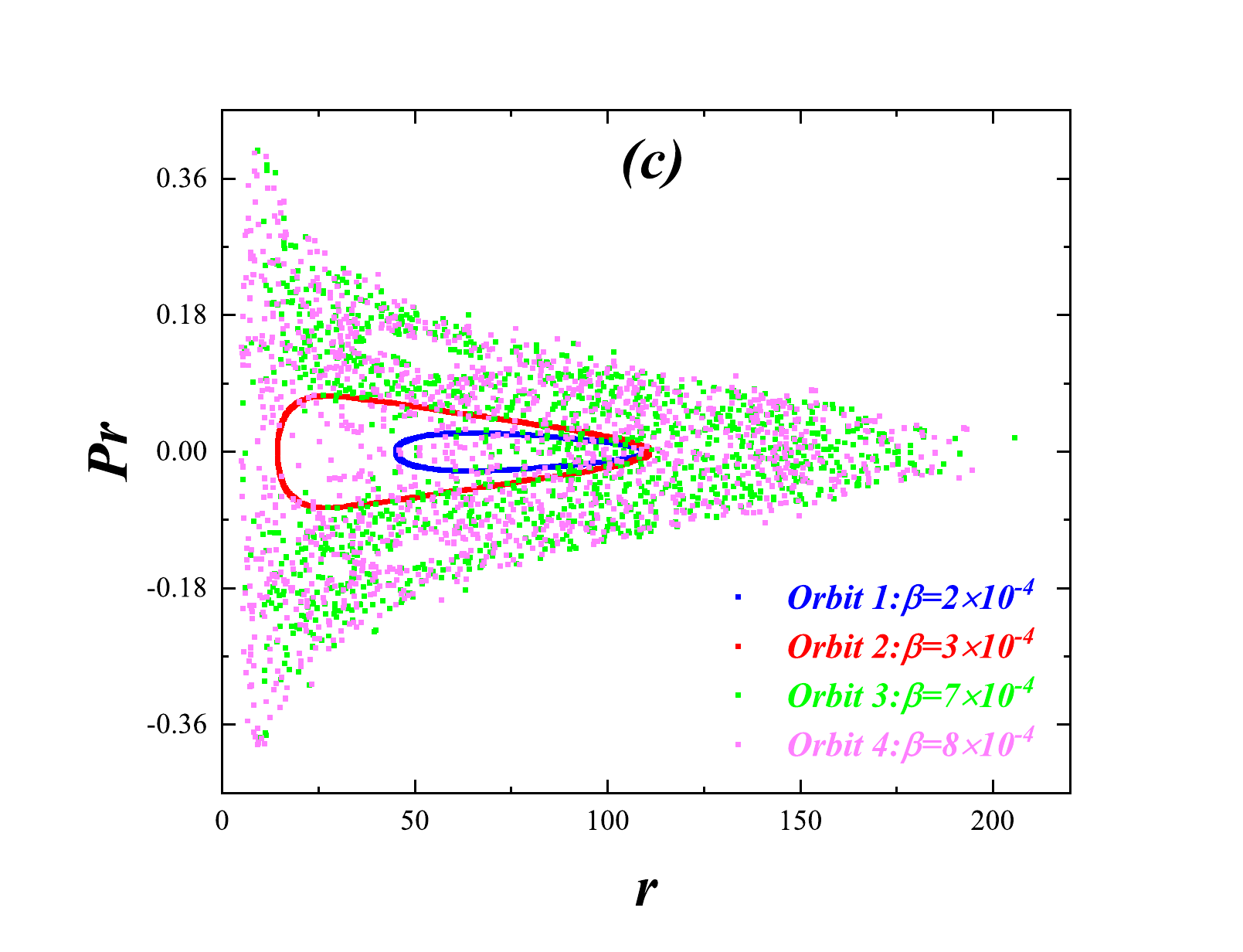} \label{fig:subfig4c}}
		\subfigure{\includegraphics[scale=0.25]{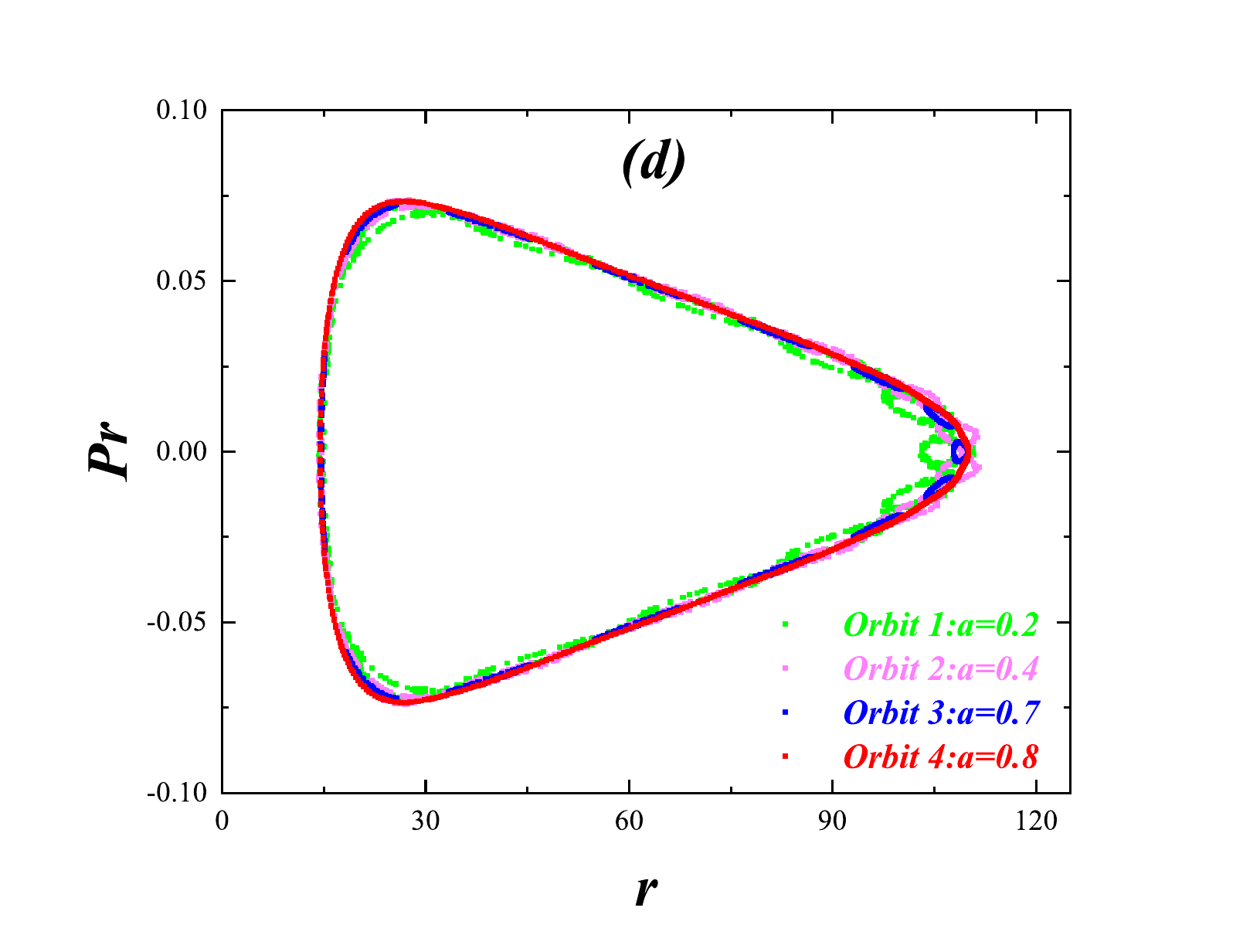} \label{fig:subfig4d}}
		\subfigure{\includegraphics[scale=0.3]{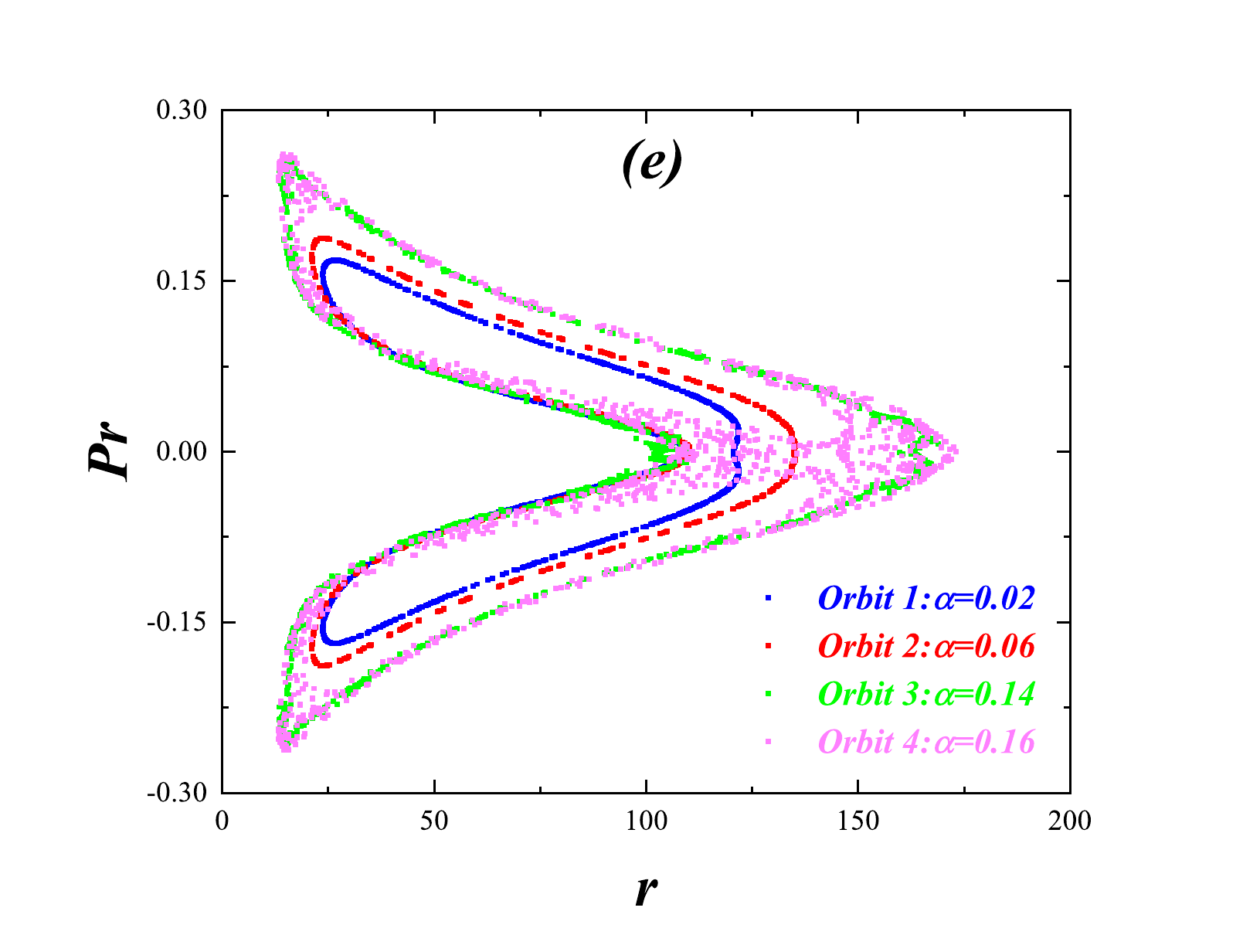} \label{fig:subfig4e}}
		\caption{\label{fig:4} Poincaré sections of the orbits at $r=110$ with different parameters. Here, each figure contains four different motion trajectories. \textbf{(a)} $L=4.6$, $\beta= 4 \times{10^{- 4}}$, $a=0.5$ and $\alpha = 0.2$. \textbf{(b)} $E=0.995$, $a=0.5$, $\alpha= 0.2$ and $\beta=3\times {10^{- 4}}$, \textbf{(c)} $E=0.995$, $L=4.6$, $a=0.5$ and $\alpha = 0.2$. \textbf{(d)} $E=0.995$, $L=4.6$, $\beta= 3\times{10^{- 4}}$ and $\alpha = 0.2$. \textbf{(e)} $E=0.995$, $L=4.6$, $\beta= 5\times {10^{- 4}}$ and $a=0.5$.}}
\end{figure*}

(2) In case of r=110. In Fig. \ref{fig:subfig4a}, the Poincaré sections of four orbits with $E=0.991$, $E=0.993$, $E=0.997$ and $E=0.998$ are analyzed, where the parameter settings are the same as those in Fig. \ref{fig:subfig2a}. The orbits are regular for $E=0.991$ and $E=0.993$, while they are chaotic for $E=0.997$ and $E=0.998$. In contrast to Fig. \ref{fig:subfig2b}, $E=0.995$ and $a=0.5$ are unchanged, while $\beta= 3\times {10^{- 4}}$, $\alpha=0.2$ and $L \in [4.0,4.6]$ in Fig. \ref{fig:subfig4b}. Here, when $L=4.6$ and $L=4.4$, the trajectories are in order. At $L=4.3$, the trajectory is quasi-periodic, while at $L=4.0$, the trajectory is evidently chaotic. Under the same parameter conditions, four different orbits are discussed in Fig. \ref{fig:subfig4c} to analyze the effect of $\beta$. The smooth curves (such as orbits 1 and 2) indicate that the motions of the orbits are regular at $\beta=2\times{10^{-4}}$ and $\beta=3\times{10^{-4}}$. On the other hand, there are many random discrete points in the phase diagram when the orbit takes value of $\beta=7\times{10^{-4}}$ and $\beta=8\times{10^{-4}}$. With respect to Fig. \ref{fig:subfig2d}, $a$ alters from 0.2 to 0.9 in Fig. \ref{fig:subfig4d}. The orbits are in order at $a = 0.2$ and $a = 0.4$, while chaos emerges at $a = 0.7$ and $a = 0.8$. In Fig. \ref{fig:subfig4e}, $E = 0.995$, $\beta=5\times{10^{-4}}$, $L=4.6$, $a=0.5$, but $\alpha$ ranges from 0 to 0.2. When $\alpha=0.02$ and $\alpha=0.06$, the trajectories exhibit saddle-shaped patterns, indicating a regular behavior. However, at $\alpha=0.14$ and $\alpha=0.16$, the trajectories clearly display chaotic characteristics. In summary, the conclusions derived from Fig. \ref{fig:4} align with those of in Fig. \ref{fig:2}.

\begin{figure*}[htbp]
	\center{
		\subfigure{\includegraphics[scale=0.25]{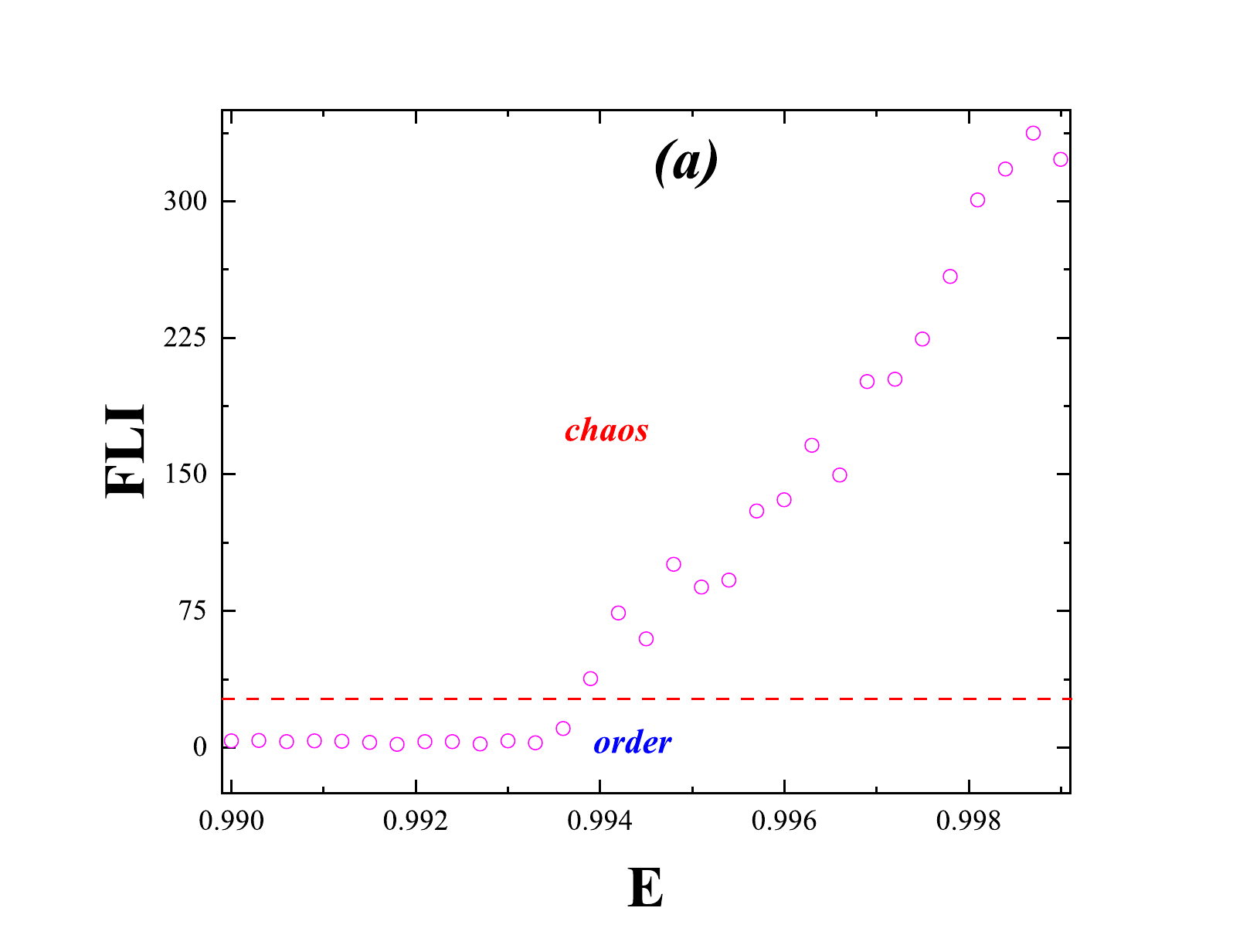} \label{fig:subfig5a}}
		\subfigure{\includegraphics[scale=0.25]{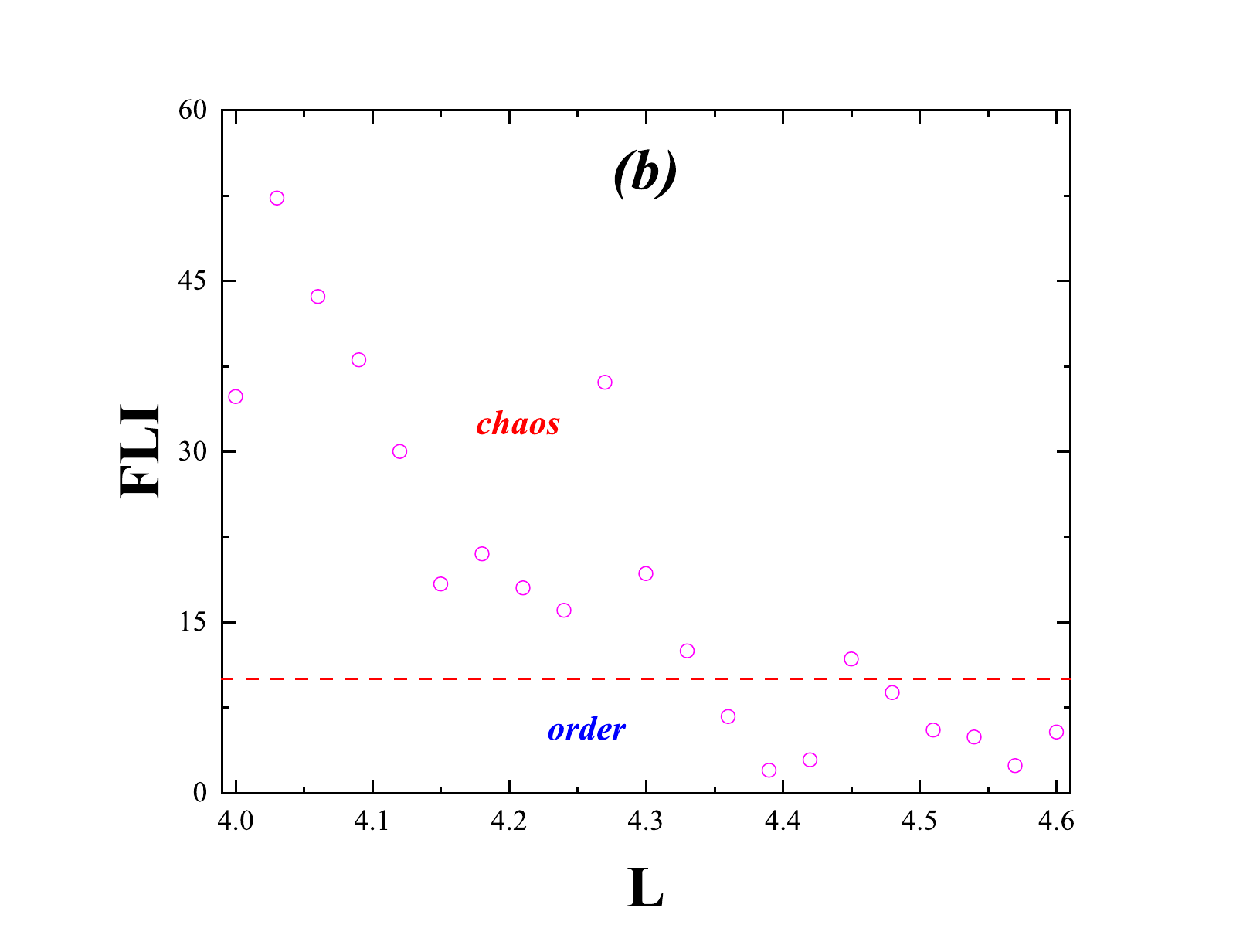} \label{fig:subfig5b}}
		\subfigure{\includegraphics[scale=0.25]{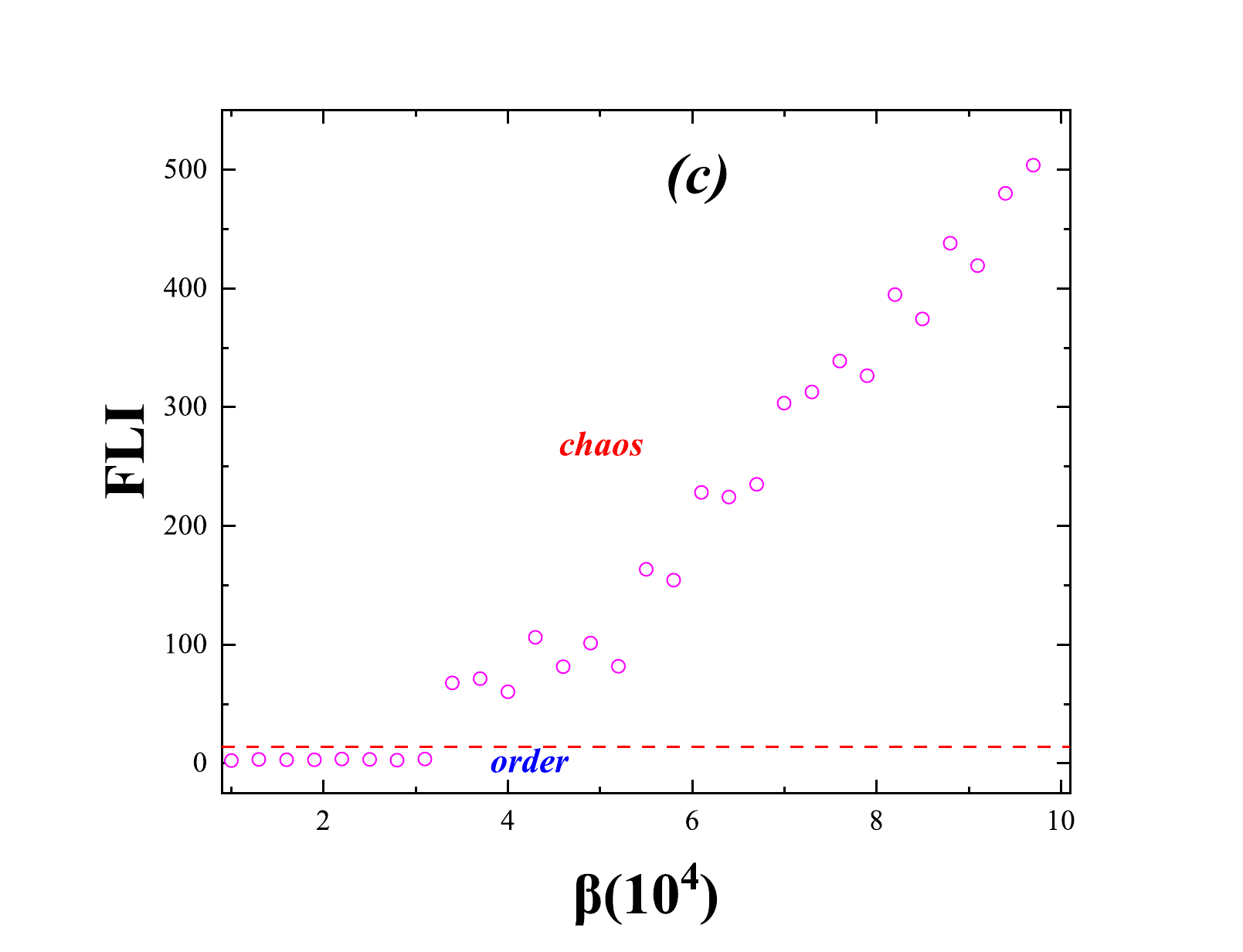} \label{fig:subfig5c}}
		\subfigure{\includegraphics[scale=0.25]{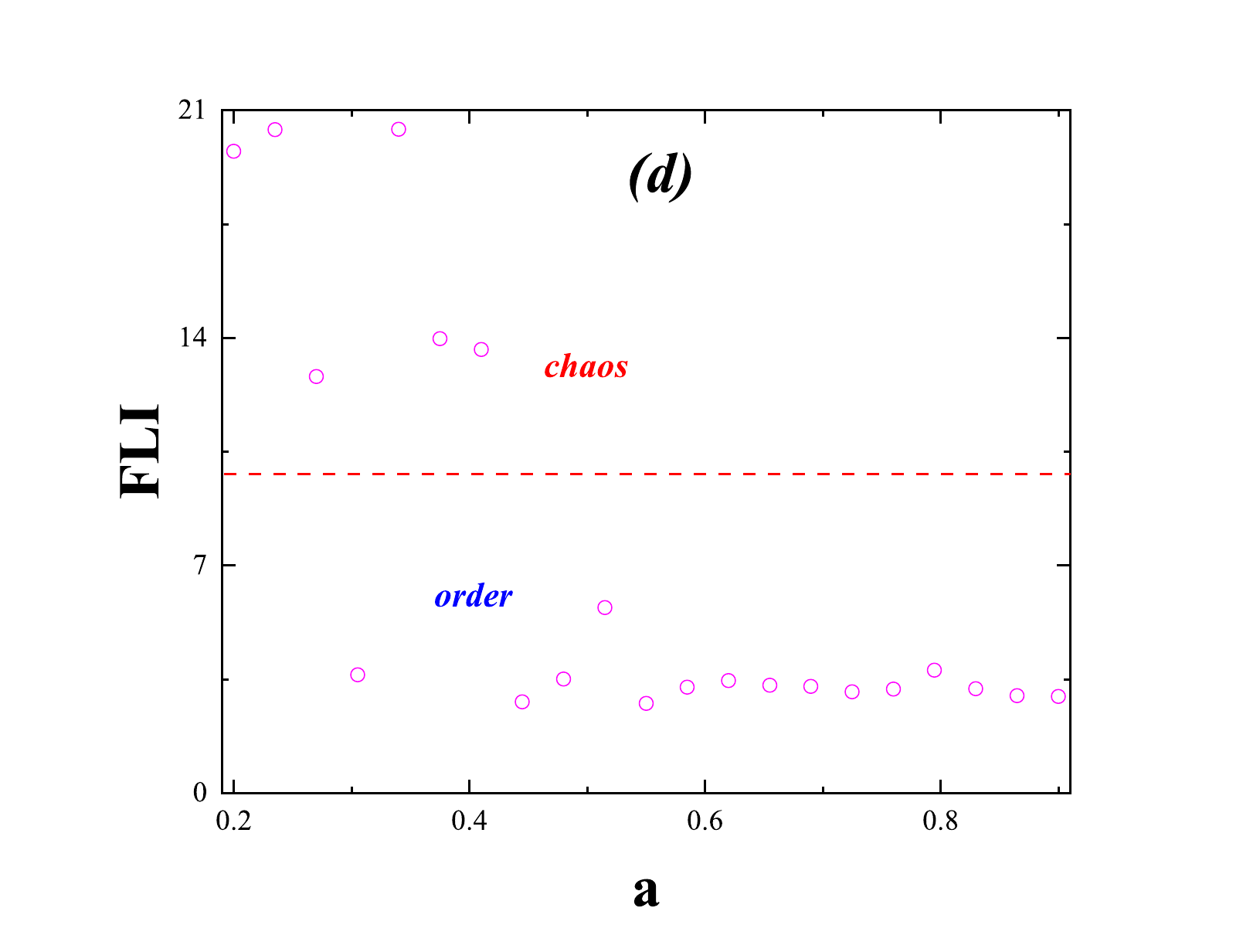} \label{fig:subfig5d}}
		\subfigure{\includegraphics[scale=0.3]{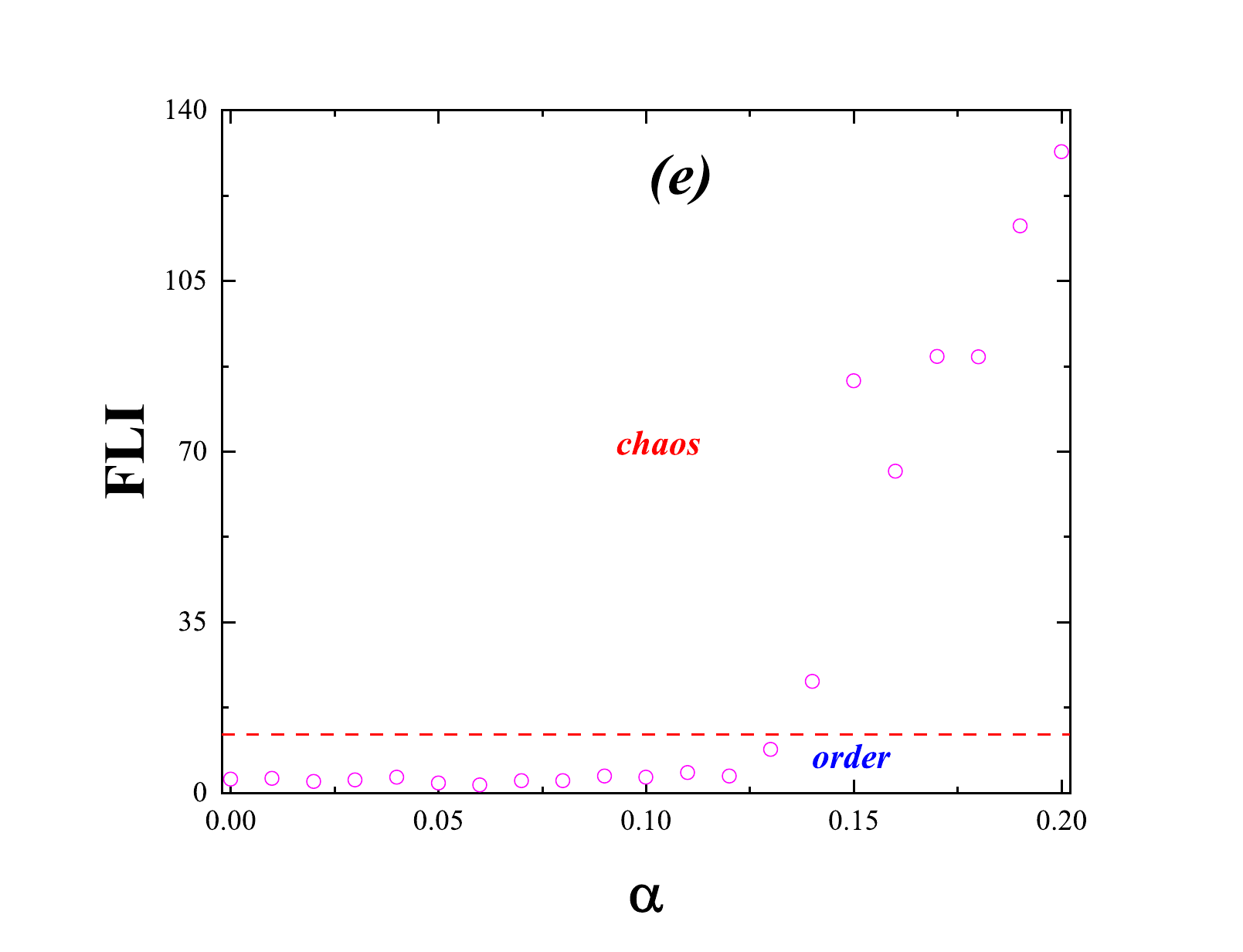} \label{fig:subfig5e}}
		\caption{\label{fig:5} The same as Figure 4, but for FLI.}}
\end{figure*}

In Fig. \ref{fig:5}, the panels (a) and (c) depict the FLIs for 30 orbits, while the panels (b), (d), and (e) show the FLIs for 20 orbits. Here, the parameters are the same as those in Fig. \ref{fig:4}. The $FLIs\ge 30$ indicate chaos, while the $FLIs<30$ indicate order in Fig. \ref{fig:subfig5a}. The change from chaos to order takes place at $E=0.9939$, when $E \ge 0.9939$, the system is mainly in a chaotic state, whereas when $E< 0.9939$, it is mostly in an ordered state. Therefore, as $E$ increases, the system transitions from order to chaos, and the degree of chaos intensifies. In Figs. \ref{fig:subfig5b}-\ref{fig:subfig5e}, $FLI = 10$ is the boundary line between ordered and chaotic regions. When the $FLIs \ge 10$, the system indicates chaos, while the $FLIs< 10$ represent order. From Fig. \ref{fig:subfig5b}, we can see that the transition from chaos to order occurs at $L=4.36$ and $L=4.48$. When $L< 4.36$, the system is in the chaotic region, whereas when $L\ge 4.48$, the system is in the ordered region. Thus, the system transitions from chaos to order as $L$ increases. In Fig. \ref{fig:subfig5c}, the system experiences a transition from order to chaos at $\beta=3.4\times{10^{-4}}$. When $\beta< 3.4\times{10^{-4}}$, the system is mostly in ordered regions. Conversely, when $\beta \ge 3.4\times{10^{-4}}$, the system is predominantly in chaotic regions. Consequently, as the parameter $\beta$ increases, the system transitions from ordered regions to chaotic regions, with the degree of chaos intensifying. In Fig. \ref{fig:subfig5d}, the shift from chaotic to ordered regions happens at $a = 0.305$ and $a = 0.445$. When $a< 0.305$, the system is chaotic, and when $a\ge 0.445$, it is regular. Thus, as $a$ increases, the system transitions from chaos to order. In Fig. \ref{fig:subfig5e}, the shift from chaotic to regular dynamics occurs at $\alpha= 0.14$. For $\alpha< 0.14$, the system is primarily in an ordered area. However, the system is largely in chaotic regimes when $\alpha \ge 0.14$. Therefore, with an increase in $\alpha$, the system transitions from order to chaos, and chaos becomes stronger.

In summary, when $E$, $\beta$, or $\alpha$ grows, the orbits have a transition from order to chaos; when $L$ or $a$ rises, the orbits have a transition from chaos to order. 

\subsection{The impact of two parameters acting simultaneously}

\begin{figure*}[htbp]
	\center{
		\subfigure{\includegraphics[scale=0.25]{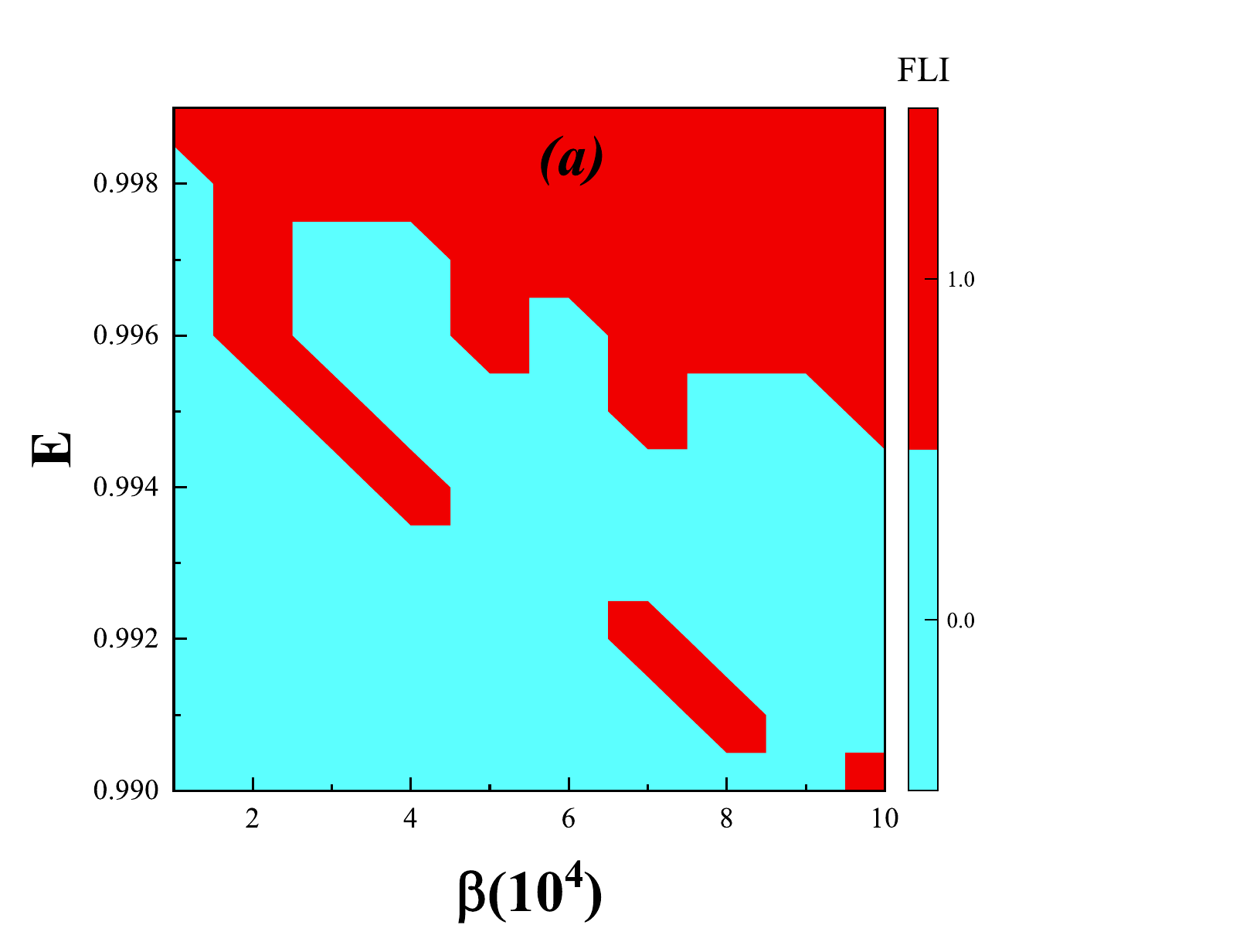} \label{fig:subfig6a}}
		\subfigure{\includegraphics[scale=0.25]{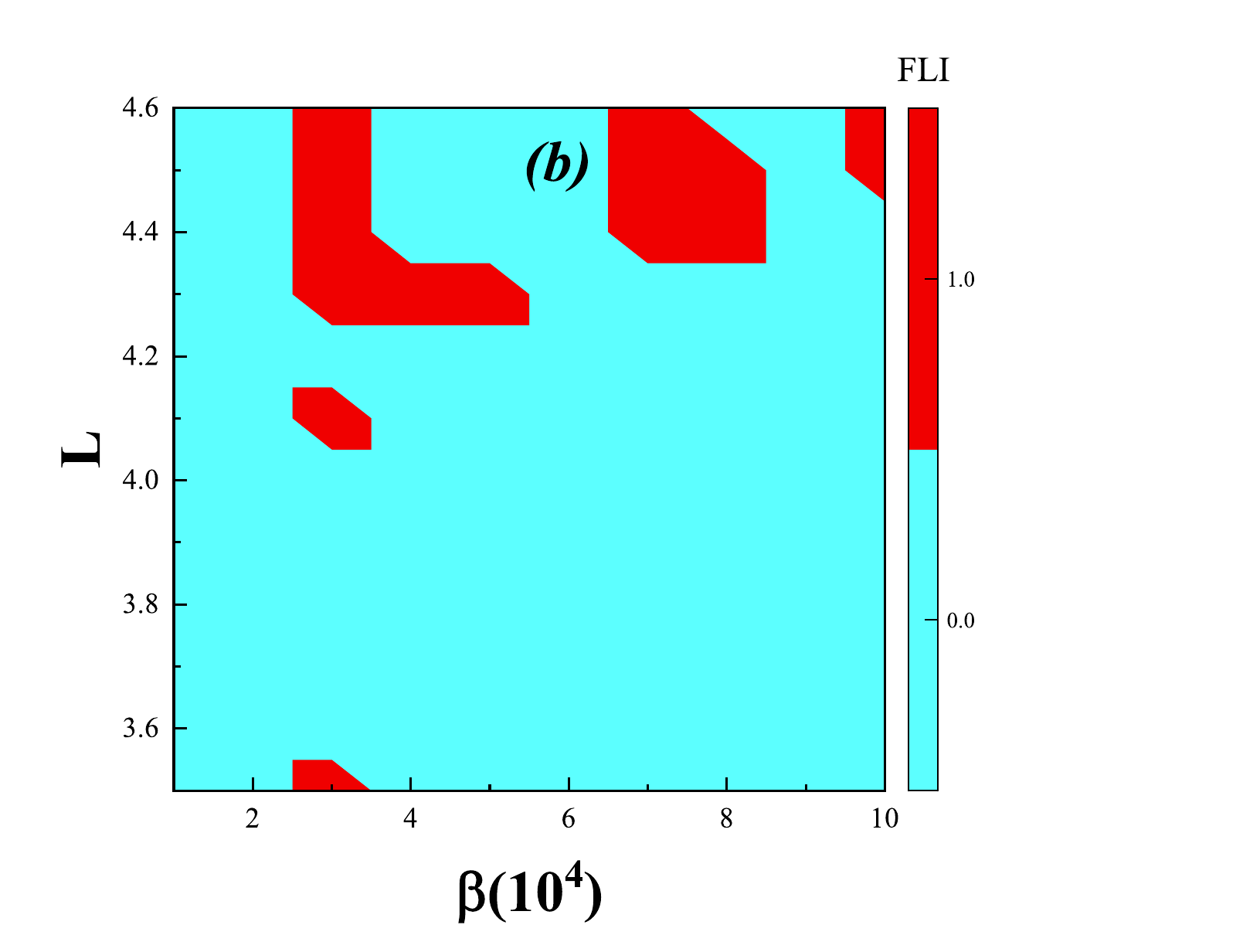} \label{fig:subfig6b}}
		\subfigure{\includegraphics[scale=0.25]{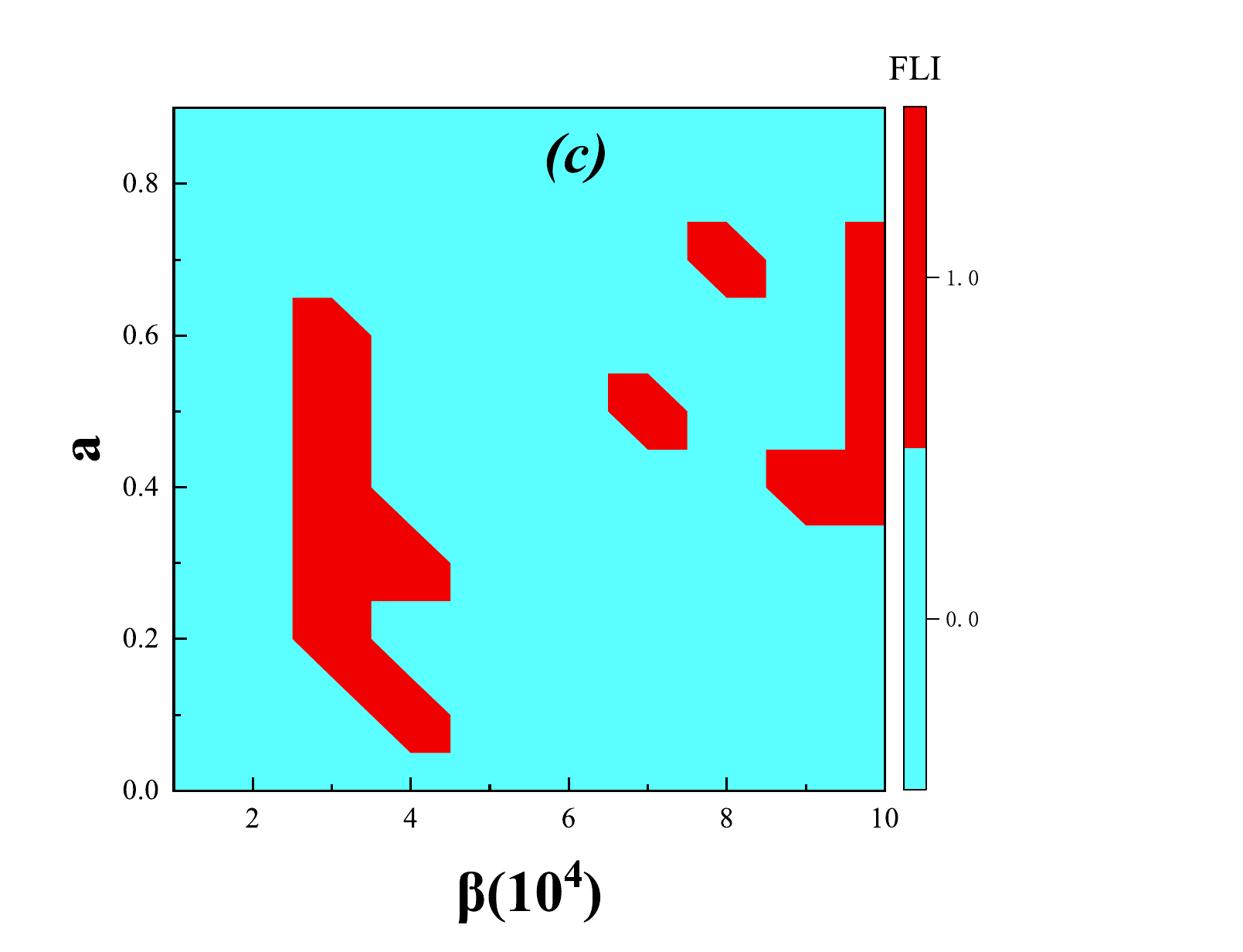} \label{fig:subfig6c}}
		\subfigure{\includegraphics[scale=0.25]{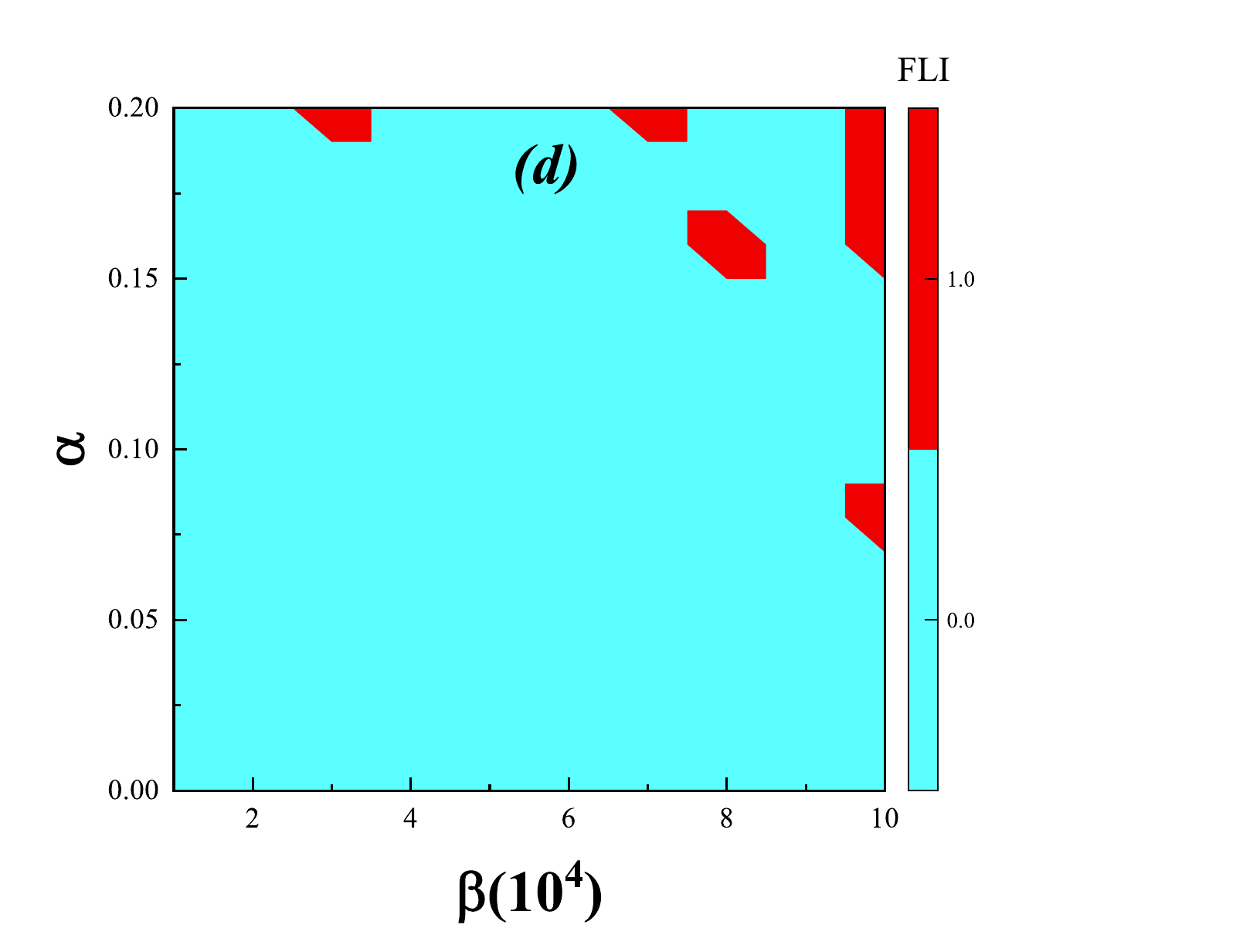} \label{fig:subfig6d}}
		\caption{\label{fig:6} The FLI distribution in two-dimensional parameter spaces when $r=11$. Here, the integration time is $w = {10^7}$, 0 and 1 denote regular and chaos, respectively. The colors in cyan and red represent the regular and chaotic regions, respectively. \textbf{(a)} The parameter space is $\left( {\beta ,E} \right)$, in which $L=4.6$, $a=0.5$ and $\alpha = 0.2$. \textbf{(b)} The parameter space is $\left( {\beta ,L} \right)$, the other parameters are $E=0.995$, $a=0.5$ and $\alpha= 0.2$. \textbf{(c)} The parameter space is $\left( {\beta ,a} \right)$, with other parameters are fixed at $E=0.995$, $L=4.6$ and $\alpha=0.2$. \textbf{(d)} The parameter space is $\left( {\beta ,\alpha} \right)$, and the rest parameters are $E=0.995$, $L=4.6$ and $a=0.5$.}}
\end{figure*}

\begin{figure*}[htbp]
	\center{
		\subfigure{\includegraphics[scale=0.25]{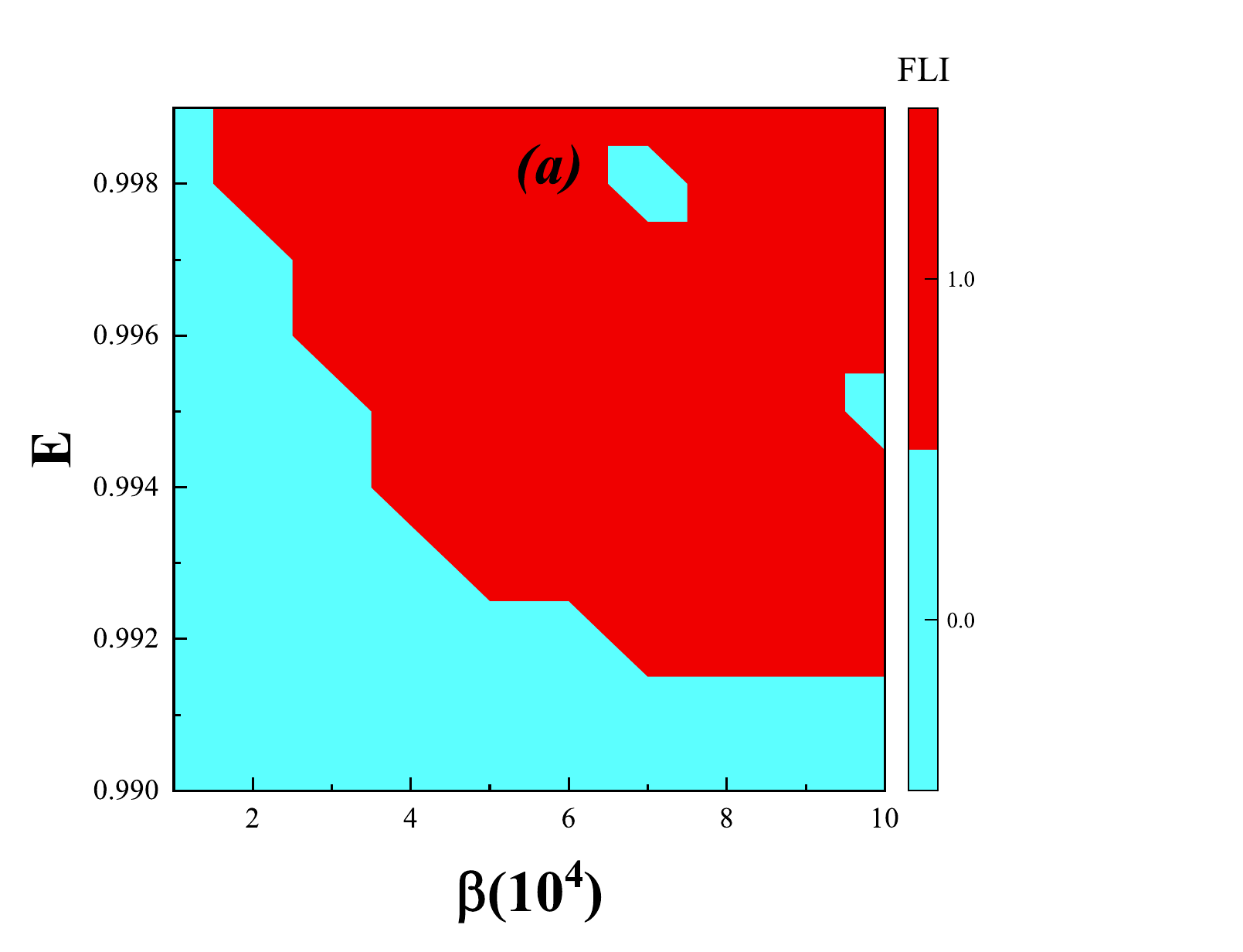} \label{fig:subfig7a}}
		\subfigure{\includegraphics[scale=0.25]{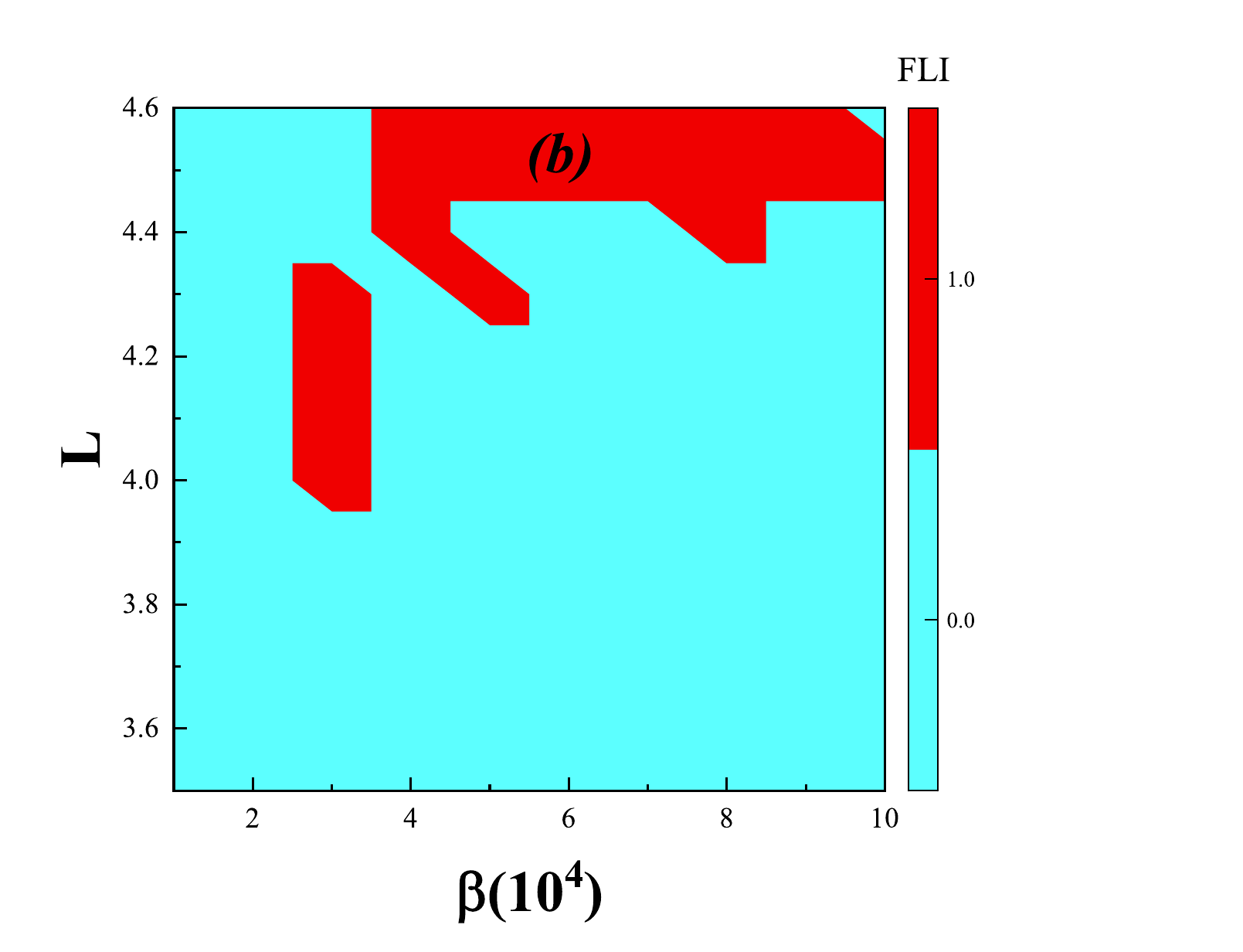} \label{fig:subfig7b}}
		\subfigure{\includegraphics[scale=0.25]{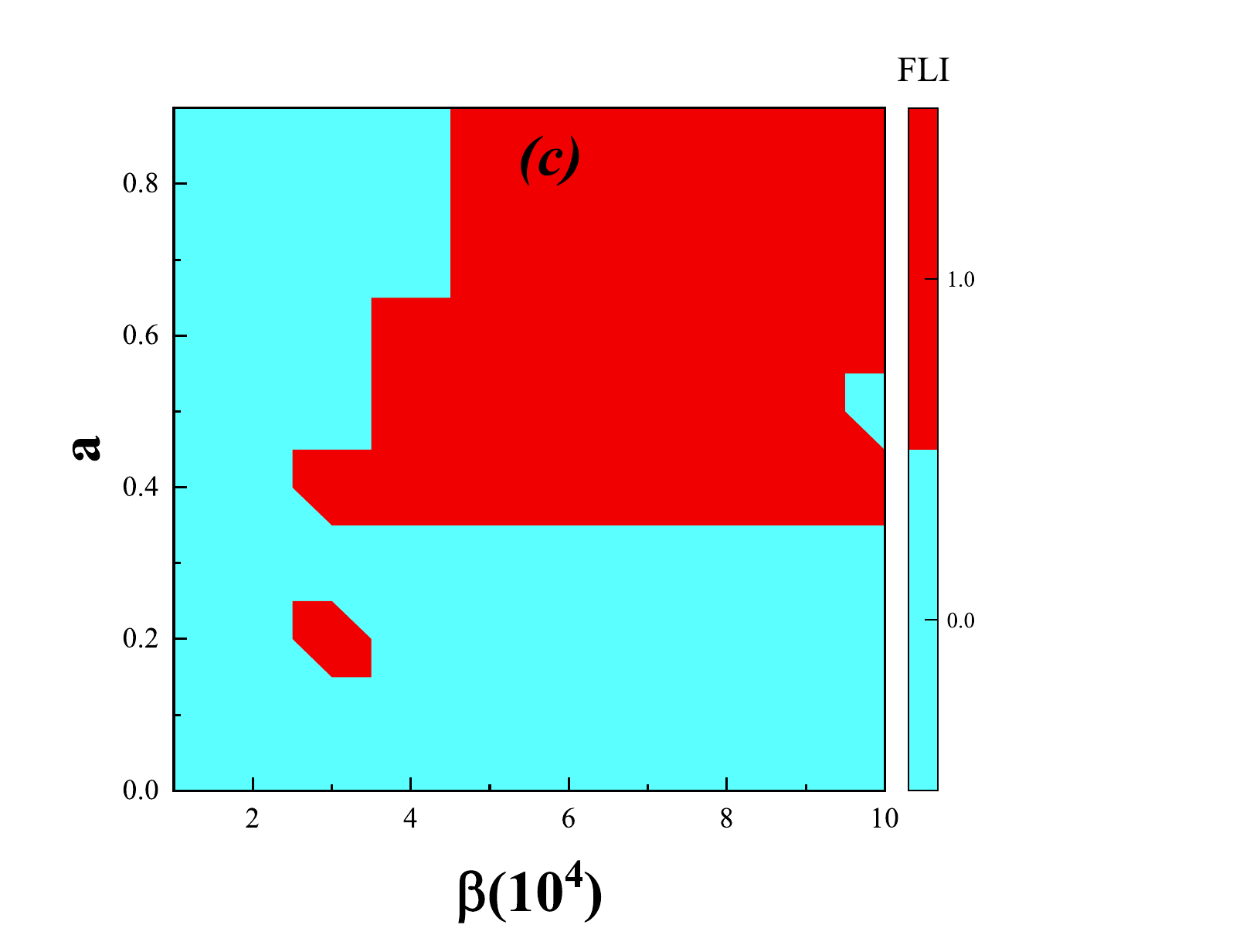} \label{fig:subfig7c}}
		\subfigure{\includegraphics[scale=0.25]{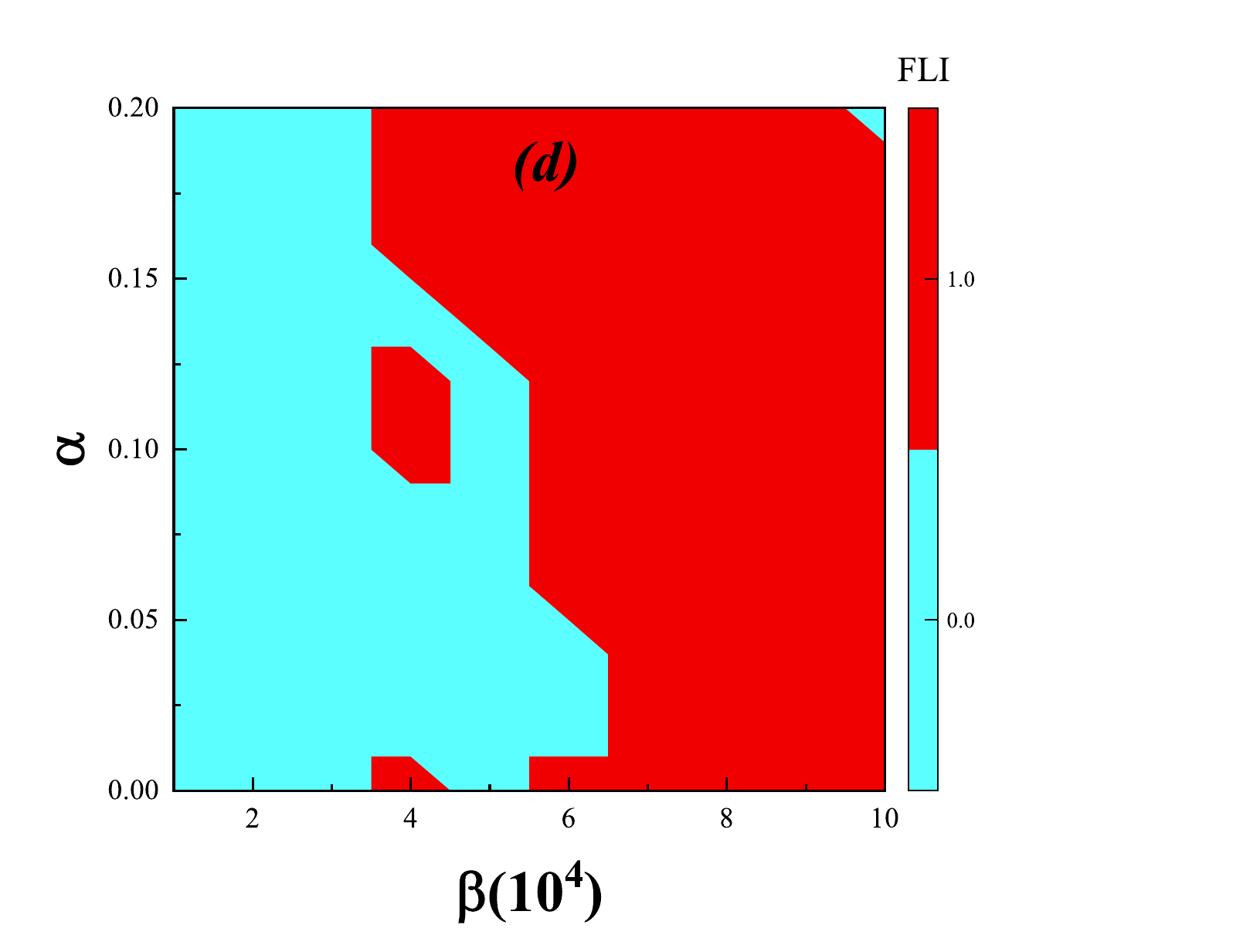} \label{fig:subfig7d}}
		\caption{\label{fig:7} The same as Figure 6, but for $r=110$.}
	}
\end{figure*}

\begin{figure*}[htbp]
	\center{
		\subfigure{\includegraphics[scale=0.25]{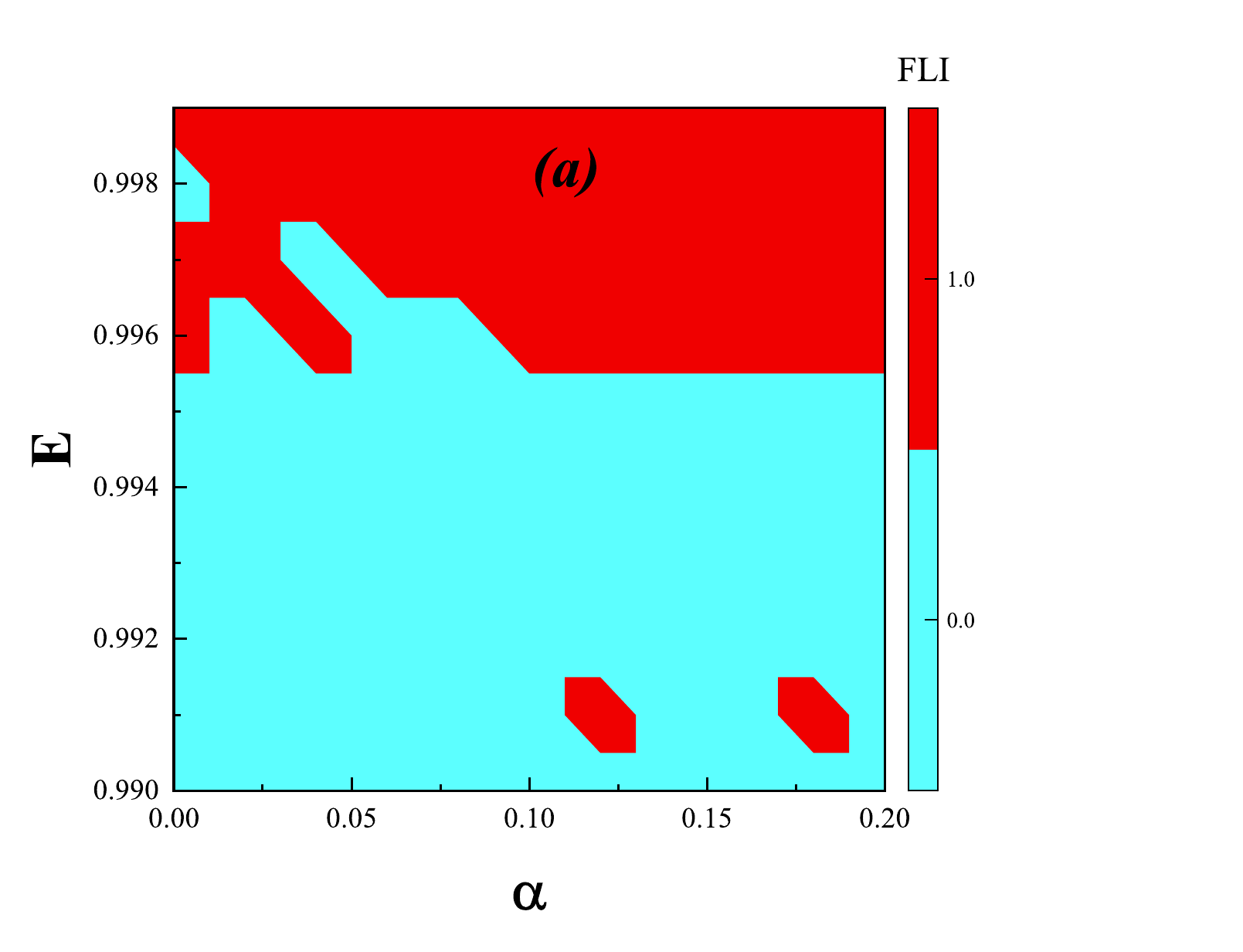} \label{fig:subfig8a}}
		\subfigure{\includegraphics[scale=0.25]{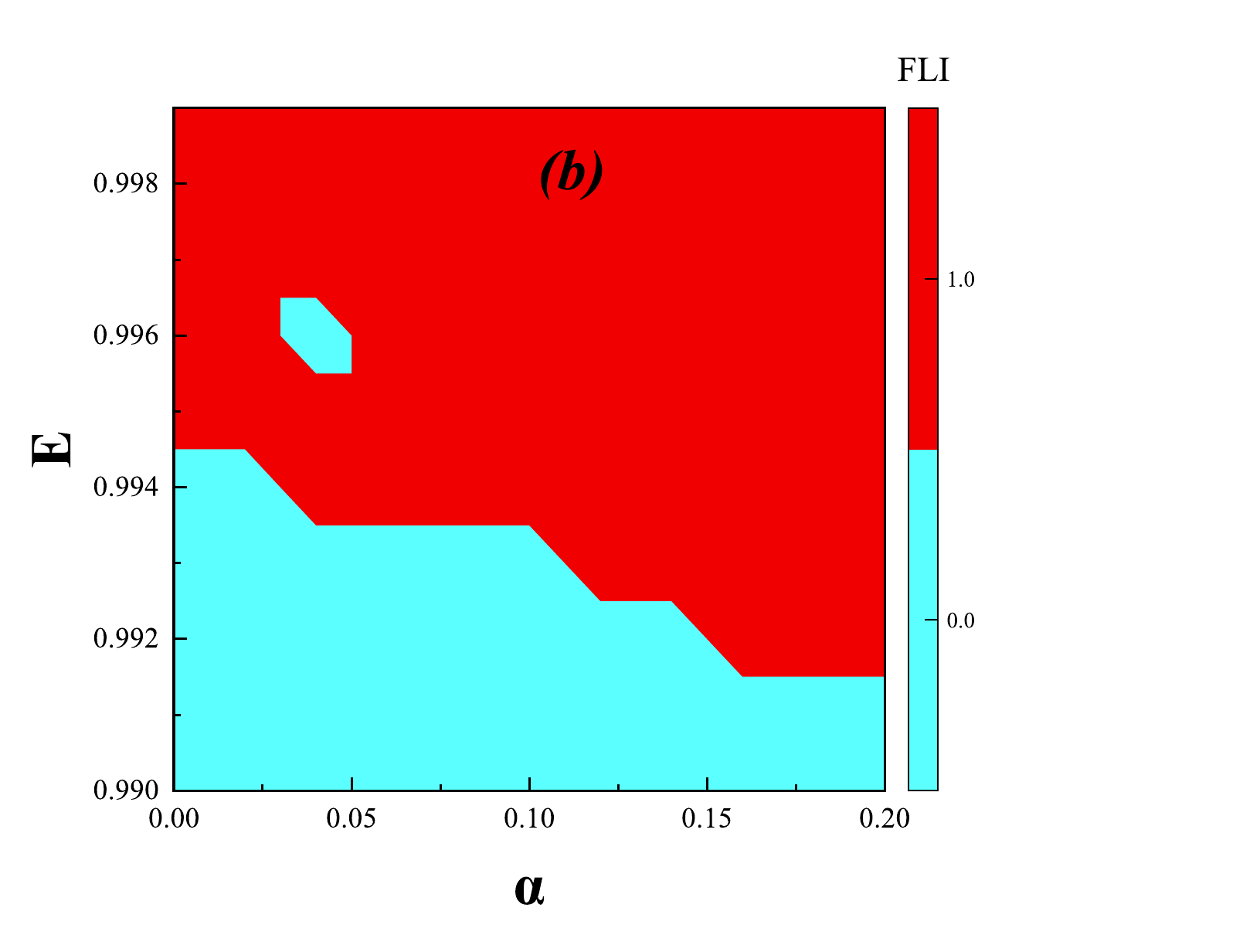} \label{fig:subfig8b}}
		\subfigure{\includegraphics[scale=0.25]{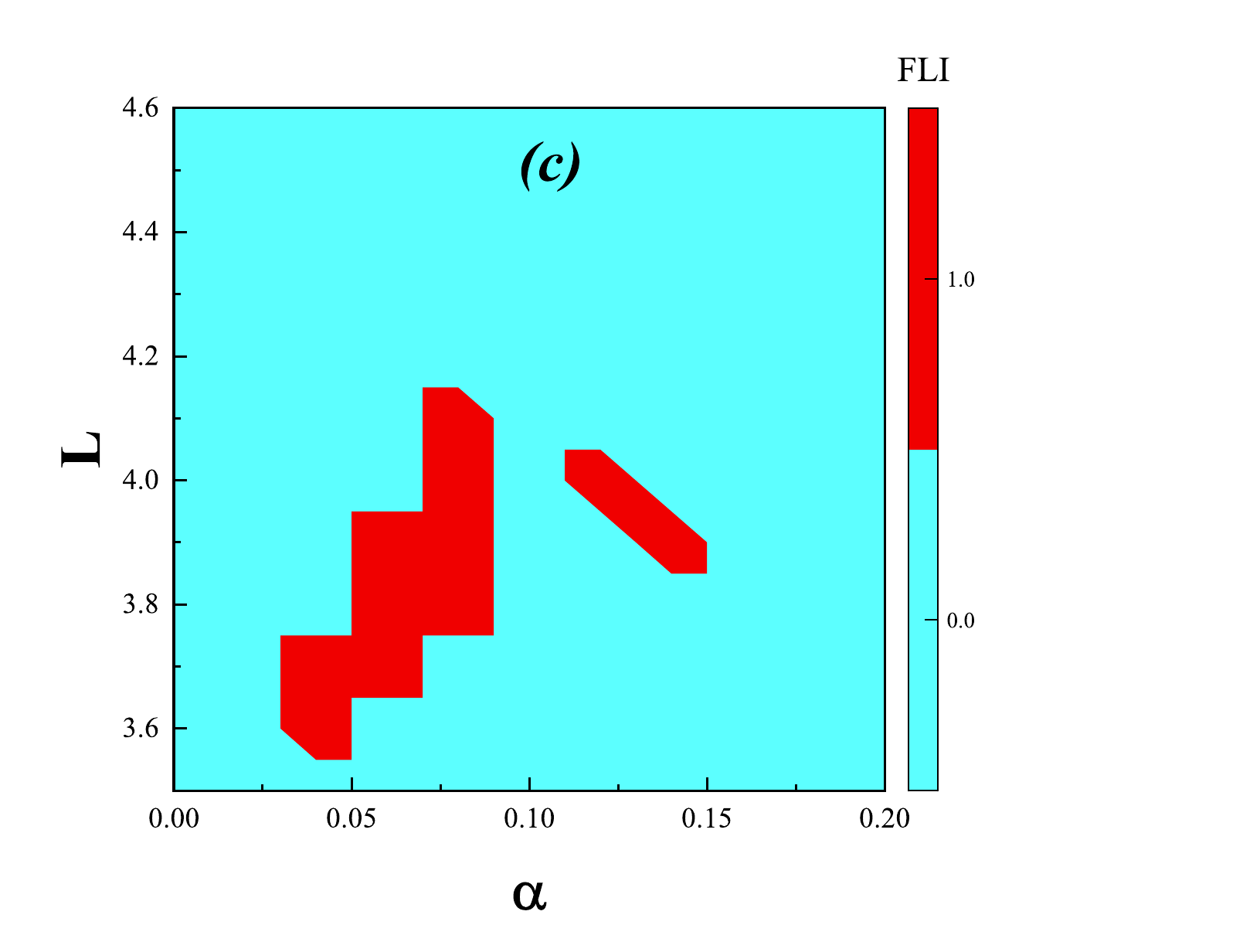} \label{fig:subfig8c}}
		\subfigure{\includegraphics[scale=0.25]{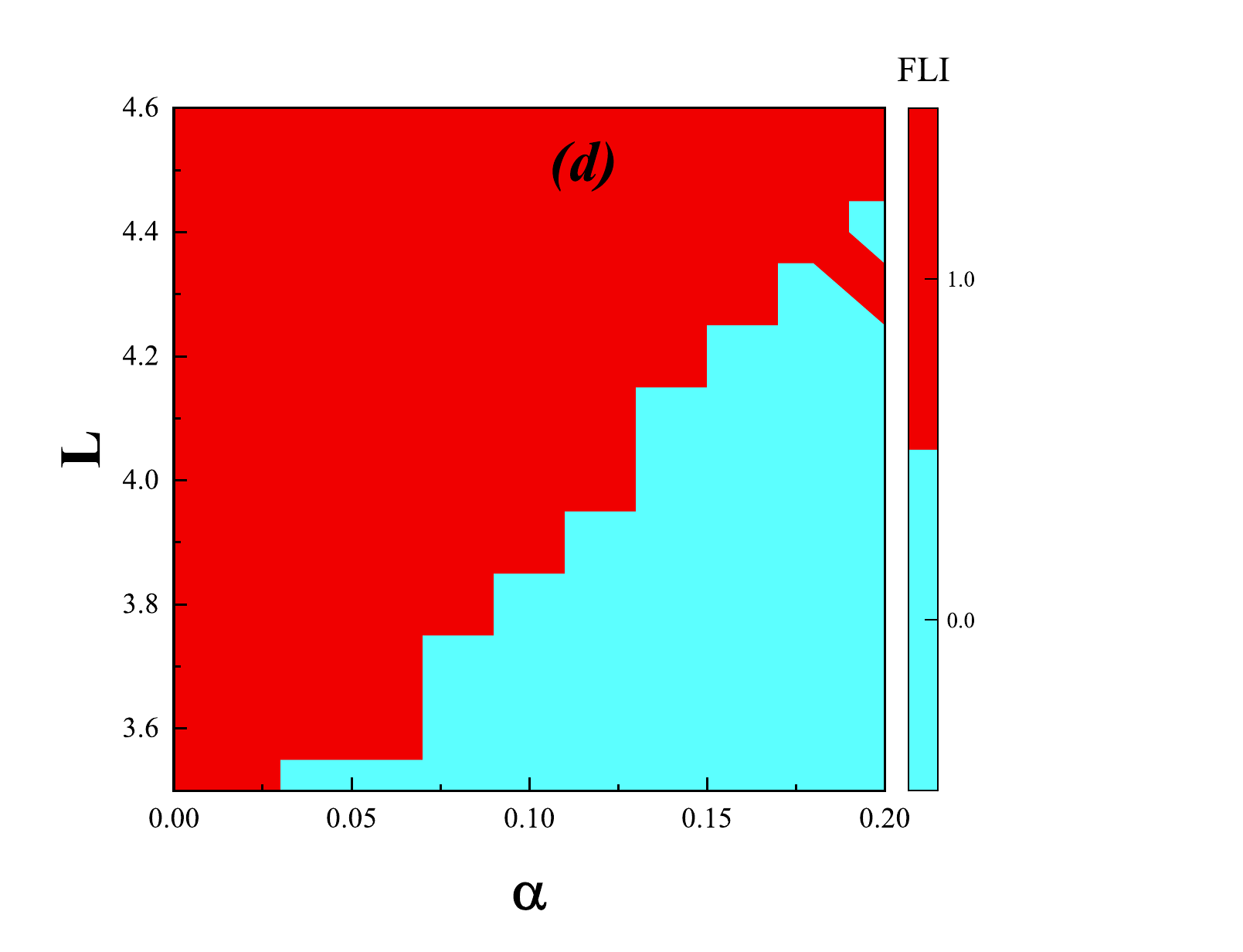} \label{fig:subfig8d}}
		\subfigure{\includegraphics[scale=0.25]{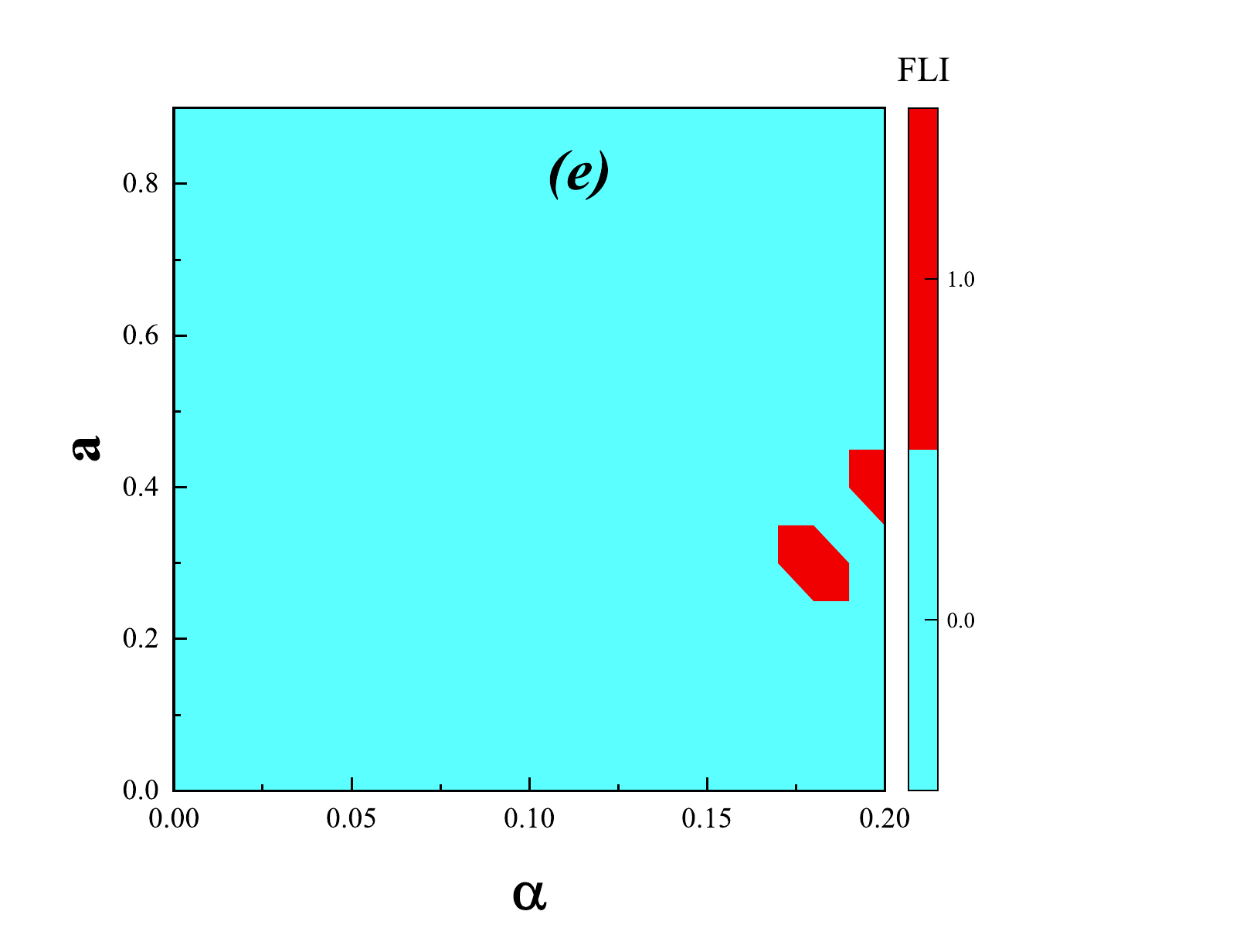} \label{fig:subfig8e}}
		\subfigure{\includegraphics[scale=0.25]{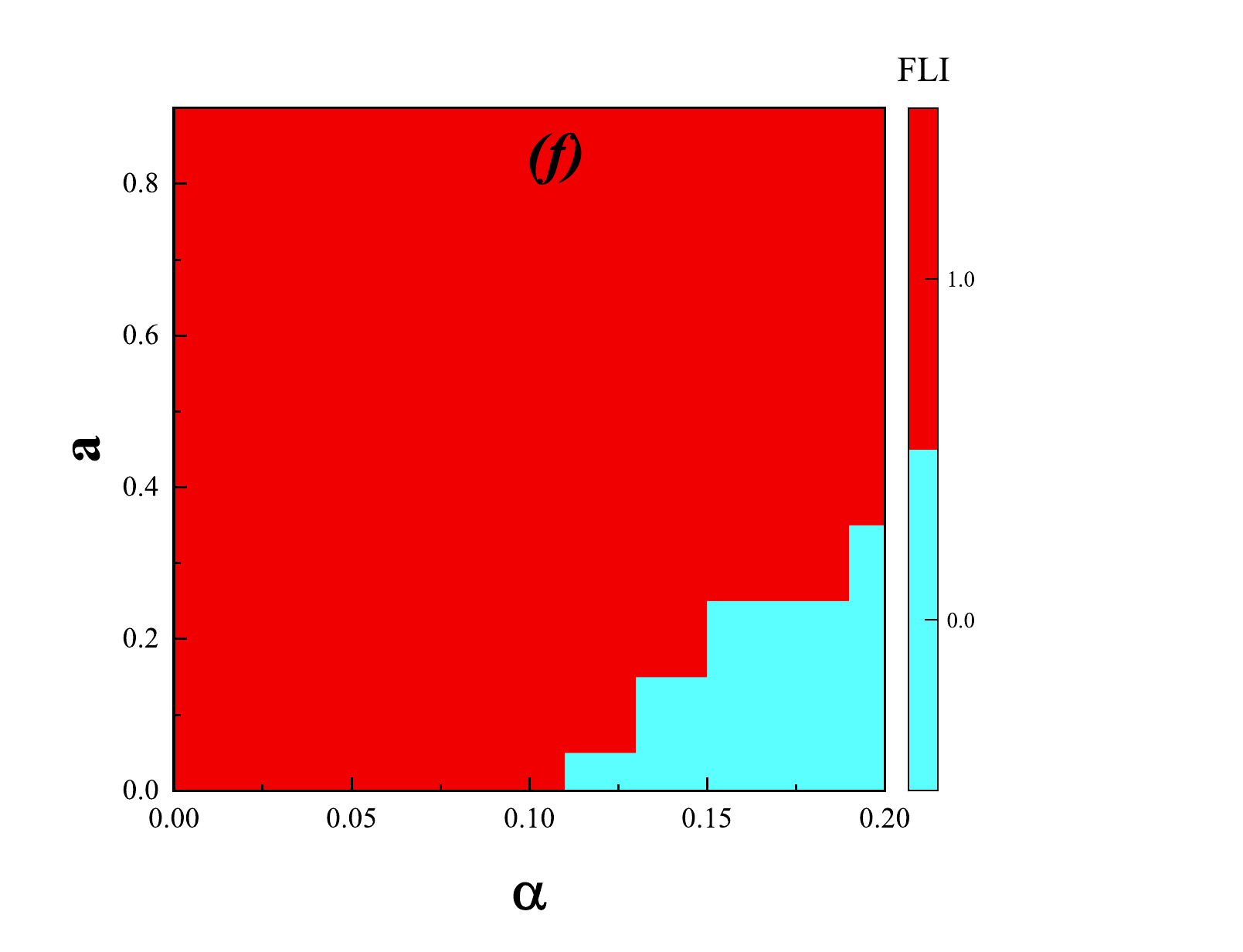} \label{fig:subfig8f}}
		\caption{\label{fig:8} The FLI distribution in two-dimensional parameter spaces. In panels a, c and e, the initial values are $r=11$ and $\theta = \pi/2$, whereas in panels b, d and f, we take $r=110$ and $\theta = \pi/2$ as the initial values. In \textbf{(a)} and \textbf{(b)}, the parameter space is $\left( {\alpha ,E} \right)$, and $L=4.6$, $\beta= 8.9\times{10^{- 4}}$ and $a=0.5$. In \textbf{(c)} and \textbf{(d)}, the parameter space is $\left( {\alpha ,L} \right)$, with $E=0.995$, $\beta= 8.9\times{10^{- 4}}$ and $a=0.5$. In \textbf{(e)} and \textbf{(f)}, the parameter space is $\left( {\alpha ,a} \right)$, and the other parameters are $E=0.995$, $L=4.6$ and $\beta= 8.9\times{10^{- 4}}$.}}
\end{figure*}

In this section, the FLIs corresponding to two parameters are shown in Figs. \ref{fig:6}-\ref{fig:8}. Here, the integration time is $w = {10^7}$, the cyan and red represent the ordered and chaotic regions, respectively. 

In Fig. \ref{fig:subfig6a}, we set $L=4.6$, $a=0.5$ and $\alpha=0.2$, a two-dimensional space $\left( {\beta ,E} \right)$ is given. Through aborative observation, we find that $FLI=8$ is the threshold. In order to clearly see the boundaries between chaos and order in figures, we use 0 and 1 to denote regular and chaotic dynamics, respectively. It is easy to find that when $E$ and $\beta$ are small, the system is in order. However, with the increase of $E$ and $\beta$, the red area expands, which indicates a rise in chaos. There are two locally chaotic regions, one is $\beta  \in [6.5 \times {10^{ - 4}}, 8.5\times {10^{ - 4}}]$ and $E \in [0.9905,0.9925]$, the other is $\beta  \in [9.8 \times {10^{ - 4}}, 1\times {10^{ - 3}}]$ and $E \in [0.990,0.9905]$. In case of $\left( {\beta ,L} \right)$, we set $E=0.995$, $a=0.5$ and $\alpha=0.2$. The results are given in Fig. \ref{fig:subfig6b}, here, the threshold is $FLI=10$. Compared to Fig. \ref{fig:subfig6a}, Fig. \ref{fig:subfig6b} does not have a complete chaotic region, the most areas are orderly, and there are only five local chaotic regions. As stated before \cite{2017RCD....22...54G}, this situation is allowed. When $E=0.995$, $L=4.6$ and $\alpha=0.2$, the parameter space $\left( {\beta ,a} \right)$ is considered in Fig. \ref{fig:subfig6c}. Here, $FLI=12$ is used as the threshold. Although most of the regions are orderly, there are still four chaotic regions. It can be seen that as $\beta$ increases and $a$ decreases, chaos gradually emerges and becomes more intense, which is the same as in Figs. \ref{fig:subfig3c}-\ref{fig:subfig3d}. Finally, the impact of the parameter space $\left( {\beta ,\alpha} \right)$ on the dynamics is analyzed in Fig. \ref{fig:subfig6d}. Here, $E=0.995$, $L=4.6$ and $a=0.5$, the threshold is $FLI=20$. There are five local chaotic regions, it reveals that the simultaneous increase of $\beta$ and $\alpha$ has little effect on the chaotic behavior of the system.

The same as in Fig. \ref{fig:6}, but the initial position of the particle is $r=110$ in Fig. \ref{fig:7}. $FLI = 30$ is selected as the boundary between order and chaos, and the two-dimensional parameter space $\left( {\beta ,E}\right)$ is shown in Fig. \ref{fig:subfig7a}. It can be observed that the system mostly remains in ordered regions when both $\beta$ and $E$ are small. As $\beta$ and $E$ gradually increase, the system has a transition from ordered to chaotic regions. Chaos starts to appear when $E=0.9915$. When $E < 0.9915$, the system is orderly regardless of how $\beta$ changes. When $E \ge 0.9915$, the red area gradually expands as $E$ and $\beta$ increase, which indicates an increase in chaos. Besides, there are two local ordered regions. The first is $\beta  \in [6.5 \times {10^{ - 4}},7.5\times {10^{ - 4}}]$ and $E \in [0.9975,0.9985]$, while the second is $\beta  \in [9.5 \times {10^{ - 4}},1\times {10^{ - 3}}]$ and $E \in [0.9945,0.9955]$. In Figs. \ref{fig:subfig7b}-\ref{fig:subfig7d}, $FLI = 10$ is the threshold. The two-dimensional parameter space $\left( {\beta ,L} \right)$ is presented in Fig. \ref{fig:subfig7b}. The system is regular when $L<3.95$, while chaos begins to appear when $L\ge 3.95$. Unlike in Fig. \ref{fig:subfig7a}, there are only two local chaotic regions, corresponding to $\beta \in [2.5 \times {10^{ - 4}},3.5\times {10^{ - 4}}]$, $L\in [3.95,4.35]$ and $\beta \in [3.5 \times {10^{ - 4}},1\times {10^{ - 3}}]$, $L\in [4.25,4.6]$. As $\beta$ increases, the red region expands, indicating a rise in chaos. However, chaos also appears when $L$ increases, which is contrary to the conclusions from Fig. \ref{fig:subfig5b}. The reason for this is that when $\beta $ and $L$ are considered simultaneously, the system is primarily influenced by $\beta $. The two-dimensional parameter space $\left( {\beta ,a} \right)$ in Fig. \ref{fig:subfig7c} reveals that the system is almost in order, such as $a<0.15$, $0.25<a<0.35$ or $\beta< 2.5\times{10^{-4}}$. The shift from order to chaos only occurs at $a=0.15$ and $a=0.35$. When $a<0.35$, there is a local chaotic region when $\beta  \in [2.5 \times {10^{ - 4}},3.5\times {10^{ - 4}}]$ and $a \in [0.15,0.25]$. For $a\ge 0.35$, it is apparent that increasing $\beta$ intensifies the chaos. In contrast, as $a$ increases, the ordered region expands and chaos weakens. As shown in Fig. \ref{fig:subfig7d}, the system is ordered when $\beta< 3.5\times{10^{-4}}$, with chaos beginning at $\beta= 3.5\times{10^{-4}}$. Moreover, for $\beta \ge 3.5\times{10^{-4}}$, increasing $\alpha$ expands the red regions, which corresponds to a stronger chaotic behavior.

In the left panels of Fig. \ref{fig:8}, the initial values are $r=11$ and $\theta = \pi/2$, whereas in the right panels in Fig. \ref{fig:8}, the initial values are $r=110$ and $\theta = \pi/2$. The parameter space $\left( {\alpha ,E} \right)$ is discussed in Figs. \ref{fig:subfig8a}-\ref{fig:subfig8b}. Here, $L=4.6$, $\beta= 8.9\times {10^{- 4}}$, $a=0.5$. In Fig. \ref{fig:subfig8a}, $FLI = 10$ is adopted as the threshold; while in Fig. \ref{fig:subfig8b}, the threshold is $FLI = 20$. In Fig. \ref{fig:subfig8a}, the system generates chaos at $E = 0.9955$. When $E< 0.9955$, the system is mostly in order, except in two intervals where local chaos exists. That is, for $E \in [0.9905,0.9915]$, there are two regions where local chaos exists approximately within $\alpha \in [0.11,0.13]$ and $\alpha \in [0.17,0.19]$. When $E\ge 0.9955$, chaos intensifies as $\alpha$ increases. However, the difference from Fig. \ref{fig:subfig8a} is that chaos appears in Fig. \ref{fig:subfig8b} when $E = 0.9915$. When $E<0.9915$, no matter how $\beta$ varies, the system is in order all the time. The results show that chaos gets enhanced with the simultaneous increase of both $E$ and $\alpha$. Figs. \ref{fig:subfig8c}-\ref{fig:subfig8d} focus on the parameter space $\left( {\alpha ,L} \right)$, with the rest parameters are $E=0.995$, $\beta= 8.9\times {10^{- 4}}$ and $a=0.5$. Here, $FLI = 20$ and $FLI = 50$ are used as the thresholds. In Fig. \ref{fig:subfig8c}, the system is mainly in order, but there are two local chaotic regions, one is $\alpha \in [0.03,0.09]$ and $L \in [3.55,4.15]$, the other is $\alpha \in [0.11,0.15]$ and $L \in [3.85,4.05]$. However, a large area of chaotic regions can be observed in Fig. \ref{fig:subfig8d}. When $\alpha< 0.03$ or $L \ge 4.45$, regardless of how the other parameter changes, the system is in a chaotic state. As for the case where $\alpha \in [0.03,0.2]$ and $L \in [3.5,4.45]$, with the simultaneous increase of  $L$ and $\alpha$ , the ordered region expands and chaos diminishes. These indicate that when $L$ and $\alpha$ change simultaneously, the system is mainly influenced by the parameter $L$. Finally, we analyze the parameter space $\left( {\alpha ,a} \right)$ in Figs. \ref{fig:subfig8e}-\ref{fig:subfig8f}, where $E=0.995$, $L=4.6$ and $\beta= 8.9\times {10^{- 4}}$. $FLI = 20$ and $FLI = 50$ are served as the thresholds. There are merely two tiny chaotic regions in Fig. \ref{fig:subfig8e}. The first occurs where $\alpha  \in [0.17,0.19]$ and $a \in [0.25,0.35]$, while the second exists when $\alpha \in [0.19,0.2]$ and $a \in [0.35,0.45]$. In Fig. \ref{fig:subfig8f}, the system is chaotic when $\alpha< 0.11$ or $a > 0.35$. For $\alpha  \in [0.11,0.2]$ and $a \in [0,0.35]$, as both $a$ and $\alpha$ increase simultaneously, the ordered region expands and chaos weakens. The results clearly demonstrate that the role of $a$ is more significant than $\alpha$.

From the chaotic dynamics of the charged particle, we find that different parameters have different impacts on the motion of the particle. When one of the parameters is taken into account, for $E$, $\beta$, or $\alpha$ increases, the system shifts from order to chaos. But for $L$ or $a$ increases, the opposite occurs. However, When any two of  them are considered simultaneously, the situation is quite complicated. For example, if $\beta$ and $L$ cooperate with each other, $\beta$ has a greater impact on the system than $L$. When $L$ and $\alpha$ function at the same time, $L$ predominates. Similarly, when $a$ and $\alpha$ interact with each other, $a$ is more important. At last, we find that the motion of the particle has a relatively small dependence on the MOG parameter $\alpha$.

\section{Conclusions} \label{section5}

The chaotic motion of the charged test particle in Kerr-MOG black hole is discussed in this paper. Due to the non-integrable of Kerr-MOG black hole in the presence of an external magnetic field, we adopt a time transformation function to convert the original Hamiltonian. The new Hamiltonian can be decomposed into five integrable parts to construct explicit symplectic algorithms. After that, three kinds of explicit symplectic methods are proposed. Numerical experiments show that $PR{K_6}4$ is the best method and has significant advantages such as the stability and accuracy in long-term numerical integration. Thus, $PR{K_6}4$ is used to explore the chaotic motion of a charged particle around Kerr-MOG black hole. With the help of the Poincaré sections and the fast Lyapunov indicators, effects of $E$, $L$, $a$, $\alpha$ and $\beta$ on chaos are discussed in detail. The results show that the chaotic area increases as $E$, $\beta$ or $\alpha$ increases, but $a$ and $L$ are not like that. When multiple parameters are applied simultaneously, we find that $a$ and $L$ play a major role. These findings are useful for understanding dynamics of the charged particles in modified gravity theories, and they provide a new perspective to analyze the chaotic motion in strong gravitational field. It should be emphasized that the accuracy and stability of numerical algorithms are crucial for discussing the chaotic motion of particles in strong gravitational fields. This is because only algorithms with high accuracy and good stability can avoid the pseudo chaos \cite{Ma:2019ewq}. Therefore, the time transformation method and Hamiltonian decomposition method are useful. 

\section*{Acknowledgments}

This work is supported by the Natural Science Foundation of China under
Grant Nos. 12073008, 12473074, 11703005.

\textbf{Data Availability Statement}: This manuscript has no
associated data or the data will not be deposited. [Author's
comment: All of the data are shown as the figures and formula. No
other associated data.]

\textbf{Code Availability Statement}: Code/software will be made
available on reasonable request. [Author's comment: The
code/software generated during and/or analysed during the current
study is available from the first author Zhenmeng Xu on reasonable
request.]

\end{document}